CCE Theses and Dissertations

College of Computing and Engineering

2020

# Predictive Accuracy of Recommender Algorithms

William Blake Noffsinger



Predictive accuracy of recommender algorithms.

by

William B. Noffsinger

A dissertation report submitted in partial fulfillment of the requirements
for the degree of Doctor of Philosophy
in
Information Systems

College of Computing and Engineering
Nova Southeastern University

2020



**We hereby certify that this dissertation, submitted by William Noffsinger conforms to acceptable standards and is fully adequate in scope and quality to fulfill the dissertation requirements for the degree of Doctor of Philosophy.**

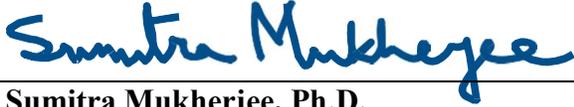

_______________________________          November 18, 2020
**Sumitra Mukherjee, Ph.D.**                        **Date**
**Chairperson of Dissertation Committee**

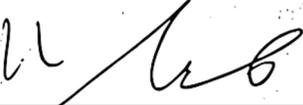

_______________________________          _______________________
**Michael J. Laszlo, Ph.D.**                         Nov. 18, 2020
**Dissertation Committee Member**                   **Date**

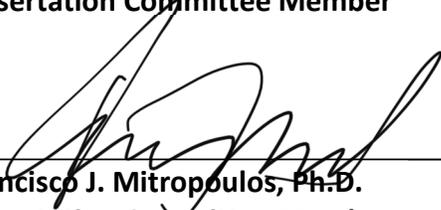

_______________________________          _______________________
**Francisco J. Mitropoulos, Ph.D.**                 Nov 18, 2020
**Dissertation Committee Member**                   **Date**

**Approved:**

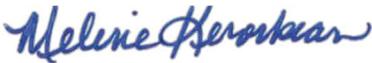

_______________________________          November 18, 2020
**Meline Kevorkian, Ed.D.**                         **Date**
**Dean, College of Computing and Engineering**

**College of Computing and
Engineering Nova Southeastern
University**

**2020**



An Abstract of a Dissertation Report Submitted to Nova Southeastern University in Partial Fulfillment of the Requirements for the Degree of Doctor of Philosophy

# Predictive Accuracy of Recommender Algorithms

By
William Noffsinger
2020


Recommender systems present a customized list of items based upon user or item characteristics with the objective of reducing a large number of possible choices to a smaller ranked set most likely to appeal to the user. A variety of algorithms for recommender systems have been developed and refined including applications of deep learning neural networks. Recent research reports point to a need to perform carefully controlled experiments to gain insights about the relative accuracy of different recommender algorithms, because studies evaluating different methods have not used a common set of benchmark data sets, baseline models, and evaluation metrics.

The dissertation used publicly available sources of ratings data with a suite of three conventional recommender algorithms and two deep learning (DL) algorithms in controlled experiments to assess their comparative accuracy. Results for the non-DL algorithms conformed well to published results and benchmarks. The two DL algorithms did not perform as well and illuminated known challenges implementing DL recommender algorithms as reported in the literature. Model overfitting is discussed as a potential explanation for the weaker performance of the DL algorithms and several regularization strategies are reviewed as possible approaches to improve predictive error. Findings justify the need for further research in the use of deep learning models for recommender systems.




## Acknowledgements

There are several people who supported and encouraged me during my dissertation and Doctoral study at Nova Southeastern University. I want to acknowledge the support of my committee chairperson Dr. Sumitra Mukherjee who guided me throughout this process and critically reviewed my work at each stage, from conception to completion. In addition, Dr. Mukherjee offered helpful encouragement and suggestions during the duration of my Doctoral course of study. I also wish to thank my other committee members, Dr. Michael Laszlo and Dr. Francisco Mitropoulos for their contributions.

My wife, Dr. Kathy Noffsinger, offered unfailing support and optimism for the duration of my graduate work and the dissertation project. I could always count on Kathy for her devotion and an understanding of the commitment required for Doctoral study.



# Table of Contents









**5.    Conclusions, Implications, Recommendations, and Summary 104**



**References 117**



# List of Figures





# List of Tables





# Chapter 1

# Introduction

Developments in e-commerce and the growing availability of consumer information on

the World Wide Web have contributed to an ever-increasing volume of information about items

as well as products (Chen, H., Chiang, R.H. and Storey, V., 2012; Fan, W., Ma, Y., Yin, D.,

Wang, J., Tang, J. and Li, Q., 2019). When shopping on e-commerce sites or making

entertainment selections, users are faced with a large volume of choices. Search facilities assist

in making these large amounts of information manageable but require the user to initiate and

guide the search process. Item selection can be greatly facilitated by recommender systems

(Jannach, D., Zanker, M., Felfernig, A. and Friedrich, G., 2011; Shalom, O.S., Jannach, D., and

Guy, L., 2019). Recommender systems (RS) present to the user a personalized list of candidate

items for selection and benefit both the consumer, by identifying selections, and the provider, by

increasing the chance a user will make selections of the provider's product or service. In 2009

Netflix offered a significant cash prize to the individual or group able to develop a collaborative

filtering recommender algorithm capable of a 10% accuracy improvement over their in-house

algorithm. The prize was won by a group using a hybrid recommender combining singular vector

decomposition (SVD) and a restricted Boltzmann machine (RBM)[1] neural network.

A large body of recent work in recommender systems has used deep learning neural

network models (Zhang, S., Yao, L. and Tay, Y., 2019). However, an examination of this

literature has revealed methodologic shortcomings in the published research (Yang, W., Lu, K.,

Yang, P. and Lin, J., 2019; Dacrema, M., Cremonesi, P and Jannach, D., 2019). The

investigation proposed herein is intended to address some of those shortcomings by proposing highly controlled and carefully designed studies. First, we will briefly review the recommender problem applied to collaborative filtering, the most common type of recommender analysis.

---

[1] See http://netflixprize.com

**The Recommender problem and an illustration**

Recommender analyses usually begin with three sources of information: (1) data about products or items, such as movies, with specified features such as genre, (2) data about users of the products with demographic or other characteristic features, and (3) rating data with ratings or rankings assigned to the products by the users.

We will illustrate the basic idea of collaborative filtering (CF) with an example. Figure 1 shows an example user-item rating matrix produced from user and item input data, for simplicity we use a Boolean "Like/Dislike" rating scale. The task is to predict whether Alice will enjoy the movie *Scarface* or not. In the table we see that the users *User3* and *User4* are similar to Alice with regard to the provided ratings, note this is not a perfect concordance but reflects similarity. Similar users are often referred to as peer users or nearest neighbors. Since both users provided a dislike statement for the target movie *Scarface*, a collaborative filtering technique will predict that Alice will also dislike the movie. Therefore, the movie *Scarface* will not be recommended to Alice. In a typical CF analysis, the number of movies and ratings will be larger, enabling a set of similarities with degrees of strength to be calculated which may then be ranked to produce a Top-N list. In this way user-based collaborative filtering techniques exploit the user preferences provided by a community of raters. This method does not consider the content of the rated movies, but only considers the similarity of user ratings on movies.



| User-based Filtering | | | | |
|---|---|---|---|---|
| **Person** | **Movie** | | | |
| | *Heat* | *Scarface* | *Amelie* | *Eat Pray Love* |
| **Alice** | dislike | ? | like | like |
| **User 2** | like | | dislike | dislike |
| **User 3** | dislike | dislike | like | |
| **User 4** | | dislike | like | like |

*Figure 1.* Example user-item rating matrix

The collaborative filtering analysis of user similarities identified the top 10 (Top-N) movies to recommend for a target user. The accuracy of the recommendation may be evaluated with metrics for root mean square error (RMSE), mean absolute error (MAE), or other measures such as receiver operating characteristic (ROC) using area-under-the-curve (AUC) analysis. In addition, the analysis can report measures of user coverage, diversity of recommendations, and novelty.

In contrast, Item-based collaborative filtering methods attempt to exploit information about the content and characteristics of rated items to develop a recommendation. An item-based CF filtering algorithm will determine similarity by measuring the degree of association between movies based upon the number of shared genre keywords. When a target user has a history of selecting movies with particular genres or has identified a preference for particular genre descriptors, movies can be recommended with rankings based upon similar descriptors.



| Content-Based Filtering | | | |
|---|---|---|---|
| *Movie Title* | Plot Keywords | | |
| *Heat* | detective | **criminal** | thief | **gangster** |
| *Scarface* | **gangster** | **criminal** | drugs | cocaine |
| *Amelie* | **love** | waitress | Garden gnome | happiness |
| *Eat Pray Love* | Divorce | India | **love** | Inner peace |

*Figure 2.* Example item-item rating matrix

In Figure 2 we see genre correspondence between *Heat* and *Scarface*, each with two shared
genre tags, and less so between *Amelie* and *Eat Pray Love*. If the target user had preferred *Heat* a
likely recommendation would include *Scarface*. Recommender analysis typically produces an
ordered output list, such as most preferred to least preferred, of recommended items for a target
user. Recommender system implementations typically use large data sets so analytic processes
are designed to accommodate large data volume. Because many recommender systems are
embedded in online information retrieval or electronic commerce applications low response
latency is a requirement.

As discussed in Da'u and Salim (2020), collaborative filtering methods are the more
commonly used techniques for recommender systems compared to content based methods. One
of the benefits of collaborative filtering models over content based approaches is the ability to
work in a domain where the content associated items are insufficiently available and in situations
where contents are difficult to process such as opinions (Isinkaye, F., Folajimi, Y. and Ojokoh,
B., 2015). Another advantage of collaborative techniques is their ability to provide serendipitous,
i.e., unexpected recommendations. However, collaborative filtering techniques are prone to a
number of well-documented weaknesses. First, the cold start problem, which generally occurs
with newly introduced items when inadequate information has been captured about an item or a



user in order to make relevant predictions (Kunaver, M. and Pozrl, T., 2017), this typically

reduces the performance of the recommender system. In practice, the profile of the new user or

new item will be sparse empty since there is little or no prior interaction with the system and as a

result, the user's preference can't be recognized by the system. Second, data sparsity is a

problem which often affects collaborative filtering methods and will occur as a result of

insufficient information in the recommender system (Kotkov, D., Wang, S. & Veijalainen, J.,

2016). This will occur when the total number of items rated by the user in the database is

relatively low, often resulting in sparse user-item matrices (Kunaver, et al., 2017). This

frequently leads to the inability to locate successful neighbors and eventually produces a poor

recommendation process. Third, scalability is another issue associated with recommender

systems. Generally, as the size of data increases it may be difficult for recommender techniques

initially designed to deal with only a limited volume of input data to function with a larger

volume. Thus, it is very important to use recommendation techniques capable of scaling up in a

successful manner as the volume of input data increases (Kunaver, et al., 2017).

    In contrast, content-based techniques rely on user item descriptions for providing item

recommendation (Wang X. and Wang Y., 2014). For producing the related user item data,

information retrieval or other web mining methods may be used (Ebesu, T. and Fang, Y., 2017).

As shown in Figure Two, content based techniques generally filter items according to the

similarity of the content the user is interested in (Lu, J., Wu, D., Mao, M., Wang, W. and Zhang,

G., 2015). Latent semantic indexing and vector space modeling are two commonly used methods

to express these items as a vector in a multidimensional space (Krishnamurthy, B., Puri, N. and

Goel, R., 2016). Different learning techniques may be used in content based filtering such as

deep learning networks (Deng, L., and Yu, D., 2014), or support vector machines or Bayesian

classifiers (Kim, D., Park, C., Oh, J., Lee, S. & Yu, H., 2016).



Content-based filtering approaches also have strong and weak points but may be applied to address some of the drawbacks that can be experienced with collaborative filtering methods. One of the advantages of content based techniques when compared to collaborative filtering techniques include user independence. In the content-based technique ratings provided by the active user are usually exploited solely to build his or her own profile, as compared to the collaborative filtering approach which depends upon neighborhood ratings. Content based filtering may also be superior in offering transparency. Transparency will provide a clear description of how explicitly listing content features relate to recommendations. These features offer information comprehendible to the user to assess the validity of recommendations. This is in contrast to collaborative filtering approaches which may be seen as black boxes since the only explanation for an item recommendation is that unknown users with common tastes liked that item. When considering new items, unlike collaborative filtering methods, content-based methods can be very effective in providing a recommendation for the items not previously rated by any user. As a consequence, content-based filtering methods are less prone to a first rater or cold start problem which is often encountered by collaborative filtering methods.

However, there are number of potential issues with content-based methods. Content-based filtering approaches may provide obvious, and hence less informative, recommendations since they rely heavily on content description (Lu et al., 2015). For example if users have never used items with particular keywords, there is no chance for that item to be recommended. This is because in many cases the analysis is always unique to the user at hand, thus the community experience from like users is not utilized and will decrease the diversity of the items to be recommended. Another significant challenge of the content-based techniques is that they may not be good for a new user's recommendation, even though they may be good for the recommendation of newly available items (Adomavicus, G. and Tuzhilin, A., 2005). This



follows because rating history is required to be present in the data for the target user (Lu et al.,

2015), so it is generally important for useful recommendation to have an adequate rating for the

target user. In summary, content-based filtering approaches also have many trade-offs when

compared to collaborative filtering approaches.

Hybrid recommender system approaches attempt to exploit the combination of two or

more recommender system approaches such as collaborative filtering and content based filtering

for generating an enhanced recommendation (Kunaver et al., 2017). The essential idea behind the

hybrid recommendation technique is that integrating different recommender system methods will

improve upon recommendations produced using only an individual technique, such as

collaborative filtering methods or content based filtering used alone. Combining techniques may

be able to minimize the drawbacks of using an individual technique. Different methods can be

used to implement hybrid recommender systems (Adomavicus, et al., 2015). This includes

feature combination, feature augmentation (Aslanian, E., Radmanesh, M. and Jalili, M., 2016),

and cascading and switching approaches (Paradarami, T., Bastian, N. & Wightman, H., 2017).

Other representative hybrid collaborative filtering techniques such as content-boosted

approaches (Melville, P., Mooney, R. and Nafgarajan, R., 2002) and "personality diagnosis"

(PD) (Pavlov, D. and Pennock, D., 2002) combine both collaborative filtering and content-based

models to circumvent limitations of either individual approach. PD is a hybrid CF approach that

combines memory based and model based collaborative filtering algorithms to leverage some

advantages of both (Pennock, D., Horvitz, E., Lawrence, S. and Giles, C., 2000). In PD the active

user is generated by choosing one of the other users uniformly at random and adding Gaussian

noise to that user's ratings. Given the active user's known ratings, we can calculate the

probability that he or she is a similar personality type (defined operationally as preferences) as

the other users. The probability the target user will like the new items can then be calculated. We



can also view PD as a clustering method with one user per cluster (Su, X. and Khoshgoftaar, T. M., 2009). In another early hybrid approach Ansari, A., Essegaier, S. and Kohli, R. (2000) proposed a Bayesian preference model to integrate multiple types of information, including user preferences, user and item features, and expert opinions. This approach used Markov chain Monte Carlo methods (MCMC) (Gelfend, A. and Smith, A., 1990) for inference and obtained better recommender performance than pure collaborative filtering methods.

**Reproducibility and Replication of Recommender Research**

Application of deep learning methods has become a focus in recent published work in recommender systems (Zhang, et al., 2019). A number of artificial neural network models have been used such as Restricted Boltzmann Machines (Salakhutdinov, R, Mnih, A. and Hinton, G., 2007), Recurrent Neural Networks (Goodfellow, I., Bengio, Y. and Courville, A., 2016; Hidasi, B., Karatzoglou, A., Baltrunas, L. and Tikk, D., 2016), Neural Autoregressive Distribution Estimation ( Zheng, Y., Liu. C., Tang, B. and Zhou, H., 2016), Adversarial Networks (Goodfellow, I., Pouget-Abadie, J., Mirza, M. and Xu, B., 2014), Deep Reinforcement Learning (Zhao, X., Xia, L., Ding, Z., Yin, D. and Tang, J., 2018), and Multilayered Auto-encoders (Sedhain, S., Menon, A., Sanner, S. and Xie, L., 2015). As with machine learning research in general, deep learning models have received a growing focus in the recommender systems literature.

Zhang et al. (2019) point out that when a new deep learning recommendation model is proposed it is expected that the publication will offer evaluation and comparisons against several baselines. However, the selection of baselines and datasets on many papers is seemingly arbitrary, and authors generally have free reign over the choice of baselines and datasets. This creates inconsistent reporting of scores, with authors reporting their own assortment of results. There is a lack of consensus or general ranking of models and as a consequence the results can



be conflicting. In related fields such as computer vision or natural language processing standard evaluation datasets are used such as the modified National Institute of Standards and Technology datasets (MNIST) for image recognition tasks. This begs the question that MNIST-type reference evaluation datasets for recommender systems would be beneficial. Zhang, et al. continue and suggest that recommender system researchers institute hidden and blinded test sets standardized for difficulty and other parameters. Similar concerns have been raised by other recent works, for example Dacrema, M., Cremonesi, P and Jannach, D. (2019); Yang, W., Lu, K., Yang, P. and Lin, J. (2019); and Lin, J. (2018).

Fouss, F. and Sacrens, M. (2008) observed that evaluating recommender systems and their algorithms is difficult because different algorithms may be better or worse on different datasets, with different number of users or items, different rating scales, variation in ratings data set density, use of different rating scales, and other data properties, also the goals of an evaluation may differ. Inconsistent availability of source code is often a challenge to reproducibility of prior recommender system research. As discussed in Ludewig, L., Mauro, N., Latifi, S. and Jannach, D. (2019), while publishing of algorithm source code is encouraged when research reports are accepted for publication often the code used for data pre-processing, train-test data splitting, hyper-parameter optimization and evaluation of results is not provided.

Dacrema, et al. (2019), identified difficulty in reproducing published work in recommender systems as a substantive issue, particularly for recent work using deep learning approaches. The authors gathered conference publications from four conference series publishing reports of recommender system research: Knowledge Discovery and Data Mining (KDD), Research and Development in Information Retrieval (SIGIR), The Web Conference, and Recommender Systems (RecSys). A paper was relevant if it proposed a deep learning method and focused on the Top-N recommendation problem. The authors identified a collection of 18



relevant papers after canvasing proceedings of the four conferences and attempted to reproduce results reported in the 18 papers by obtaining the algorithm source code and data. A paper was considered to be reproducible if source code was available with at least one data set used in the original paper and either the originally used train/test data splits were available or could be reconstructed based on the information in the published paper. This resulted in only about one third of the original identified works as being reproducible. Notably, of the four conferences examined the RecSys conference was found to have the lowest reproducible rate of approximately 14%.

These methodological concerns can be illustrated by several representative examples from recent recommender systems literature. Dacrema, et al. (2019) compared the performance of deep learning algorithms to the baseline performance of several non-neural algorithms. The non-neural baseline algorithms included traditional KNN User-User Collaborative Filtering, ItemKNN Collaborative Filtering, and ItemKNN-CBF with neighborhood content based filtering, and ItemKNN-CFCBF with hybrid collaborative filtering and content based filtering. In Dacrema's work these baseline methods were compared with a sample of deep learning methods:

*Collaborative Memory Networks (CMN).* This deep learning method combines memory networks and neural attention with latent factor and neighborhood models (Ebesu, T., Shen, B., and Fang, F., 2018). The evaluation metrics included nDCG and Hit Rate. When comparing this CMN method with the baseline methods results showed no instance in which CMN produced more accurate results than any of the baseline methods on any of the input datasets.

*Metapath based context for Recommendation (MCRec).* This method (Hu, B., Shi, C., Zhao, W., and Yu, P., 2018) is a meta-path based approach that used information including movie genres for recommendation. The evaluation metrics included Precision, Recall, and nDCG. When comparting this method with the baseline methods results showed the traditional



non-neural ItemKNN collaborative filtering outperformed MCRec on all of the evaluation metrics.

*Collaborative Variational Autoencoder (CVAE)*. The CVAE is a hybrid technique that incorporates both content as well as rating information (Li, X. and She, J., 2017). The CVAE learns deep latent representations from the content data using unsupervised learning and learns implicit relationships between items and users from both ratings and content. Recall was the evaluation metric. Results showed that for returned recommendation lists CVAE was outperformed by the majority of traditional non-neural collaborative filtering methods. Only for long recommendation list lengths (length 100 and over) was CVAE able to outperform traditional baseline methods, however the long list lengths were not justified by Li et al. in their discussion.

*Collaborative Deep Learning (CDL)*. CDL is a probabilistic feed-forward model for joint learning with stacked denoising autoencoders and collaborative filtering (Wang, H., Wang, N., and Yeung, D., 2015). The denoising feature allows for analysis of noisy, partially incomplete, or corrupted input data. The CDL method uses deep learning techniques to jointly learn a deep representation of both content as well as collaborative information. As with the CVAE method the evaluation metric was recommendation list Recall. For CDL results also showed that for returned recommendation lists the method was outperformed by the majority of traditional non-neural collaborative filtering methods, again with exceptions for long (100 or over) returned list lengths.

*Neural Collaborative Filtering (NCF)*. Neural network-based collaborative filtering (He, X., Liao, L., Zhang, H., Nie, L., Hu, X., and Chua, T., 2017) is a variation of matrix factorization and replaces the matrix inner product with a neural architecture able to learn an arbitrary function. In this work the evaluation metrics were nDCG and Hit Rate. In a comparison study of



NCF versus baseline methods Dacrema et al. determined the NCG algorithm in He et al. had not

been properly implemented; After corrections to the NCF algorithm results showed baseline non

neural methods outperformed NCF on some, but not all, input datasets.

Another example of reproducibility limitations in the recommender systems literature is

discussed in Yang, W., Lu, K., Yang, P. and Lin, J. (2019) which describes a meta-analysis of

research reports in information retrieval that used deep learning methods applied to the TREC

Robust04 test data collection. Yang, et al. did not find evidence of an increase in information

retrieval effectiveness of deep learning approaches over time. In summary the authors state that

"While there appears to be merit to neural IR approaches, [of which deep learning recommender

methods are a subset] at least some of the gains reported in the literature appear illusory".

The methodological concerns about the state of deep learning research raised by these

recent research reports point to a need to complete additional carefully designed research studies.

### Problem Statement

As discussed above, Dacrema, et al., (2019) and Zhang, et al., (2019), among others, have

raised concerns about the state of recommender systems research methodology, citing many

instances of irreproducibility of published research, or failure to identify clear baselines for

comparison when evaluating new deep learning models. Lin (2018) pointed out instances in

which new deep learning methods do not consistently outperform existing baseline approaches

when the baseline methods are carefully implemented or fine-tuned. Different recommender

algorithms may perform better or worse on multiple datasets and on different tasks due to

differences across the datasets and differences in the nature of the recommendation task. As a

consequence, we see in some cases deep learning models being pursued as a type of *method du*

*jour* with a focus upon the newness of the models and less concern for proper experimental



design or reliance upon comparisons with weaker baseline methods. These observations point to a need to perform carefully controlled experiments to gain insights about the relative merits of different approaches.

## Dissertation Goal

This dissertation aims to deploy well-designed experiments to evaluate alternate models for recommendation systems on specific tasks. Given training data consisting of a set of objects with specified features, a set of users with specified characteristics, and ratings assigned by users to objects, the goal is to evaluate models that can predict the rating assigned by a user, with specified characteristics, to an object with specified features.

The proposed investigation uses three conventional recommender algorithms: K Nearest Neighbor (KNN) User-User collaborative filtering, Feature-based biased matrix factorization trained with alternating least squares (BiasedMF), and Feature-based filtering using singular value decomposition (SVD). In addition, two deep-learning inspired algorithms are used: a two-layer restricted Boltzmann machine (RBM) for collaborative filtering and a multilayer autoencoder network. The predictive accuracy of the models are evaluated based on the mean absolute error (MAE) and root mean square error (RMSE), appropriate for regression related models. The time to train the model and the time required for a trained model to predict a rating are also recorded. Initially the models are evaluated using large input sources of ratings data such as the *MovieLens* data sets available from Kaggle.com and other sources. As the investigation progressed no additional data sources were used. The publicly available MovieLens (ML) datasets are collected and distributed by the GroupLens Project (Kluver, D., Ekstrand, M. and Konstan, A., 2018) at the University of Minnesota (http://grouplens.org/). The MovieLens data



sources contain information of movie users, movie items, and user preference content in the form

of movie ratings and descriptive content data; these data sources also include time stamps.

## Relevance and Significance

The proposed work is offered as a research effort to address many of the methodological

concerns reflected in the Introduction and Problem Statement, citing critiques of current

recommender research. A common theme expressed in these critiques may be summed up as a

lack of empirical rigor (Lin, 2018) and a willingness on the part of some researchers in the

recommender system space to embrace the latest analytic models, such as deep multilayer

networks, but in the absence of carefully designed and executed experimental studies or with a

reliance upon comparison with weaker baseline methods. While these methodological limitations

certainly do not apply to all work in the deep learning recommender research space, there are

sufficient instances to deserve attention as documented above. In an important article, Ioannidis

(2005) asserted that the more celebrated a research field or method the less likely the published

research findings are in fact true. This suggests that some claimed research findings may be

measures of a prevailing bias in favor of "hot" methods, showing an inverse relation between

popularity and reproducibility in the scientific literature. Work by Pfeiffer and Hoffmann (2009)

reported empirical support for this inverse relation. Ioannidis was writing about molecular

biology and not machine learning, but his observations are spot-on in the context of the

burgeoning recommender system literature and the current surge in investigations using deep

learning techniques. The proposed research demonstrated carefully designed experiments with

clear definitions of both measured outcomes, such as model accuracy, and the manipulation of

experimental parameters associated with those outcomes. Accuracy has been identified as the

most discussed challenge of recommender systems (Batmaz, Z., Yurekli, A., Bilge, A. and



Kaleli, C., 2019) and is often assessed in terms of accuracy of rating predictions, usage

predictions, and ranking of recommended items (Shani, G. and Gunawardana, A., 2011).

**Definitions of Terms**

**Autoencoder Neural Network:** An unsupervised ANN used to learn a representation, at a lower

dimension, of an input then to later generate an output reconstruction as close as possible

to the original input.

**Classification:** The process of assigning a set of data points into distinct groups, categories, or

labels.

**Convex Function:** A convex function has a single minimum and can be solved by gradient

descent optimization; a line joining any two points of a convex function will be on or

above the graph of the function.

**Cross Entropy Loss:** In a classification problem the degree of divergence between a predicted

classification and an observed classification, often expressed as log loss.

**Deep Learning:** A form of machine learning enabling representation learning using multilayer

neural networks.

**Deep Neural Network:** A multilayer neural network of three or more layers.

**Dot Product Based Model:** A model based upon dot product calculation which is the sum of the

products of corresponding elements of two equal-size vectors.

**Feedforward Neural Network:** An ANN with multiple layers having no feedback cycles to

previous layers in the network.



**Gradient Descent (GD):** Function optimization used to minimize a loss (error) function by

iteratively moving in the direction of steepest descent toward the minimum of the

function.

**Hidden Layer:** One or more layers of an ANN that are neither exposed to accept input nor able

to supply output.

**Input Layer:** In a neural network the exposed layer that receives input information.

**Learning Rate:** The degree of change in network weights at each iteration as a model converges

toward the minimum of a loss function. If the rate is too low the model will take longer to

converge, if too high the minimum may be overshot.

**Log Loss:** Also referred to as cross-entropy loss, is a metric to assess the performance of a

classification model with output expressed as a probability between 0 and 1.

**Logistic Regression:** In logistic regression the dependent variable is a binary outcome rather

than being continuously scaled.

**Long Short-Term Memory (LSTM):** An enhancement to the simple recurrent network (SRN)

able to learn time series sequences by using gating processes to control acceptance of

input to the node, persistence of the nodes internal state, and when the internal state is

output.

**Loss Function:** A function that quantifies the extent observed output ($y$) of a model differs from

the predicted output ($\hat{y}$). RMSE and MAE are loss functions.

**Mean Absolute Error (MAE):** A metric to measure the average absolute deviation between the

predicted rating of the recommender $\hat{y}$ and a withheld user's actual ratings $y$.

**Net Discounted Cumulative Gain (nDCG):** A metric suitable for evaluating the rank of each

recommended item within a recommendation list. Produces a higher value when a more



relevant item is recommended at higher position in the recommendation list than less

relevant ones

**Neural Network:** A collection of processing nodes, each one analogous to a biological neuron

able to receive a number of inputs then produce an output as governed by a transfer

function.

**Objective Function:** In optimization problems, the function sought to minimize or maximize

subject to one or more constraints.

**Output Layer:** In an ANN a visible layer exposing the output of the network.

**Overfitting:** An error in modeling in which a solution is too closely fit to a limited and non-

representative set of data points. As a result the model is fit to the noise in the data and

produces poor prediction and low generalizability. Overfitting can occur in both

regression as well as deep learning models.

**Principal Component Analysis (PCA):** A dimension reduction technique in which a square

matrix of input bivariate correlations is expressed as orthogonal linear combinations

(eigenvectors) of the original variables. The ordering of the principal components

corresponds to the magnitude of their eigenvalues.

**Receiver Operating Characteristic (ROC):** Originated in signal detection theory, showing the

ratio between recognition of true positives (signal) versus true negatives (noise). Also

appropriate as a metric measuring output accuracy of a classifier.

**Recurrent Neural Network:** A multilayer neural network containing backward connections to

previous layers and able to model temporal sequential representations.

**Regression:** An application of the general linear model (GLM), univariate regression models

one dependent variable as a function of one or multiple independent variables;



multivariate regression models two or more dependent variables as a function of multiple

    independent variables. Regression uses least squares analysis to minimize an error term.

**Reinforcement Learning:** Model outputs in an environment that improve cumulative gain will

    be strengthened (made more likely) while outputs that decrease cumulative gain will be

    weakened. Often contrasted with both supervised learning and unsupervised learning.

**Restricted Boltzmann Machine:** A multilayer neural network having no intra-layer

    connections.

**Root Mean Square Error (RMSE):** In a general linear model, such as regression, the deviation

    between a predicted value and an observed value.

**Singular Value Decomposition (SVD):** A dimension reduction technique wherein a matrix $A$ is

    factored as the product of three matrices such that $A = UDV^{\mathrm{T}}$. The values of $D$ are

    analogous to principal components.

**Softmax:** A standardization algorithm to rescale the outputs of a model to a sum of 1.0, thereby

    setting the output values equivalent to probabilities.

**Stochastic Gradient Descent (SGD):** A variation of gradient descent in which a single data

    point, or small set, is sampled and used for each iteration of model training. The sampling

    can produce notable improvements in efficiency compared to simple gradient descent,

    which uses the full data set of points.

**Training Data set:** A data set used to train a machine learning algorithm. Data to be used by the

    model are frequently partitioned into training and validation data sets.

**Transfer Function:** The mathematical representation of the relation between the inputs to a

    process and the output produced by the process.

**Universal Approximation Algorithm:** A feed-forward neural network with a single hidden

    layer and a finite number of nodes can approximate any monotonic function.



**Validation Data set:** A data set used to evaluate the performance of a machine-learning

algorithm following training. Validation and training data sets are usually partitions of

data from the same source.

## Summary

Recommender systems have found increasing use given the growth of electronic

commerce and online information systems with much of the impetus for recommenders being

driven by requirements for personalization. In this chapter recommender applications were

introduced, specifically collaborative filtering, the largest class of recommender applications.

User-based and content-based methods comprise the most often deployed approaches to

collaborative filtering. Recommender systems research is a burgeoning area following the

resurgence in neural network applications, deep learning in particular. Examination of recent

research reports shows a growing concern with reproducibility and replication of reported works,

pointing to a need for systematic replication of both conventional as well as deep learning

recommender algorithms. The literature identified recommendation accuracy as a salient

measure of recommender value and utility. This chapter continued with the dissertation problem

statement, the dissertation goal, a statement of relevance, and concluded with a definition of

terms.

The following chapters of this proposal are arranged in this manner: Chapter 2 offers a

review of the literature most pertinent to the planned research. The area of recommender systems

has a rapidly growing literature, so the focus of the literature review is upon work most germane

to the dissertation problem and the algorithms chosen for investigation. Chapter 3 discusses the

design of planned experiments using the five selected recommender algorithms. Chapter 4

summarizes results of the experiments with tabular and graphic presentations for each



experimental replication of each algorithm. Chapter 5 offers conclusions, implications of the

investigation, recommendations for future research, and a summary of this report.



# Chapter 2

# Literature Review

Park, D. H., Kim, H. K., Choi, I. Y., and Kim, J. K. (2012) produced a literature review and classification of recommender systems prior to the recent introduction of deep learning methods. The authors reviewed 210 articles discussing recommender systems from 46 journals published between 2001 and 2010 and classified the articles by year of publication, the journals in which the work appeared, and the application area. The works were categorized into eight application fields and eight data mining techniques. The application areas were books, documents, images, movie, music, shopping, TV programs, and others. The data mining techniques mentioned in this work include association rules, clustering, decision trees, K nearest neighbor, link analysis, neural network models, regression methods, and other heuristic approaches. Ricci, F., Rokach, L., Shapira, B. and Kantor, (2015) also offers extensive coverage of recommender systems techniques and algorithms outside of deep learning.

**Collaborative Filtering**

Varieties of collaborative filtering (CF) comprise the largest single fraction of algorithms for implementing recommender systems. Su and Khoshgoftaar (2009) offer a concise survey of techniques used to implement CF. In their overview, collaborative filtering techniques can be grouped into several categories: memory-based collaborative filtering, model-based collaborative filtering, content-based recommenders, and collaborative filtering using hybrid recommenders. Memory-based approaches retain the complete datasets of user ratings and movie information used to conduct the recommender analysis within program memory as the analysis is completed. Typical techniques for memory-based CF include nearest neighbor-based CF, often using item based (Sarwar, B., Karypis, G., Konstan, J., and Reidl, 2001) or user-based CF algorithms



evaluated with metrics based upon Pearson correlation or vector cosine similarity. Other nearest

neighbor-based CF approaches include item-based or user-based Top-N recommenders.

Advantages of memory-based CF are a relatively straightforward implementation, an ability to

add new data incrementally, no need to consider the content of the items being recommended,

and an ability to scale well with co-rated items. Some disadvantages of memory-based CF are

dependence upon human ratings, declining performance as data sparseness increases, decreased

ability to develop recommendations for new users or newly added items (the "cold start"

problem), and limited scalability for large data size.

Model-based methods use initial analysis of user ratings data to develop a recommender

model with the model being used in a later pass to make recommendations. Representative

techniques for model-based CF are Bayesian belief nets (BBN), cluster-based CF, Markov

decision process (MDP) based CF, latent semantic CF, factor analysis of sparse matrices, and CF

with other dimensionality reduction methods such as singular value decomposition (SVD),

principal component analysis (PCA), or matrix factorization. Some advantages of model-based

CF are a better ability to deal with sparseness and scale, improved prediction performance, and a

capability to deliver an intuitive rationale for explaining recommendations to users.

Disadvantages of model-based CF are computing expense (time and resource) of model

generation, a trade-off between prediction performance and scalability, and often a loss of

information due to the nature of dimension reduction techniques.

Typical implementation techniques for hybrid recommenders (Burke, 2002) include

content-based CF recommenders, content-boosted CF, and hybrid CF incorporating both

memory-based and model-based CF algorithms. Some advantages of a hybrid implementation

are an ability to deal with limitations of CF and content-based recommenders, improved

prediction performance, and a capacity to deal with data sparsity and non-representative cases.



Disadvantages are increased complexity of implementation, and often a need for external

validation information which may be unavailable.

In prior work Adomavicius and Zhang (2014) investigated two scalable, general-purpose

meta-algorithmic approaches – based on "bagging", a form of bootstrap aggregation (Breiman,

L., 1996), and iterative smoothing. Iterative smoothing is introduced in this work and has close

correspondence with prior work using classification methods of the collective inference

paradigm (Macsassy and Provost, 2007; Neville and Jensen, 2000). When used in conjunction

with different traditional recommendation algorithms these meta-algorithmic approaches were

found to improve prediction stability. Experimental results on actual rating data demonstrated

that both approaches were able to demonstrate higher stability as compared to the original

recommendation algorithms. In addition, it was found that use of the meta-algorithmic

approaches did not sacrifice predictive accuracy in order to improve recommendation stability,

but were able to provide additional accuracy improvements.

Adomavicius and Zhang (2010) introduced stability as a new criterion of recommender

systems performance. In general, a recommendation algorithm is said to be "stable" if its

predictions for the same items are consistent over a period of time. This definition assumes any

new rating data acquired by the recommender system over the same period of time are in

substantive agreement with system's prior predictions. Their paper advocates that consistency

and stability should be a desired property of recommendation algorithms, since unstable

recommendations (Ekstrand, M. and Riedl, J., 2012) can lead to user confusion, hence reduce

trust in recommender systems. Trust has been shown to be a desirable attribute of recommenders

(Gedikli, F., Jannach, D., and Ge, M. Z., 2014). This work empirically evaluated stability of

several popular recommendation algorithms with results suggesting that model-based

recommendation techniques demonstrate higher stability than memory-based collaborative



filtering heuristics. It was also found that the stability measure for recommendation algorithms is influenced by many factors, including matrix sparsity of the initial rating data, the number of new incoming ratings (a function of the length of the time period over which the stability is being measured), the distribution of the newly added rating values, and the rating normalization procedures employed by the recommendation algorithms.

**Neural Networks**

The artificial neural network (ANN) is essentially a mathematical model where the nodes of the ANN are analogous to biologic neurons as a basic processing unit. The biologic neuron can be visualized as a cell body with projections to receive electrochemical signals as input from adjacent neurons and then develop an output electrochemical output signal as a result. Figure 3 illustrates a typical neuron and an analogous ANN node.

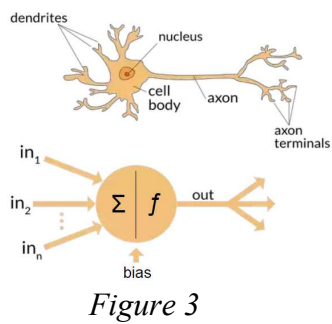

*Figure 3*

In a computational neural network the relation between the inputs and output is governed by a transfer function, typically a sigmoid activation function calculated as a mathematical dot product. Mculloch and Pitts (1943) are credited with the initial introduction of the computational neural network. Their fundamental algorithm specified the calculation of a single neuron as a weighted sum of the neuron's inputs, the relation between those inputs and an output is governed by the characteristics of the transfer function. A number of transfer functions have been employed in ANN's such as the Logistic/Sigmoid, Hyperbolic Tangent, Rectified Linear Unit



(ReLU), Linear, and Softmax. In contrast to the original ANN architecture, deep neural networks are elaborations of the basic neural network model with many more neurons per layer and potentially multiple layers of neurons. As the neural network responds to incoming stimuli patterns of inputs and outputs that frequently occur together are strengthened. Hebb's rule (Hebb, 1949) described how the connection strengths among biological neurons is altered as a function of learning. This characteristic is often stated as Hebb's rule: "the neurons that fire together wire together". Hebb's rule has been confirmed by neurophysiological evidence – during learning, reinforcement increases the density of inter-neuron connections (white matter) at a cellular level while over time unreinforced connections will decrease in density and ultimately be reabsorbed by metabolic processes.

The artificial neural network was refined by the work of Rosenblatt (1958) in his introduction of the perceptron, a two-layer ANN with an input layer and an output layer employing a simple sigmoid transfer function between the two layers. The perceptron demonstrated success with well-bounded classes of problems, as in recognition tasks for letters or numbers, but important work by Minsky and Pappert (1969) critical of the perceptron had a chilling effect upon work with ANN's. Minsky and Pappert showed Rosenblatt's perceptron was unable to learn exclusive-or (XOR) operators, the XOR being a non-linearly separable problem. Following Minsky and Pappert's critique, ushering in what is often referred to as the "AI winter", work on ANN's was continued as a topic in parallel-distributed processing (PDP) (Rumelhart and McClelland, 1986). The PDP connectionist model defined an alternative approach to artificial intelligence in contrast to the then dominant symbolic rule-based approach of AI, exemplified in Newell and Simon's Logic Theorist (Newell, Shaw and Simon, 1958). One consequence of these divergent views is that the connectionist and now deep learning models of AI are often implemented with more of a black box approach making explanations of the



relations between inputs and outputs more difficult. In comparison, the rule-based approaches of AI are, by their nature, constructed from collections of a priori rules and much easier to explain to human observers. However, recent work in ANN's and deep learning has demonstrated progress in developing models better able to present explanations to users.

Over the course of ANN development much attention has been devoted to the analytic techniques able to train neural models. Werbos (1974) suggested that *gradient descent* (GD) was suitable for training of ANN's and offered a means to apply iterative adjustments to the network node weights. Prior to this, gradient descent was frequently seen more generally in optimization problems to find minima of functions. We can view gradient descent as an optimization procedure able to converge upon the values of function parameters so as to minimize a loss (error) function. Gradient descent is useful when the function's parameters cannot be estimated in a single batch calculation using typical dot-product methods and instead requires an iterative solution able to converge upon the minimum of a convex function. Successive iterations calculate the derivative of the loss function at each step and ultimately stop at the minimum without overshoot. A learning rate parameter is selected to determine how much the coefficient of the loss function is allowed to change on each iteration. For large data sets gradient descent can be a time consuming computationally inefficient process and s*tochastic gradient descent* (SGD) is able to improve upon the performance of gradient descent. Gradient descent is a more exhaustive process requiring use of all the data points in the training set to calculate an update for the loss function in each iteration. Stochastic gradient descent samples the input data set to use only single, or a small subset, of points to estimate the loss function parameter in each iteration, when the sample subset approach is used this is termed minibatch stochastic gradient descent. As a result, if the number of training data points is large gradient descent will be computationally more demanding since each function update requires use of all data points in the



training data set. In contrast, stochastic gradient descent, or minibatch stochastic gradient descent, will be much faster and converge upon a solution sooner.

Important work by Rumelhart, Hinton, and Williams (1986) further demonstrated the initial application of gradient descent to ANN training and defined backpropagation. In backpropagation the initial, often randomly derived, weights of all nodes are adjusted at each network iteration. The gradient of each node weight is computed and used to determine a change in weight for each training iteration. Each weight gradient is calculated as the partial derivative of a loss function for a weight while all other weights are held constant (Rumelhart, et al., 1986). Later, Bengio (2009) reported that backpropagation was not effective for training ANN's of over two layers. However, more recent work (Adigun and Kosko, 2020) has extended backpropagation in a bidirectional manner with successful application for training of deep multilayer ANN's.

Research on neural networks continued through the first decade of the 2000's and neural networks have found application in a variety of settings such as real-time speech and image recognition. However, during this time network training methods for larger multilayered networks were frequently found to be ineffective and again skepticism about neural networks increased, as a result research in the area stagnated. Enthusiasm returned to neural network research following work on deep belief networks by Hinton, G., Osindero, S. and Teh, Y., (2006). Also, recent developments in computational infrastructures such tensor processing units (TPU) (Jouppi, N., Young, C., Patil, N. and Patterson, D., 2018) offer platforms able to better accommodate the computational demands of large multilayered deep learning networks.



## Deep Neural Network Architectures

Recent attention in RS research has focused upon application of neural networks and deep learning models**.** Zhang, S., Yao, L. and Tay, Y. (2019) offer a thorough survey of deep learning based recommender systems and identify nine variations of deep learning architectures. Deep learning ANN's are able to learn levels of abstractions and representations from input data. For their purpose an architecture is regarded as deep learning if it optimizes a *differentiable* objective function using Stochastic Gradient Descent (SGD). These neural architectures have shown excellent success in both supervised and unsupervised learning tasks as well as reinforcement learning. After Hinton, G., Osindero, S. and Teh, Y. (2006) described an efficient way of training deep neural models and Bengio (2009) demonstrated the potential of deep architectures in complex artificial intelligence tasks, deep learning has become an emerging approach in machine learning. Deep learning approaches have produced many state of the art solutions in computer vision, natural language processing, and speech recognition (Deng, L. and Yu, D., 2014). It has also been demonstrated that multilayer feedforward ANN's can function as universal approximators (Hornik, K., Stinchcombe, M. and White, H., 1989) enabling them to learn nonlinear as well as linear functions.

The *Multilayer Perceptron* (MLP), is a feed-forward neural network with multiple hidden layers between the input and output layers. Unlike the basic perceptron the MLP can use an arbitrary activation function so will not always be limited to a binary classifier. We can view the MLP as stacked layers of nonlinear transformations able to learn hierarchical feature representations. An MLP can also be viewed as an example of a universal approximator (Hornik, et al., 1989). As discussed in Hastie, T.,Tibshirani, R. and Friedman, J. (2009), the term MLP does not refer to a single perceptron of multiple layers, instead it contains many perceptron's organized into layers so could be described as a multilayer perceptron network. As a result the



MLP in a strict sense is not a classic perceptron. The classic perceptron is a special case of ANN that will have a threshold activation function, e.g. the Heaviside step function (Liu, Q. and Wang, J., 2011). In contrast MLP's can employ arbitrary activation functions. While a true perceptron performs *binary* classification, an MLP ANN is free to either perform classification or regression, depending upon its activation function.

Wide and deep learning (Cheng, H., Koe, L., Harmsen, J., Shaked, T., Chandra, T., Aradhye, H., Anderson, G., Corrado, G., Chai, W., Ispir, M. & others., 2016) is an example of a hybrid MLP model combining both a classic perceptron and a MLP so is able to solve both regression and classification recommendation problems. The wide learning component is the classic single layer perceptron which can also be regarded as a generalized linear model. The deep learning component is a nonlinear model implemented as an MLP. The motivation for combining these two techniques is to enable a recommender to solve both memorization and generalization tasks. Memorization is achieved by the wide learning classic perceptron component and enables the capability of modeling direct features from historical data. Concurrently, the deep learning component enables generalization by modeling more general and abstract representations. As a result this model can improve the accuracy as well as the diversity of recommendation.

The *Autoencoder* (AE), is an unsupervised dimension reduction model that attempts to represent its input data in the output layer. The autoencoder is a type of feedforward neural network, which is trained to encode the input into some representation, so that the input can be reconstructed from that representation (Hinton and Salakhutdinov, 2006). Minimally, an autoencoder consists of 3 layers, the input layer, a hidden layer, and an output layer. In a typical autoencoder the number of neurons in the input layer is equal to the number of neurons in the output layer. The autoencoder reconstructs the input layer at the output layer by using the



reduced representation produced in the hidden layer. During the learning process, the network

uses 2 mappings, one each for the encoder and decoder. The encoder maps the data from the

input layer to the hidden layer and the decoder maps the encoded data from the hidden layer to

the output layer.

The autoencoder reconstruction strategy may fail to extract useful features and the

resulting model may produce uninteresting solutions, or it may provide a direct copy of the

original input. To avoid such problems, a denoising factor may be used and applied to the

original input data. A denoising auto encoder (DAE) is a variant of the basic autoencoder that is

trained to reconstruct the original input in the presence of potentially corrupted or noisy input

data. The denoising factor makes autoencoders more stable and better able to deal with potential

data corruption (Vincent, P., Larochelle, H., Lajote, I., Bengio, Y. and Manzagol, P., 2010). This

feature makes DAE's suitable for use in recommender systems to predict missing values in the

presence of corrupted data (Wu, Y., DuBois, C., Zheng, A., and Ester, M., 2016).

The *Convolutional Neural Network* (CNN), is a feed-forward network with convolution

layers (Goodfellow, I., Bengio, Y. and Courville, A., 2016). We can view the CNN as a variation

of feed-forward neural network applying convolution in place of typical matrix multiplication in

one or more of its layers. A typical CNN consists of convolutional layers, pooling layers, and

fully connected layers. Convolutions identify features from the input using convolution filters

with mathematical operations. To model nonlinearities a rectified linear unit (RelU) may be

applied after each convolution operation. Pooling is used to reduce dimensionality of the feature

maps produced by the convolution layer then output from the convolution and pooling layers can

be used within the fully connected layers to derive classifications. The CNN can be very

effective for processing unstructured multimedia information using its convolution and pool

processes and CNN recommendation models are useful for feature extraction.



Wang, S., Wang, Y., Tang, J., Shu, K., Ranganath, S. and Liu, H. (2017) studied the influences of visual features for point of interest (POI) recommendation and developed a visual content enhanced POI recommendation system (VPOI). The VPOI used CNN's to extract image features. With VPOI the recommendation model is built by exploring the interactions between visual content with latent user factors and visual content with latent location factors. Using a CNN for text feature extraction, DeepCoNN (Zheng, L., Noroozi, V. and Yu, P., 2017), uses two CNN's in parallel to model user behaviors and item properties from review texts. This approach reduces the data sparsity problem and enhances the model's interpretability by exploiting semantic information from text. By using information from the texts and improving interpretability this approach addresses one of the often-cited limitations of deep or connectionist learning models. The approach uses a word imbedding technique to map the review text into a lower dimensional representation as well as retain the word sequence information. The extracted review representations then pass through a CNN convolutional layer with different kernels, followed by a max-pooling layer and finally a fully connected layer. The output of the user network and item network are finally concatenated as the input to the prediction layer where a factorization machine is used to capture the interactions for rating prediction.

CNN's have also been used for audio and video feature extraction. Van den Oord, A., Dieleman, S., & Schrauwen, B. (2013) used a CNN to extract features from audio signals. In their model the convolutional kernels and pooling layers of the CNN enable processes at multiple times scales. As a result this content-based model can in many cases alleviate the cold start problem when the target music has not yet been consumed by a sufficient body of users. In a similar vein, Lee, J., Abu-El-Haija, S., Varadarajan, B. and Natsev, A. (2018) suggested extracting audio features using a CNN model. With this model the recommendation is executed in a collaborative metric learning framework.



A *Recurrent Neural Network* (RNN), has an embedded memory suitable for modeling sequential data. The RNN is a type of ANN that is able to make use of sequential information in the input stream (Donahue, J., Hendricks, A., Guadarrama, S., Rohrbach, M., Venugopalan, S., Saenko, K. and Darrell, T., 2015). Sequential information will be captured within-session as an information system user navigates a site. The RNN is designed to process a sequence of values generated by a user click-stream and its output depends upon successive computations in the sequence. It is known that basic RNN's may fall victim to a vanishing or exploding gradient (Hochreiter and Schmidhuber, 1997). Backpropagation is often used to train RNN's and when the gradient is passed back through many steps it can approach zero or grow exponentially. A Long Short-Term Memory (LSTM) architecture may be used as a solution (Kumar, V., Kbattar, D., Gupta, S. and Varma, V., 2017). Because RNN's are suitable for sequential data processing they have become a frequent choice for working with temporal dynamics of interactions and sequential patterns of user behaviors while dealing with text, video, and audio under session-based sequential conditions. This approach also enables side information to be integrated into recommendations thereby assisting with the interpretability of recommender results.

In many practical applications or websites, the system often does not recognize users so therefore will have no access to the user's identity or long-term consumption habits or interests. Often session or cookie mechanisms enable systems to get a user's short term preferences but this can be a difficult task for recommender systems due to the extreme sparsity of training data. However, recent advances have demonstrated the efficiency of RNN's in minimizing this issue (Hidasi, B., Karatzoglou, A., Baltrunas, L. and Tikk, D., 2015). These authors proposed an RNN model *GRU4Rec*, which is a session-based model. In this model the input is the actual state of a session with 1-of-N encoding, were N is a number of items. In this encoding a coordinate will be 1 when the corresponding item is active in the session, and otherwise zero. In this model the



output is the likelihood of an item being next in the sequence. To train a *GRU4Rec* network, the authors proposed a parallel session many-batches algorithm and a sampling method for the output.

Although RNN's have achieved success with session-based recommendation, Jannach, D., and Ludewig, M., 2017, have indicated some trivial conventional methods, for example simple neighborhood collaborative filtering methods, may achieve similar or better recommendation results while also being simpler and more computationally efficient. Combining conventional and RNN models in an ensemble approach often leads to better performance. As noted above in the motivations for this dissertation research, these cases are not accidental in the deep learning and machine learning literature and it is suggested that more experiments should be done and systematic approaches such as ablation studies and hyper parameter studies should be conducted to justify the proposed deep approaches and baselines.

The *Restricted Boltzmann Machine* (RBM), is a two-layer network with a visible layer and a hidden layer but no intra-layer communications in either layer. The restricted Boltzmann machine is a particular type of Boltzmann machine, which has two layers of units with the first layer consisting of visible units while the second layer includes hidden units. In this restricted architecture there are no connections between the units in a layer. In the RBM the visible units in the model correspond to the components of observation, and the hidden units represent the dependencies between the components of the observations (Ciresan, D., Meir, U., Gambardella, L. and Sehmidhuber, J., 2010).

The Boltzmann machine was introduced when Hinton, G., and Sejnowski, T. (1986) added a mathematical function reminiscent of metallurgical annealing to a Hopfield network, an early feed forward neural network, resulting in the Boltzmann Machine. In this approach local minima are avoided by inserting randomness to the process of energy minimization, so that when



the network converges toward a local minimum it has an opportunity to defer. To implement this process the Boltzmann machine updates the binary states of individual neurons by stochastic processes instead of deterministic rules.

Boltzmann machines were originally introduced as a general connectionist approach to learning arbitrary probability distributions over two-state vectors (Fahlman, S., Hinton, G. and Sejnowski, T., 1983). Variations of the Boltzmann machine that include other kinds of variables have surpassed the popularity of the original Boltzmann configuration. The Boltzmann machine also becomes more powerful when not all the variables are observed. In this case latent variables can behave similarly to hidden units in the multilayer perceptron and model higher-order interactions of the visible units. A Boltzmann machine with hidden units is no longer limited to modeling linear relationships between variables and instead approaches a universal approximator (Hornik, et al., 1989) of probability mass functions over discrete variables (Le Roux, N. and Bengio, Y., 2008).

The *Deep Belief Network (DBN)* a variation of the RBM, a multilayer deep architecture that uses a group of stacked RBM's. In this design, the hidden layer of one subnetwork serves as a visible layer for each successive RBM subnetwork (Hinton, 2009). Training in the DBN uses the first layer to receive the original data into its visible units then the hidden units of the first RBM are used as the input into the second RBM layer. The training process continues until reaching the top of the stack of subnetworks to produce a suitable model used to extract features from the input. Since the learning process is unsupervised, it is common to add an additional network of supervised learning at the output of the DBN (Batmaz, Z., Yurekll, A., Bilge, A. & Kaleli, C., 2019). This enables it to be used in a supervised learning task such as classification or regression.



A *Neural Autoregressive Distribution Estimation* (NADE) network is an unsupervised network built with an autoregressive model and a feed-forward network (Uria, B., Cote, M., Gregor, K., Murray, I. and Larochelle, H., 2016). The motivation for NADE arose from the restricted Boltzmann (RBM) being computationally infeasible with some data sets. As an alternative to RBM the NADE can provide a more computationally feasible solution for collaborative filtering. In contrast to RBM, NADE does not calculate any latent variables where expensive inference would be required, instead it can be optimized efficiently by use of backpropagation. The NADE model has achieved successful results on many machine-learning tasks (Uria, B., Murray, I., and Larochelle, H., 2014).

The *Generative Adversarial Network* (GAN), is a generative network consisting of a discriminator and a generator. With the introduction of the GAN model (Goodfellow, I., Pouget-Abadie, J., Mirza, M., Xu, B., Warde-Farley, D., Ozair, S., Courville, A. and Bengio, Y., 2014) successive studies have proposed to exploit the model for developing and improving recommender systems. Cai, X., Han, J. and Yang, I. (2017) used a GAN to introduce a deep network model that combines content and network structure into a unified architecture to produce a personalized citation recommender. The GAN represents different types of content in the heterogeneous network in a continuous vector space, the distributed representation is then used to calculate similarity scores. The model consists of two primary components, one component which aims to learn an effective feature representation able to preserve both the network structure and content information, and a generative adversarial bibliographic component. Wang, J., Yu, L., Zhang, W., Gong, Y., Xu, Y., Wang, B., Zhang, P. and Zhang, D. (2017) also used a GAN to propose a model for information retrieval. The deployed model demonstrated capability in three different information retrieval tasks including question answering, web search, and item recommendation. This model sought to combine both



generative and discriminative processes into a unified implementation able to complete a

minimax game similar to the discriminator and generator in a GAN. In similar work He, X., He,

Z., Du, X. and Chua, T. (2018) introduce an adversarial personalized ranking method which uses

adversarial training for improving Bayesian personalized ranking (BPR). Their model is capable

of playing a minimax game between the original BPR loss and an adversary which adds

permutation for maximizing the objective function. Wang, Q., Lian, D., and Wang, H. (2018)

proposed a GAN model which generates negative samples for a memory network-based

streaming recommender. This work demonstrated that the model performance could be approved

by a GAN based sampler.

*Attentional Models* (AM), are differentiable neural architectures able to discriminate the

importance of different segments of a target sequence or memory. These models were originally

used in the natural language processing (NLP) and computer vision domains but are now

appearing in recommender systems applications. In deep learning the attention mechanism is

motivated by an analogy to human visual or auditory attention; people only need to focus on

component parts of visual inputs to understand or recognize them - the gestalt of the whole may

be recalled or recognized by perception of a part of the object. Similarly, in the case of auditory

processing previously encountered auditory sequences may be recalled or remembered by

exposure to fractions of the original auditory stream. The attention mechanism is capable of

disregarding uninformative or noisy features and minimizing the influence of noisy or irrelevant

data. This has been an effective technique and has garnered considerable attention over many

years across diverse areas such as computer natural language processing, visual image

processing, and speech recognition.

Neural attention can be applied to multiple deep models including NLP, CNN's, and

RNN's. As an example, attention-based CNN's are capable of capturing many of the more



informative elements of complex inputs (Seo, S., Huang, J., Yang, H., and Liu, Y., 2017). By

Applying an attention mechanism to a recommender system, that mechanism can be used to filter

out noisy content and select the most essential or salient items. This will also assist

interpretability by facilitating the generation of recommender explanations for users. In attention-

based deep methods, inputs are usually weighted according to some form of attention index or

weight. Recent work as in Gong, Y. and Zhang, Q. (2016) and Chen, J., Zhang, H., He, X., Nie,

L., Liu, W. and Chua, T. (2017), demonstrate the capability of an attention mechanism to

improve recommender performance. Work by Li, Y., Liu, T., Jiang, J. and Zhang, L. (2018) has

also suggested that including an attention mechanism into RNN's may significantly improve

their performance. This work developed an attention-based LSTM model for Recommendation

to take advantage of both RNN's and attention mechanisms to capture information from blogs.

*Deep Reinforcement Learning* (DRL), architectures at basis implement a reinforcement-

learning model. Reinforcement learning has its origins in human behavioral psychology

(Skinner, B.F., 1938). Behavior can be viewed as an ongoing sequential interaction with an

environment such that behaviors followed by a successful outcome, similar in many ways to

supervised learning, are strengthened by reinforcement and are more likely to be repeated in the

future. Conversely, behaviors that are not followed by a successful outcome in the environment

are not reinforced and hence have a lower probability of being repeated. The same framework

can be applied to the outputs of analytic processes and ANN's in machine learning (Noffsinger,

W.B., 1989). Most previous work with RS models has viewed the recommendation process as

being static, especially in the case of off-line or batch recommendation, and hence less able to

model a user's temporal or sequential interactions. In response, deep reinforcement learning

(DRL) models have achieved success in making personalized recommendations in the context of

ongoing sequential interactions. A good example of this approach is found in Zhao, X., Zhang,



L., Ding, Z., Xia, L., Tang, J., and Yin, D. (2018) who propose a DRL method which can model the recommendation stream as a Markov Decision Process. This model uses DRL to learn optimal recommendation strategies. Zhao et al. discuss DRL based recommender systems as having two primary advantages, first the capability to continuously update strategies in real time during the interaction streams, and second, the capacity to learn a strategy that maximizes the long-term cumulative reward from users.

Generative deep architectures attempt to model the higher order association properties of input data for pattern synthesis (Deng, L., 2014a) and are essentially applications of unsupervised learning methods (Schmidhuber, J., 2015). In an unsupervised architecture, specific known-outcome information such as class labels are not available during the learning process. Most of the deep networks in this class typically use the network model to generate samples and are thus referred to as generative models. Typical examples of such deep network models include autoencoders (AE), restricted Boltzmann machines (RBM), deep belief networks (DBN), deep Boltzmann machines, and the recently introduced generative adversarial network (GAN) architecture (Goodfellow, I., Pouget-Abadie, J., Mirza, M., Xu, B., Warde-Farley, D., Ozair, S., Courville, A. and Bengio, Y., 2014). The GAN has been shown to be highly successful in many settings. In the GAN two networks are in an adversarial architecture. Given a training set the GAN learns to generate new data with the same statistics as the training set and the GAN develops a discriminator as a classifier with a sigmoid cross entropy loss function. Later work by Mao, X., Li., Q., Xie, H., Lau, R., Wang, Z., and Smolley, S. (2017) improved upon the original GAN.

**Selected recommender algorithms**

The proposed investigation will use five recommender algorithms in a series of experiments. Three conventional algorithms will be used which are essentially applications of



linear models or members of the multivariate linear family (He, X. and TaT-Seng, C., 2017). These are: KNN User-User collaborative filtering, Feature-based biased matrix factorization trained with alternating least squares (BiasedMF), and Feature-based filtering using singular value decomposition (SVD). In addition, two deep-learning inspired algorithms will be used: a multilayer restricted Boltzmann machine (RBM) for collaborative filtering, and a multilayer autoencoder network (EN). In contrast to the linear models underlying the conventional algorithms, deep neural networks are capable of modeling the nonlinearity in data with nonlinear activation functions. As discussed in Zhang and other sources this property makes it possible for the deep-learning models to more adequately capture complex and intricate user/item interaction patterns. The decision to select these five algorithms is based upon the availability of existing code implementation frameworks (e.g., Ekstrand, 2019; Kane, 2018) and access to processing infrastructure, including TensorFlow and Keras, able to serve as a basis for this investigation. The five selected algorithms are also a representative set of conventional linear recommender algorithms as discussed in Park, et al. (2012) and nonlinear deep learning algorithms discussed in Zhang, et al. (2019).

**Conventional Recommender Algorithms**

*KNN User-User Collaborative Filtering* (Ekstrand, 2019; Herlocker, J., Konstan, J. and Reidl, J., 2002; Resnik, P., Iacovou, N., Suchak, M., Bergstrom, P. and Riedl, J., 1994): This user-based CF algorithm implements a nearest neighbor approach, setting the neighborhood size at run time and using a measure of similarity applied to user-mean-normalized ratings. This algorithm will also output a ranked Top-N list. Given a target user and their positively rated items, this algorithm will identify the k-most similar users of the target user. Many user-based CF algorithms use a Pearson correlation measure to assess the degree of similarity between the item preferences of users, another often-used measure of similarity is cosine similarity. Cosine



similarity is a measure of the cosine of an angle between two vectors. We can visualize these two vectors as being regression lines (lines of best fit) between two sets of items or scores and the two regression lines will project in a Euclidian space with an angle. As the angle of these two vectors approaches zero (cosine one) the similarity will be measure one and indicate complete agreement or similarity. As the angle approaches 90 degrees (cosine zero) the similarity will be measure zero and indicate compete orthogonality (independence) between the two vectors. Mean absolute error (MAE) and root mean square error (RMSE) are metrics frequency used to assess the predictive accuracy of these algorithms. Coba, Symeonidis, and Zanker (2018) document the *Most Frequent Item in the Neighborhood* (MFIN) algorithm used in both user-based and item-based CF:

```
Vector Top_N_MFIN(u, NN(u), I, N)

//User u
//Set NN(u) nearest neighbors of user u
//int I domain of all items
//int N size of recommendation list
//P_t threshold for positive ratings
//r_{v,I} the rating of user v on item i

begin
    For i = 1 to I
        F[i] = 0;
    For each v element of NN(u)
        For i = 1 to I
            If r[v][i] >= P_t
                f[i] = f[i] + 1;
    Sort(f); //descending order
    n = 0; i = 1;
    While (n < N and f[i] > 0)
        A[n] = i;
        n = n + 1;
    return A;
end.
```



The goal of collaborative filtering algorithms is to identify new items or predict the value of an item for a particular user based on that user's previous likings in conjunction with the opinions of other users with similar preferences. In a typical collaborative filtering analysis (Sarwar, B., Karypis, G., Konstan, J. and Riedl, J., 2001), we begin with a list of users $U = \{u_1, u_2, ..., u_m\}$ and a list of n items $I = \{i_1, i_2, ..., i_n\}$. Each user $u_i$ will have a list of the items $I_{u1}$ about which they have rated their preference. Preferences may be explicitly given by the user as a rating score, generally using an ordinal numeric scale, or can be implicitly derived from the user's selection history using data mining techniques applied to timing logs or other user session-based information. The user's set of items will be some or all the elements of $I$ or could be empty. In a recommendation there will be an active or target user for whom the task of the collaborative filtering algorithm will be to determine a recommendation prediction and to return a recommendation list. We can express prediction as a numeric value, $P_{a,j}$, expressing the estimated probability of item $i_j$ and not among items the target users have rated or selected. This predicted value will be within the same ordinal scale (e.g., 1 to 5) as the preference values provided by $u_a$. The recommendation will be a list of $N$ items, $I_r$ estimated to be most attractive to the target user. As a rule the recommendation list will be for items not already consumed or viewed by the target user. This output list produced by CF algorithms is also called the Top-N recommendation.

CF algorithms model the full set of $m \times n$ user-item data as a ratings matrix, $A$. Each entry $a_{i,j}$ in $A$ corresponds to the preference score (preference ratings) of an $ith$ user on a $jth$ item; each individual rating is based upon a numeric ordinal scale and may be 0 when that user has not yet rated that item. The commonly used collaborative filtering algorithms can be grouped into two primary categories: *Memory-based* (user-based) and *Model-based* (item-based) algorithms.



Memory-based Collaborative Filtering Algorithms use the entire user-item data set to produce a prediction. These approaches use statistical techniques to identify a set of users, e.g., nearest neighbors, that have a history of being similar to the target user and will have either rated diverse items in a similar manner or were likely to buy a similar set of items. Once a neighborhood of users is formed, these systems use different algorithms to combine the preferences of neighbors to produce a prediction or top-N recommendation for the active user. The techniques, also known as nearest neighbor or user-based collaborative filtering, have been popular and widely used in practice.

Model-based Collaborative Filtering Algorithms produce item recommendations by initially developing a full model of user ratings. Algorithms of this type usually employ a probabilistic approach and see the collaborative filtering process as computing the expected value of a user prediction, given that user's prior ratings on other items. The model building analysis may be performed by different machine learning algorithms such as Bayesian networks, clustering, or rule-based approaches (Sarwar, et al., 2001). A Bayesian network approach, as in Breese, J., Heckerman, D., and Kadie, C. (1998), develops a Bayesian probabilistic model for collaborative filtering recommendation. Clustering approaches treat the collaborative filtering problem as a classification problem. Such approaches work by grouping similar users in the same class and then estimate the likelihood that a particular user is in a particular class $C$ to compute the conditional probability of ratings. Rule-based approaches apply methods of association rule discovery to derive rules based upon the strength of association between items.

User-based collaborative filtering systems have been widely deployed but after a substantial record of use have been shown to have a number of significant limitations such as susceptibility to data sparsity and scalability issues. As an example, commercial recommender systems have been used to evaluate very large data sets, e.g., Amazon.com and other large e-



commerce sites. Typically, in these systems even the more frequent users may have viewed or

selected well under 1% of the items. As a consequence, the resulting very sparse matrices may

produce an intractable computational problem for a recommender system using typical nearest

neighbor algorithms, as a result the accuracy and utility of recommendations may be poor.

Computational scalability is another problem. Many nearest neighbor algorithms require

computation that grows with both the number of users and the number of items. The

computational complexity of these problems is such that with millions of users and items, a

typical web-based recommender system running existing algorithms my encounter serious

scalability problems.

*Feature-based biased matrix factorization* BiasedMF (Zhou, Y., Wilkinson, D.,

Schreiber, R. and Pan, R., 2008): This is an alternating least squares (ALS) implementation of

matrix factorization using parallelization and is suitable for large datasets. Matrix factorization

encompasses a body of dimension reduction procedures to predict a personalized ranking or a set

of items for an individual user with similarities among users and items. The idea behind matrix

factorization is to represent sparse and high dimensional user/item matrices in a lower

dimensional latent space. Factors in the latent space are estimated as a result of the matrix

factorization method, in much the same way that principal component analysis (PCA) can

identify latent factors to express a correlation matrix of continuously scaled variables in reduced

dimensions. Since the initial work by Funk in 2006 a number of matrix factorization approaches

have been proposed for recommender systems.

When possible, it is best to account for the user and item measurement biases often

present in collaborative filtering (Hu, Y., Koren, C. and Volinsky, C., 2008). As an example, in

the context of predicting movie preference rankings, if Sam can be described as having a critical

temperament and consistently gives movies an average preference ranking of 2 stars this will



reflect a bias toward a lower ranking. Conversely, if Sally is less critical in nature and

consistently delivers preference rankings of 4 stars this will reflect a bias toward a higher

ranking. As a result a ranking of 3 from Sam is likely very different than a ranking of 3 from

Sally. This sort of measurement discrepancy is very common in the domain of preferences and

perception. For example, in the medical setting a "stoic" patient stating his or her pain is 2 on a

scale of 1 to 10 may not reflect the same underlying scale as an "emotionally reactive" patient

stating their pain is a 7 since the former statement from the stoic is likely to be an attenuated

measure while the later statement from the emotionally reactive patient is likely to be an

emphasized measure.

The same logic applies to potential biases of item rankings. If the movie *Dumb and

Dumber* gets an average rating (measured across all users) of 1 star while the movie *Apocalypse

Now* is given an average ranking of 5 stars then a 3 star average rating is very different for *Dumb

and Dumber* than for *Apocalypse Now*. Of course it is possible that the full population of raters

actually sees *Apocalypse Now* as the better movie but that is unlikely when measured across a

large and diverse sample of viewers. As a result, if we can attempt to take into account the

average user rating for each movie (the item bias), and incorporate that information into our

move preference predictions the recommendations may have higher validity.

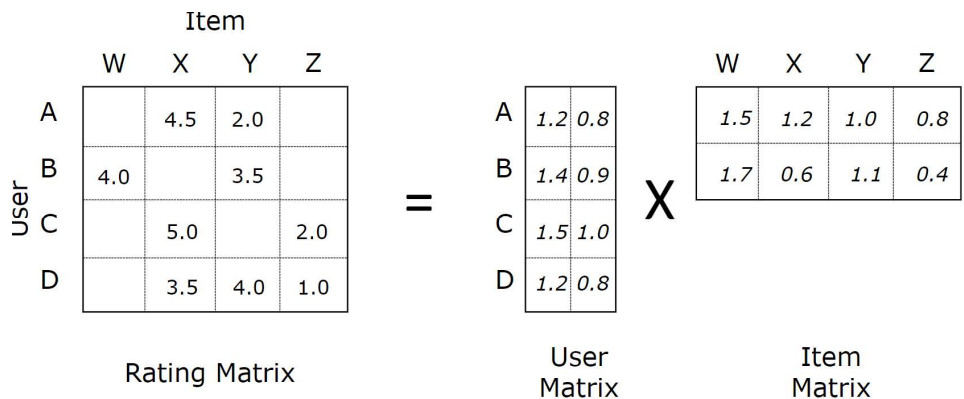

*Figure 4*



As shown in Figure 4, the ALS algorithm factors the user-item rating matrix $M$ as the product of two lower dimensional matrices $U$ and $P$, where the first one has a row for each user, while the second has a column for each item being rated. The row or column associated to a specific user or item is identified as a latent factor; note that in the ALS form of matrix factorization singular value decomposition (SVD) is not applied. The model may be tuned by altering the number of latent factors. It has been shown (Jannach, D., Lerche, L., Gedikli, F. and Bonnin, G., 2013) that a matrix factorization with a single latent factor is equivalent to a top popular recommender, one without user personalization. In alternating least squares (ALS) the user-item matrix $M$ is decomposed into the two matrices $U$ and $P$ and for a solution the algorithm will learn two types of variables, those for $U$ and those for $P$. If we hold $P$ constant and solve for $U$ alone, the problem is then reduced to linear regression. Alternating least squares is a two-step iterative solution applied to $U$ and $P$, in each iteration the algorithm initially fixes $P$ and then solves for $U$, then it fixes $U$ and solves for $P$. Since the typical ordinary least squares solution is unique and results in a minimal mean squire error (MSE), in each step the error will either decrease or stay constant but never increase. In ALS the alternation between the two steps guarantees reduction of the error until a solution is found. Similar to gradient descent, the ALS approach is certain to find a local minimum that ultimately depends on the initial values for $U$ or $P$. Another advantage is that each step can potentially be solved by applying large-scale parallelization.

*Feature-based SVD* (Funk, 2006; Paterek, 2007): This variation of singular value decomposition (SVD) is another often-used dimension reduction approach in recommender algorithms based upon feature analysis using matrix factorization with stochastic gradient descent. Most SVD implementations allow the size of the feature set as well as the number of training iterations to be set at run time. The SVD analysis identifies a latent set of features which



may be used to describe items as well as distinctive attributes of users. The resulting vectors of item features and user features may then be used to match items with users according to commonality of the features. We can also view SVD as similar to other dimension reduction techniques such as principle component analysis (PCA) or factor analysis but using discrete (1, 0, or ordinal) values for input versus the continuous values often used with parametric implementations of PCA. Ideally, PCA will produce a set of orthogonal component factors such that all components are fully independent of one another and share no common variance. As a result, this set of orthogonal components will produce a parsimonious representation of the original input matrix. SVD has the capability to produce the principle component representation when the input data matrix is sparse with many missing values, typical for user-item rating matrices in recommender analysis. Rather than setting the missing values to zero or attempting to use other corrections to the sparse matrix, Funk's SVD algorithm simply ignores the missing values; empirical investigation confirmed the effectiveness of this approach.

$$
\hat{X} \approx U \quad\quad S \qu\quad V^{\mathsf{T}}
$$

$$
\begin{pmatrix} x_{11} & x_{12} & \cdots & x_{1n} \\ x_{21} & x_{22} & & \\ \vdots & \vdots & \ddots & \\ x_{m1} & & & x_{mn} \end{pmatrix}_{m \times n} \approx \begin{pmatrix} u_{11} & \cdots & u_{1r} \\ \vdots & \ddots & \\ u_{m1} & & u_{mr} \end{pmatrix}_{m \times r} \begin{pmatrix} s_{11} & 0 & \cdots \\ 0 & \ddots & \\ \vdots & & s_{rr} \end{pmatrix}_{r \times r} \begin{pmatrix} v_{11} & \cdots & v_{1n} \\ \vdots & \ddots & \\ v_{r1} & & v_{rn} \end{pmatrix}_{r \times n}
$$

*Figure 5.*

Singular value decomposition (SVD) (Sarwar, B., Karypis, G., Konstan, J. and Reidl, J. ,2000) is a common matrix factorization technique able to express an $M$ x $N$ user Rating/Item matrix $X$ as three component matrices where $X = U \cdot S \cdot V^{\text{transpose}}$ as in Figure 5. $U$ is a singular orthogonal matrix representing the loading of users upon latent factors, $S$ is a diagonal matrix representing the magnitude of each latent factor, and $V^{\mathsf{T}}$ is another singular orthogonal matrix representing the loading of items upon the latent factors. All the entries of matrix $S$ are positive and stored in decreasing order of their magnitude, reflecting the strength of each latent factor. We say a factor



is latent because it is not directly measured but is determined by weighted linear combinations of the measured input data; as a result the latent factors are abstractions of the empirical observations. The matrices obtained by performing SVD are very useful for dimension reduction since SVD provides the best lower-rank approximations of the original matrix *R*. We can also reduce *S* to have only *k* largest diagonal values thereby determining the number of calculated latent factors. Most often we calculate SVD in recommender systems to perform two different tasks: First, we can use SVD to estimate latent relationships between both customers and products, allowing us to compute the predicted attractiveness of a given product to a particular customer, and second we can use the technique to derive a *lower-dimensional* representation of the original customer-product information, then estimate its neighborhood in the reduced dimensions. We can then use those results to produce a list of *top-N* product recommendations for users.

It is helpful to compare dimension reduction techniques such as PCA and SVD with feature selection methods (Khalid, S., Khalil, T. and Nasreen, S., 2014). Feature selection includes or excludes features but does not change them, thereby reducing the set of features. In contrast we have seen how dimension reduction techniques will transform measured features to a reduced model, often mapping measured input variables onto a smaller set of latent factors or components. In comparison, representative feature section techniques include removing features with missing values, eliminating features with low variance, excluding highly inter-correlated features, using univariate feature selection, and employing recursive feature elimination.

The weakness of conventional CF algorithms for large, sparse databases led Sarwar, et al. (2001) to explore alternative recommender system algorithms. In their work, latent semantic indexing (LSI) was used to reduce the dimensionality of the large and sparse user-item ratings using SVD. LSI is functionally another term for SVD and a dimensionality reduction technique



that has been broadly used in information retrieval (IR) methods. As we have seen, LSI, which relies upon singular value decomposition (SVD) as its underlying dimensionality reduction algorithm, maps well onto CF recommender algorithms since SVD is demonstrated to successfully produce lower-rank approximations.

**Deep Learning Models**

*Restricted Boltzmann Machine* (RBM) (Salakhutdinov, R., Mnih, A. and Hinton, G., 2007; Zhang, 2019): The original RBM configuration is a two layer neural network but can be easily stacked into a deep net. *Restricted* as used here means that there are no intra-layer communications in the visible or hidden layers. The RBM is used to extract latent features of user preferences or item ratings in recommender systems (Deng, S., Huang, L., Xu, G., Wu, X. and Wu, Z., 2017). The RBM is used for collaborative filtering, and is an implementation of an undirected graphical model able to build recommendations from tabular data such as user's ratings of movies.

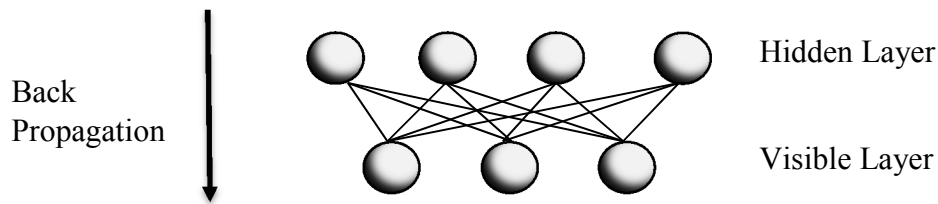

*Figure 6*. A two-layer Restricted Boltzmann Machine

As shown in Figure 6 the outer (visible) layer receives input data in a forward pass through the network then training weights and biases are updated between the multiple layers during backpropagation. RBM's came into prominence in recommender systems after Hinton, et al. (2006) introduced RBM fast learning algorithms. RBM's have applications in dimensionality reduction, classification, and collaborative filtering and can be trained in either supervised or



unsupervised ways. As their name suggests, RBM's are a variant of the Boltzmann Machine (a form of recurrent neural network) with the requirement that their nodes form a bipartite graph such that a pair of nodes from each of the two groups of units (often referred to as the visible and hidden units respectively) will have a symmetric connection between them with no connections between nodes within a group. By contrast, unrestricted Boltzmann machines may have connections between hidden units. This restriction of the RBM allows for more efficient training algorithms than are available for the general class of Boltzmann machines, and restricted Boltzmann machines can also be configured as multilayer deep learning networks. Deep belief networks (Hinton, 2009) are formed by "stacking" RBM's and fine-tuning the resulting deep network with gradient descent and backpropagation.

In the context of recommendation, RBM's may be used for jointly modeling correlations between a user's rated items and other users who rated a particular item to improve the accuracy of a recommender system (Georgiev, K. and Nakov, P., 2013). RBM's have also been used in group-based recommendation systems to model group preferences through jointly modeling collective features and group profiles (Hu, L., Cao, J., Xu, G., Cao, L., Gu, Z. and Cao, W., 2014). RBM's are used primarily to produce a low-rank representation of user preferences and for integrating correlations between the user or item pairs and neighborhoods within visible network layers. Using a hybrid model it is possible to combine both user-user and item-item correlations in which hidden layers are connected to two visible layers, with one for items and one for users (Georgiev, et al., 2013). RBM's have simpler parameterization and are more scalable than Boltzmann Machines, hence are preferred when the pairwise user and item correlations are considered. RBM's will also handle large datasets and can enable integrating other auxiliary information from different data sources into the recommendation.



Wang J. and Kawagoe K. (2018) used a RBM model for web page based recommender systems so users could be provided with the most interesting items predicted by the system based upon previous user/item interactions. Du, Y., Yao, C., Huo, S. and Liu, J. (2017) and Liu, X., Ouyang, Y., Rong, W. and Xiong, Z. (2015) utilized a RBM method based on a multilayer network architecture for better feature extraction and to address the cold start problem. These authors developed a model called Item Category Aware Conditional Restricted Boltzmann Machine Frame Model (IC-CRBMF) to integrate item category information as the conditional layer for optimizing the model parameters and improving the accuracy of the recommender. This model combines two different components based on the difference in the visible layer's presentation: an IC-CRBMF item based component and an IC-CRBMF user based component.

*Multilayer Autoencoder* (AE) (Cao, S., Yang, N. and Liu, Z., 2017). The basic form of an autoencoder is a feedforward, non-recurrent neural network having an input layer, an output layer, and one or more hidden layers which may be viewed as the latent space, as shown in Figure 7. In its basic form the output layer has the same number of nodes as the input layer, with the purpose of reconstructing its inputs by minimizing the difference between the input and the output. Autoencoders are unsupervised learning models and do not require labeled inputs. The autoencoder has a dimension reduction function, developing a parsimonious representation of its input feature set.



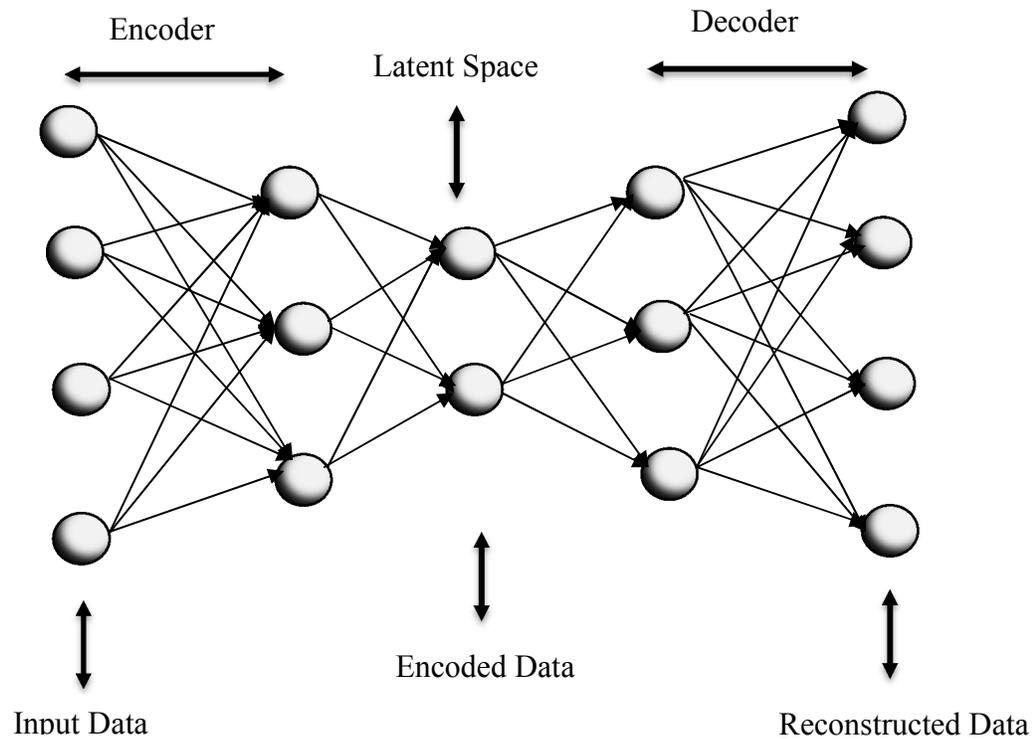

*Figure 7.* A multilayer deep learning autoencoder network

The autoencoder is a feedforward non-recurrent neural network similar to a single layer

perceptron that can be combined in a multilayer perceptron (MLP), having an input layer, an

output layer, and one or more hidden layers, where the output layer has the same number of

nodes (neurons) as the input layer. This network has the purpose of reconstructing its inputs

(minimizing the difference between the input and the output) instead of predicting the target

value $y$ given inputs $x$. Therefore, autoencoders are unsupervised learning models and do not

require labeled inputs for training.

Autoencoders (AE) were initially used to learn a reduced representation of the data

(encoding). The autoencoder has an encoder function, which reduces the data using one or more

hidden layers, and the decoder, to transform the encoding back into its original form. As there is

dimension reduction, the autoencoder will need to learn a representation in a lower dimension of



the input (the latent space) to be able the reconstruct the input. In the context of RS, an autoencoder can be used be used to predict new recommendations. To do so, the input and the output are both based upon user click vectors (it is usual for an AE that the input and the output are the same) and we will have reduced information after the input layer. This means that the model will have to reconstruct the click vector when some element from the input will be missing, hence learning to predict the recommendation for a given click vector.

In rating-based collaborative filtering, we have $m$ users, $n$ items, and a partially observed user-item rating matrix. With the autoencoder each user can be represented by a partially observed vector $r^{(u)}$. Similarly, each item can be represented by a partially observed vector $r^{(i)}$. The autoencoder can take as input each partially observed $r^{(i)}$ $(r^{(u)})$, project it into a low-dimensional latent (hidden) space, and then reconstruct $r^{(i)}$ $(r^{(u)})$ in the output space to predict missing ratings for purposes of recommendation (Sedhain, S., Menon, A., Sanner, S. and Xie, L. , 2015).

## Metrics for evaluating recommender performance

For the proposed investigation six metrics have been identified: Mean Absolute Error (MAE), Root Mean Square Error (RMSE), Net Discounted Cumulative Gain (nDCG), Precision, Recall, and Receiver Operating Characteristic (ROC). Accuracy addresses the question of how closely a recommendation list corresponds to how a user actually did, or would, rate an item (Gunawardana, A. and Shani, G., 2009). Partition of rating data into training and testing sets with a "hold one out" technique (Gedikli, F., 2013) is able to compare how a user did rate an item versus how the recommender rated that item in a Top N list. The train / test split can also be enhanced with the use of k-fold cross validation (Polatidis, N., Kapetanakis, S., Pimenidis, E. and Kosmidis, K., 2018) in which a number of randomly assigned training sets, or folds, are partitioned from the original data source. Recommender accuracy is then assessed using the test



set, showing how closely the recommender rated items the user had already seen and rated but were withheld from the training set. This comparison produces MAE and RMSE measures of recommendation accuracy, both suited for regression problems and algorithms where observed versus predicted outcomes can be compared.

The mean absolute error (MAE) is a widely accepted evaluation metric in recommender systems research (Coba, L., Symeonidis, P. and Zanker, M., 2018) and is used to measure the predictive accuracy of recommender algorithms.

$$\boldsymbol{MAE} = \frac{1}{N} \sum_{i=1}^{N} |yi - \hat{y}i| \qquad (1)$$

The MAE metric measures the average absolute deviation between the predicted rating of the recommender $\hat{y}$ and withheld user's actual ratings $y$. The lower the MAE value the better a recommender can predict a person's evaluation of an item. The root mean squared error RMSE is another often use metric in recommender research.

$$\boldsymbol{RMSE} = \sqrt{\frac{1}{nT} \sum_{i,t} (\hat{y}_{i,t} - y_{i,t})^2} \qquad i = 1,...,n \quad t = 1,...,T \qquad (2)$$

Monotonically related to MAE, RMSE is also used to measure the predictive accuracy of recommender algorithms and is a statistical accuracy metric which gained popularity during the Netflix recommender prize competition. The RMSE metric measures the average absolute deviation between the prediction rating by a recommender and a user's real but again withheld rating, however since the RMSE is an averaged value it puts more emphasis on larger prediction errors. Measure of MAE and RMSE are suitable for regression-based analysis with a goodness of fit between estimated outcomes compared with observed outcomes.



In addition to accuracy measures of MAE and RMSE suitable for regression based analyses, net discounted cumulative gain (nDCG), prediction recall, and prediction precision of the recommender can also be measured for classification problems. The nDCG is used for evaluating the effectiveness of the list of recommended items and is a utility score for a ranked list of items.

$$nDCG = \frac{1}{Z} \sum_{k=1}^{N} (2^{\text{rel}_k} - 1 / \log_2(k+1)) \qquad (3)$$

This metric is suitable for evaluating the rank of each recommended item inside the recommendation list and gives more reward, a higher value, when a more relevant item is recommended at higher position in the recommendation list than less relevant ones (Coba, et al., 2018). The nDCG is motivated by the purpose of recommendation being to identify the most relevant items for a given user.

Precision and recall are two other classification metrics used in recommendation analysis and are also useful for measuring the quality of information retrieval tasks in general. Both of these measures (Jannach, D., Zanker, M., Felfernig, A. and Friedrich, G., 2011) are computed as fractions of hits, the number of correctly recommended items for a given user. The precision metric relates number of relevant hits to the total number of recommended items. In contrast the recall metric computes the ratio of relevant hits to the maximum number of relevant hits given by the testing set size. According to McLaughlin and Herlocher (2004) accuracy measurement based on precision and recall more closely reflects the actual user experience than accuracy measures such as MAE and RMSE alone.

In the context of recommendation systems we are most likely interested in recommending top-N items to the user so we can compute precision and recall metrics for the first N items



instead of all the items, thus the notion of *precision and recall at k* where k is a user definable

integer that is set by the user to match the top-N recommendations objective.

$$\textbf{\textit{Precision@k}} = \frac{n\ of\ recommeded\ items\ that\ are\ relevant}{n\ of\ recommeded\ items\ @k\ that\ are\ relevant} \qquad (4)$$

$$\textbf{\textit{Recall@k}} = \frac{n\ of\ recommeded\ items\ @k\ that\ are\ relevant}{total\ n\ of\ relevent\ items} \qquad (5)$$

The Receiver Operating Characteristic (ROC) is a sensitivity measure originally defined

by Green, D., and Swets, J. (1966) in their work on signal detection theory and psychophysics.

The ROC curve is a 2 dimensional representation of classifier performance with the TPR (true

positive rate) plotted on the Y-axis and FPR (false positive rate) plotted on the X-axis. This

follows well from a confusion matrix, Figure 8, where we have the TPR being equal to True

Positive/ Total Positive and the FPR being equal to False Positive/Total Negative.

| Observed | Predicted | |
|----------|-----------|-----------|
| | **Positive** | **Negative** |
| **Positive** | True Positive (TPR) | False Negative |
| **Negative** | False Positive (FPR) | True Negative |

*Figure 8.* Confusion Matrix

By tuning a threshold value, all the items ranked above the curve are observed (detected) and

items below the curve are unobserved (undetected). As a result different threshold values will

produce different detection values. The ROC curve is produced by plotting the different

threshold and detection values. In the context of recommender system evaluation ROC is

discussed in Herlocker, J., Konstan, J., Terveen, L., and Riedl, J. (2004). The area under an ROC

curve (AUC), also known as Swet's *A* measure, is a useful metric of a system's ability to

discriminate between desired and undesired items. The area underneath the ROC curve is

equivalent to the likelihood the system will be able to discriminate correctly between two

alternatives, one randomly sampled from the set of undesired items, and one randomly sampled



from the set of desired items (Hand, D. and Till, R., 2001). We can view ROC sensitivity as a

measure of diagnostic power. In the case of binary (yes/no) classification the AUC will be the

area under the ROC curve. For multiple levels of classification the AUC can be estimated as the

weighted average of AUC's produced by taking each class as a reference class, setting it as class

0 and all other classes as class 1. The AUC is functionally an indicator of the quality of a ranking

so recommender performance increases for increasing values of the AUC.

**Summary**

The literature review began with a discussion focusing upon conventional collaborative

filtering methods, followed by a brief review of neural networks and then deep learning

(multilayer) neural networks. Next, the review emphasized the five algorithms selected for

replication experiments. This included three conventional algorithms: KNN user-user

collaborative filtering, feature-based biased matrix factorization using alternating least squares

(ALS), and feature-based singular value decomposition (SVD). Two deep learning algorithms

are also selected: a restricted Boltzmann machine and an autoencoder network. The chapter

concluded with a discussion of metrics selected for accuracy evaluation of the five algorithms:

mean absolute error (MAE), root mean square error (RMSE), net discounted cumulative gain

(nDCG), precision, recall, and receiver operating characteristic (ROC). Being focused upon

prediction accuracy, the present investigations will rely upon RMSE and MAE.



# Chapter 3

# Methodology

The proposed research used five recommender algorithms in a series of experiments to
investigate the comparative accuracy of a suite of collaborative filtering methods, with both
classic as well as deep neural approaches applied to large-data sources of movie rating data.
Initially, models were developed and tested with smaller datasets from the MovieLens (ML)
collection then applied to larger MovieLens datasets. As the investigation progressed, other
partitions of the MovieLens ratings were created to provide a series of progressively sized ratings
datasets. Head to head comparisons of the five algorithms were completed. The candidate
parameters that could be varied for each algorithm during the investigation included:

(1) *KNN User-User Collaborative Filtering.*

    Size of the input ratings dataset
    Number of retained recommendations
    Minimum similarity

(2) *Feature-based Biased Matrix Factorization* BiasedMF

    Size of the input ratings dataset
    Number of features to train
    Number of iterations to train
    The regularization factor
    Damping factor for the underlying mean
    Bias option
    Solver method

(3) *Feature-based SVD – Singular Vector Decomposition*

    Size of input ratings dataset
    Number of features to train
    Number of iterations to train each feature
    The learning rate
    The regularization factor
    Damping factor for the underlying mean
    The underlying Bias model to fit
    The minimum and maximum rating values for ratings



(4) *Restricted Boltzmann Machine* (RBM)

   Size of input ratings dataset
   Epochs
   Visible dimensions
   Hidden dimensions
   Rating values
   Learning rate
   Batch size

(5) *Multilayer Autoencoder* (AE)

   Size of input ratings dataset
   Epochs
   Visible dimensions
   Hidden dimensions
   Learning rate
   Batch size

During execution of the recommender algorithms the models were trained and tested with train/test data partitions. Crossfold methods packaged with the recommender algorithms were applied to ratings data, or user data, to produce (train, test) data pairs to use for model validation. For example, a 5-fold partition split resulted in 5 splits, each of which selected 20% of input rows for testing and 80% or rows for training. Data Crossfold preparation can take as input the rating data to produce a row-based partition or may take user data as input to result in a user-based partition.

Algorithm performance during each phase of the investigation was assessed using metrics defined previously: Mean Absolute Error (MAE), Root Mean Square Error (RMSE), Net Discounted Cumulative Gain (nDCG), Precision, Recall, and Receiver Operating Characteristic (ROC). Evaluation of model goodness of fit (predicted versus observed) and accuracy of the investigated algorithms are primarily addressed with MAE and RMSE to make comparative evaluations of predictive accuracy.



**Experiment One: KNN User-User Collaborative Filtering.**

　　User-based collaborative filtering is the original form of collaborative filtering (Resnick, et al., 1994) and attempts to identify users with preferences similar to the active user then to recommend items preferred by those users. This can be accomplished by identifying user neighborhoods to arrive at an estimate of the active user's preference for each item, often using a weighted mean of the neighbors' ratings, then returning the highest predicted items as a recommendation. The user-user CF algorithm used in this experiment is limited to using explicit ratings obtained from the user item rating input data. The central class in the implementation framework (Ekstrand, 2019) is a user-user item scorer and results in a user-user collaborative filter. The user-user item scorer uses a user-event data access object (DAO) to obtain the user data from the input rating data set, a user-vector normalizer to standardize user data from the active user as well as potential neighbors, then a method to identify user neighbors. The neighborhood finder employs a single method that receives as input a user profile and a set of items to be scored and returns a set of candidate neighbors with each neighbor object corresponding to a user with a rating vector and having similarity to the active user. The item scorer then uses the set of candidate neighbors to score the items as shown below where $u$ denotes the standardized version of a user $u$ or a rating value:

$$\text{Score}(u,i) = \text{denorm} \left( \left( \Sigma_{v \in N(u,I)} \, \text{sim}(u, \, v) r_{vi} \, / \, \Sigma_{v \in N(u,i)} \, |\text{sim}(u,v)| \right); \, u \right) \qquad (6)$$

If users are standardized by mean-centering of their ratings, this equation corresponds to the same calculation discussed in Resnick, et al. (1994).

　　The user-user collaborative filtering algorithm processes the input ratings matrix for candidate neighbors and only users who have rated at least one of the items to be scored are retained as neighbors. Also, given the sparse similarity function, users who have not explicitly



rated any of the same items rated by the active user were not be retained as neighbors. To optimize the identification of neighbors, the neighborhood finder takes the smaller of the active user's set of rated items plus the set of target items, and considers all users who have rated at least one item among them.

The implementation framework also includes an optional neighborhood finder function using the same logic as the default finder but working instead with a capture of the rating data stored in memory. As a result, this is more efficient to process than using an external ratings file for identifying neighbors and can make this algorithm more practical to use. When using this optional neighborhood finder, the active user's most recent ratings will still be used, but their set of candidate neighbors are fixed as of the time the sample was taken. If this algorithm were used in off-line production, this would happen in batch mode. The user-user collaborative filtering algorithm supports a range of user similarity functions via a similarity interface allowing cosine, MSD, Pearson, and Pearson Baseline methods to be selected.

*Experimental Design*

The first experiment applied the KNN user-user collaborative filtering algorithm to a MovieLens ratings data set. The implementation of the algorithm has four parameters that may be varied: The size of the ratings data set, selection of test subjects to be provided recommendations, the number of retained recommendations for output, and a minimum similarity value as the neighbors are scored.

*Presentation of Results*

Results from this experiment are presented in a tabular format and included columns for the input data set size, number of retained recommendations, similarity level, and algorithm accuracy (RMSE or MAE as appropriate). The results appear similar to Table 1.



| Input Data Size | # Retained Recommendations | Similarity Level | Accuracy |
|---|---|---|---|
| ####### | ### | ##### | ##### |
| ####### | ### | ##### | ##### |

*Table 1*        Experiment One KNN User-User Collaborative Filtering

**Experiment Two: Feature-Based Biased Matrix Factorization with ALS (BiasedMF).**

The implementation framework supports collaborative filtering using weighted lambda regularization with alternating least squares (Zhou, Y., et al., 2008), and subsequent work. The input user-rating matrix *R* contains ratings data but also much noise due to the sparse nature of the matrix, hence it is important to minimize the effect of the missing values and use the recovered information to predict missing ratings. *Singular Value Decomposition* (SVD), demonstrated in Experiment Three, is a commonly used dimension reduction technique to approximate the raw user-movie rating matrix *R* as the product of two lower dimension matrices. However, because there are many missing elements in the rating matrix *R*, typical SVD algorithms cannot solve for *U* and *M.,* fortunately, the alternating least squares (ALS) approach is able to solve the lower rank matrix factorization problem. A typical ALS solution uses the following steps:

**Step 1.** Initialize matrix *M* by assigning the average rating for that movie as the first row, and small random numbers for the remaining entries.
**Step 2.** Hold *M* constant then solve for *U* by minimizing the objective function;
**Step 3.** Hold *U* constant then solve for *M* by minimizing the objective function in a similar manner.
**Step 4.** Repeat Steps 2 and 3 until a stopping criterion is reached.

Within the ALS algorithm when the regularization matrices $\Gamma(U,M)$ are nonsingular, Steps 2 and 3 of the algorithm have a unique solution. The sequence of calculated errors is monotonically non-increasing and bounded, so the sequence converges. Rather than proceeding all the way to convergence, ALS uses a stopping criterion based on the observed RMSE on values in the probe dataset. After each cycle of updating *U* and *M*, if the delta of RMSE on the



probe dataset is less than a preset minimum, the iterations halt and we can use the obtained $U$

and $M$ to make final predictions on the test dataset. There are many free parameters in the model

so in the absence of regularization, ALS might result in overfitting. A standard adjustment is to

use Tikhonov regularization (Tikhonov, A., Goncharsky, A., Stepanov, V. & Yagola, A., 1995),

a form of ridge regression also useful for reducing multicollinearity in regression models, which

helps the algorithm to reduce the parameter set. The ALS algorithm in the framework used the

following weighted lambda regularization so as to not overfit the test data as the number of

features or iterations increase:

$$f(U, M) = \Sigma_{(I,j) \in I}\, (r_{ij} - u^T_i m_j)^2 + \lambda\, (\Sigma_i\, n_{ui}\, \|u_i\|^2 + \Sigma_j\, n_{mj}\, \|m_j\|^2) \qquad (7)$$

where $n_{ui}$ and $n_{mj}$ denote the number of ratings of user $i$ and movie $j$ respectively.

*Experimental Design*

The second experiment applied biased matrix factorization with ALS to a MovieLens

ratings data set. The implementation of this algorithm has seven parameters that may be varied:

the size of the input ratings dataset, number of features to train, number of iterations to train,

regularization factor, damping factor for the underlying mean, the bias option, and the method

used by the solver.

*Presentation of Results*

Results from this experiment are presented in a tabular format and included columns for

the input data set size, number of features, iterations, regularization factor, damping factor, bias

option, solver method, and algorithm accuracy (RMSE). The results appear similar to Table 2.



| Input data size | # Features | # Iterations | Regularization | Damping | Bias option | Solver method | Accuracy |
|---|---|---|---|---|---|---|---|
| #### | ### | ### | ##### | ##### | ## | ## | ##### |
| #### | ### | ### | ##### | ##### | ## | ## | ##### |

*Table 2*          Experiment Two Matrix Factorization with ALS

## Experiment Three: Feature-Based Singular Value Decomposition (SVD).

The framework used for this experiment (Kane, 2018) includes an implementation of Funk's SVD algorithm (Funk, S., 2006), introduced during the Netflix Prize competition. This version of the SVD algorithm uses classes from Scikit Surprise and also includes an option to compute probabilistic matrix factorization but in this experiment SVD is computed. The prediction $r^{\wedge}{}_{ui}$ is calculated as:

$$r^{\wedge}{}_{ui} = \mu + b_u + b_i + q^T_i p_u \qquad (8)$$

If user $u$ is unidentified then the bias $b_u$ and the factors $p_u$ can be assumed to be zero, this also applies for item $i$ with $b_i$ and $q_i$. To estimate the unknowns the algorithm minimizes the regularized squared error:

$$\sum r_{ui} \in R_{train}(r_{ui} - r^{\wedge}{}_{ui})^2 + \lambda(b^2 i + b^2 u + \|qi\|^2 + \|pu\|^2) \qquad (9)$$

The minimization is computed using stochastic gradient descent executed over all the ratings of the training data set and repeated for a number of epochs. Baselines are initialized to zero while user and item factors are randomly initialized per a normal distribution, which may be modified using the mean and standardization parameters. Parameters may also be set for learning rate $\gamma$ and the regularization term $\lambda$ and both can be set to different values for each type of parameter.



Default execution of the algorithm sets parameters for learning rates to 0.005 and sets

regularization to 0.02.

*Experimental Design*

The third experiment applied both original singular value decomposition (SVD) and

SVD++, a variation of SVD, to a MovieLens ratings data set. The SVD++ algorithm (Koren,

2009) is an extension of SVD taking into account implicit ratings. The SVD++ variation takes

into account item factors that capture implicit ratings, where an implicit rating describes the fact

that a user $u$ rated or viewed an item $j$, regardless of the rating value. The implementation of

these algorithms in Scikit Surprise has many parameters that may be varied but initially the

experiment was executed with default values for all parameters, with the exception of number of

epochs, number of factors, and the algorithm global learning rate. Varied parameters included the

size of the input ratings dataset, number of epochs, number of factors, and the learning rate.

*Presentation of Results*

Results from this experiment are presented in a tabular format and included columns for

the input data set size, SVD model variation, number of epochs, and number of factors, learning

rate, and the algorithm accuracy (RMSE and MAE). The results appear similar to Table 3.

| Input data size | Model | # Epochs | # Factors | Learning rate | Accuracy |
|---|---|---|---|---|---|
| #### | SVD | ### | ### | ##### | ##### |
| #### | SVD++ | ### | ### | ##### | ##### |

*Table 3*        Experiment Three Singular Value Decomposition

**Experiment Four: Multilayer Restricted Boltzmann Machine (RBM).**

By and large**,** the original computational approaches to collaborative filtering could not

accommodate the very large data sets which became increasingly prevalent as Internet

ecommerce and social media grew in prominence. Salakhutdinov, R., Mnih, A., and Hinton, G.



(2007) demonstrated the class of two-layer neural network Restricted Boltzmann Machines (RBM's) could be used to model large collections of data, such as user movie ratings. This paper was instrumental in launching the resurgent work in neural networks now designated as deep learning. In their 2007 paper the authors developed computationally tractable learning and inference procedures for an RBM and showed it could be applied to the Netflix Prize test data set. The data set contained over 100 million user movie ratings, and the RBM was able to produce notable improvements over SVD models.

In collaborative filtering applications the raw user ratings matrix is often very sparse. If $N$ users each rate $M$ movies the ratings matrix will have a majority of missing values since any given user will only have rated a portion, often small, of the full set of movies. Other CF approaches such as SVD use a variety of methods to handle the missing values such as assigning random values to the missing points based upon an underlying distribution, estimating the missing values by prediction, or ignoring the missing values in initial iterations of the algorithm. In the RBM approach of Salakhutdinov et al. (2007), a separate RBM is trained for each user with binary hidden units and softmax visible units – a softmax function rescales the values as probabilities summing to 1.0 using a conditional multinomial distribution. All RBMs in the implementation have an equal number of binary hidden units, but each individual RBM has visible softmax units only for the items rated by a particular user. Notably, all the weights and biases are shared among the RBMs, as a result if two users had rated the same item, their individual RBMs would use the same weights between the visible unit for that item and the hidden units. The model is learned by gradient descent. Forward and backward passes are made in the RBM to predict ratings for a target user and return top-N recommendations.

The implementation framework (Kane, F., 2018) supports an RBM solution using TensorFlow and taking input parameters for visible dimensions (product of the number of



movies times the number of distinct rating values), RBM epochs, hidden dimensions (hidden

nodes), rating values, learning rate, and batch size. Accuracy metrics are output as root mean

square error (RMSE), mean absolute error (MAE), hit rate (HR), cumulative hit rate (cHR),

average reciprocal hit rate (ARHR), coverage, diversity, and novelty.

*Experimental Design*

The fourth experiment created a restricted Boltzmann machine (RBM) multilayer neural

network to process a MovieLens ratings data set for collaborative filtering. The implementation

of this algorithm used TensorFlow and has a number of parameters that may be varied. The

experiment will be executed with varying parameters for the size of the input ratings dataset,

number of visible dimensions, epochs, number of hidden nodes, and number of distinct rating

values, the learning rate, and batch size.

*Presentation of Results*

Results from this experiment are presented in a tabular format and included columns for

the input data set size, number of visible dimensions, number of epochs, number of hidden

dimensions, number of  distinct rating values, learning rate, batch size and algorithm accuracy

metrics. The results appear similar to Table 4.

| Input Data Size | # Visible Dimensions | # Epochs | # Hidden Dimensions | # Rating Values | Learning Rate | Batch Size | RMSE | MAE | HR |
|---|---|---|---|---|---|---|---|---|---|
| ### | ### | ### | ### | # | #### | ### | ##### | #### | #### |
| ### | ### | ### | ### | # | #### | ### | ##### | #### | #### |

*Table 4*                Experiment Four Restricted Boltzmann Machine



Two variations of the RBM were tested. One version used a grid-search method to determine an

optimal model of RBM parameters of hidden dimensions and learning rate prior to a solution

while the other model did not use any preprocessing.

**Experiment Five: Multilayer Autoencoder Network (EN).**

In its simplest form an Autoencoder is a neural network designed to copy its input to its

output using a lower dimension representation. A basic autoencoder has a hidden layer $E$ that is

encoded to represent the input; the autoencoder network may be seen as consisting of two parts,

the encoder function $E = f(x)$ and a decoder that produces a reconstruction $R = g(E)$. If an

autoencoder learns only $g(f(x)) = x$ completely then it will have little informative value. To

avoid this, autoencoders are implemented so as to encode the input representation with a lower

dimension. The lower dimension representation results in the autoencoder encoding but not fully

matching the input training data. Because the model will select and prioritize which aspects of

the input will be represented it often learns useful properties of the data.

Autoencoders offer another deep learning alternative to collaborative filtering algorithms.

When contrasted with an RBM-based collaborative filtering model we see several differences.

First, the RBM is a generative, probabilistic model and an autoencoder is a discriminative model

(Sedhain, S., Menon, A., Sanner, S. and Xie, L., 2015). Second, the RBM estimates the solution

by maximizing a log likelihood function, while an autoencoder minimizes the loss function

RMSE, a standard accuracy metric in evaluating prediction models. Also, training an RBM often

requires the use of contrastive divergence, a gradient technique to approximate the slope

representing the relationship between a network's weights and its error, whereas in training an

autoencoder gradient-based backpropagation may be used, which is often faster. Finally, most

implementations of the RBM are only suitable for discrete ratings, and estimate a separate set of

parameters for each rating value resulting in large models with many terms. For $r$ possible



ratings, this means there will be *nkr* parameters for user or item based RBM. By comparison, an autoencoder is independent to *r* and so will require fewer parameters. Having fewer parameters allows an autoencoder model to have lower memory requirements and to be less prone to overfitting. Compared to matrix factorization approaches, which embed both users and items into the same latent space, an autoencoder model only embeds items into the latent space. In addition, matrix factorization models learn a linear latent representation, but an autoencoder can use different activation functions to learn nonlinear latent representations.

The implementation framework (Kane, F., 2018) implements an autoencoder (EN) model also using TensorFlow and built upon an elaboration of the RBM algorithm implementation. The EN model takes input parameters for visible dimensions (product of the number of movies times the number of distinct rating values), epochs, hidden dimensions (hidden nodes), learning rate, and batch size. Accuracy metrics paralleling those for RBM are output and include root mean square error (RMSE), mean absolute error (MAE), hit rate (HR), cumulative hit rate (cHR), average reciprocal hit rate (ARHR), coverage, diversity, and novelty.

*Experimental Design*

The fifth experiment created an autoencoder (EN) neural network to process a MovieLens ratings data set for collaborative filtering. The implementation of this algorithm used TensorFlow and has a number of parameters that may be varied. The experiment was executed with varying parameters for the size of the input ratings dataset, number of visible dimensions, epochs, number of hidden nodes, the learning rate, and batch size.

*Presentation of Results*

Results from this experiment are presented in a tabular format and includes columns for the input data set size, number of visible dimensions, number of epochs, number of hidden



dimensions, number of distinct rating values, learning rate, batch size and algorithm accuracy

metrics. The results appear similar to Table 5.

| Input Data Size | # Visible Dimensions | # Epochs | # Hidden Dimensions | # Rating Values | Learning Rate | Batch Size | RMSE | MAE | HR |
|---|---|---|---|---|---|---|---|---|---|
| ### | ### | ### | ### | # | #### | ### | ##### | #### | #### |
| ### | ### | ### | ### | # | #### | ### | ##### | #### | #### |

*Table 5*                              Experiment Five Autoencoder Network

## Resources

Given the size of the MovieLens and other input data sets plus use of deep learning

models, sufficient computation resources were required to complete the planned experiments.

Depending upon the breadth and depth of neural architectures the hyper-parameter tuning and

network training typically require long periods of processing time. Google Colab offers the

enhanced resources in an open-source environment, including optional access to GPU and TPU

(Tensor Processing Units) for computation. A discussion of the TPU origins and motivation for

development can be found in Jouppi, et al. (2018). Briefly, the TPU is able to run deep neural

network (DNN) inference 15-30 times faster and with 30-80 times better energy efficiency than

contemporaneous CPUs and GPUs. Computational Python libraries such as Numpy, Pandas,

Tensorflow, and Keras are also available within the Colab environment. Colab Plus, available for

a nominal monthly fee, offered a longer uninterrupted connection time to the Colab virtual

machine environment and access to TPU execution with a 35GB memory allocation.

## Summary

This chapter examined more closely the implementation details of the five algorithms

selected for the planned experiments, discussed the planned experimental design of each



experiment, the parameters, and conditions to be varied across replications within each

experiment, and an anticipated presentation format of the results.



# Chapter 4

# Results

## Introduction

All replications of the five experiments were executed within the Google Colab Plus environment with 35GB of memory and allocation of a Tensor Processing Unit (TPU). Each experiment included multiple replications and exercised one of the proposed recommender system algorithms. As each experiment progressed selected model parameters and input data size were modified across successive replications in response to the behavior of preceding replications. Uninterrupted connection times, memory availability, and overall clock-time execution speed is not guaranteed in the Colab Pro environment. In spite of the 35GB allocation of memory and computational resources offered by Colab Pro the more complex deep learning algorithms, e.g. the Restricted Boltzmann Machine and the Autoencoder, would not reliably run to completion on the larger input data sizes, such as 1MB and above, attempted in some replications. When memory requirements exceeded the 35GB allocation the Colab virtual environment would halt and terminate with no capability to resume execution. Some completed runs had long execution times, such as ten or eleven hours of elapsed clock time but were able to run to completion. In reaction, a series of smaller sized ratings datasets were extracted from the 20M ratings data source found on grouplens.org. These smaller datasets were sized 100K, 200K, 300K, 400K, and 500K ratings and enabled the more computationally demanding deep learning algorithms to run to completion.

Seventy-four experiments were successfully completed on the Google Colab Pro platform using Python 3.6 code and the *Surprise* Python recommender framework (Hug, N., 2020).The framework is written primarily in Python, with some computational routines written using



Cython (Behnel, S., Bradshaw, R., Citro, C., Dalcin, L., Seljebotn, D. and Smith, K., 2011).

*Surprise* closely follows the *scikit-learn* API and implements tools for model evaluation such as

cross-validation iterators, train-test data partitioning, and evaluation metrics. The framework

includes scikit-learn functions for model section and automatic hyper-parameter search, such as

grid search and randomized search. All source code for the investigations will be available at the

author's GitHub account (github.com/wbnoffsinger).

**Experiment One KNN User-User Collaborative filtering**

Results of the first experiment, using a traditional User-User collaborative filtering

algorithm, are displayed in Table Six. Seventeen replications were completed using Google

Colab Pro. Depending upon the input data size the algorithm execution time varied from 79 to

1174 seconds and returned predictive accuracy metrics of RMSE and MAE. The Scikit Surprise

framework (http://surpriselib.com) allowed four different methods for calculation of the user-

ratings similarity matrix: cosine similarity, MSD similarity, Pearson similarity and Pearson-

baseline similarity. Cosine similarity computes the similarity of ratings vectors between all pairs

of users. For this calculation only the ratings between common users are taken into account.

Fully dissimilar (independent and orthogonal) vectors will produce a cosine value of 0 while

completely similar vectors will have a cosine value of 1. Intermediate ranges of similarity or

independence will range between 1.0 and 0. The MSD similarity option calculates similarity

using the mean squared difference between all pairs of users. For MSD only ratings between

common users are taken into account. The Pearson similarity option uses the familiar Pearson

correlation calculation between two sets of ratings. With the Pearson method only user ratings in

common between users are included. The Pearson correlation coefficient can be seen as a mean-

centered cosine similarity. The Pearson-baseline similarity option computes an adjusted Pearson

correlation coefficient between all pairs of users using baselines for centering instead of means.



A shrinkage parameter helps minimize overfitting when only few ratings are available, a

shrinkage parameter of 0 is equivalent to no shrinkage. As with the other similarity methods, if

there are no common users similarity will be 0 (and not -1). The rationale for the Pearson-

baseline method is discussed in Ricci, F., Rokach, L., Shapira, B., and Kantor, P. (2015). For all

replications the number of retained recommendations in the Top-N list was set to 10 and the

maximum number of neighbors to take into account for aggregation was set to 40. Results for all

KNN User-User replications are shown in Table 6.

| Experiment Run Number | Input Data Size (# Ratings) | Retained Recommendations | Similarity Calculation | RMSE | MAE | Clock Time (secs) |
|---|---|---|---|---|---|---|
| 6 | 1.25M | 10 | cosine | .9486 | .7300 | 1174 |
| 7 | 1.25M | 10 | MSD | .8941 | .6834 | 923 |
| 8 | 1.25M | 10 | Pearson | .9357 | .7252 | 851 |
| 9 | 1.25M | 10 | Pearson-baseline | .8984 | .6957 | 1173 |
| 18 | 1M | 10 | cosine | .9517 | .7316 | 722 |
| 19 | 1M | 10 | MSD | .8982 | .6864 | 542 |
| 20 | 1M | 10 | Pearson | .9414 | .7290 | 828 |
| 21 | 1M | 10 | Pearson-baseline | .9080 | .7030 | 707 |
| 13 | 500K | 10 | cosine | .9600 | .7348 | 342 |
| 12 | 500K | 10 | MSD | .9158 | .6974 | 242 |
| 11 | 500K | 10 | Pearson | .9540 | .7362 | 386 |
| 10 | 500K | 10 | Pearson-baseline | .9296 | .7155 | 309 |
| 14 | 250K | 10 | cosine | .9778 | .7524 | 136 |
| 15 | 250K | 10 | MSD | .9402 | .7203 | 82 |
| 16 | 250K | 10 | Pearson | .9749 | .7555 | 164 |
| 17 | 250K | 10 | Pearson-baseline | .9589 | .7398 | 127 |

*Table 6.* Experiment One KNN (K=40) User-User Collaborative Filtering

Values of RMSE by size of the input User Ratings dataset and similarity calculation

method are displayed in Figure 9. As can be seen, there is a monotonically decreasing value of

RMSE for all four similarity calculation methods as the size of the input dataset increases, and

with no interaction between increasing data size and similarity method. As is true for general

linear and regression models, it follows that increasing sample sizes will be associated with



decreasing error. This is to be expected given the calculation of RMSE in the KNN algorithm corresponds to error calculations in other regression-type models.

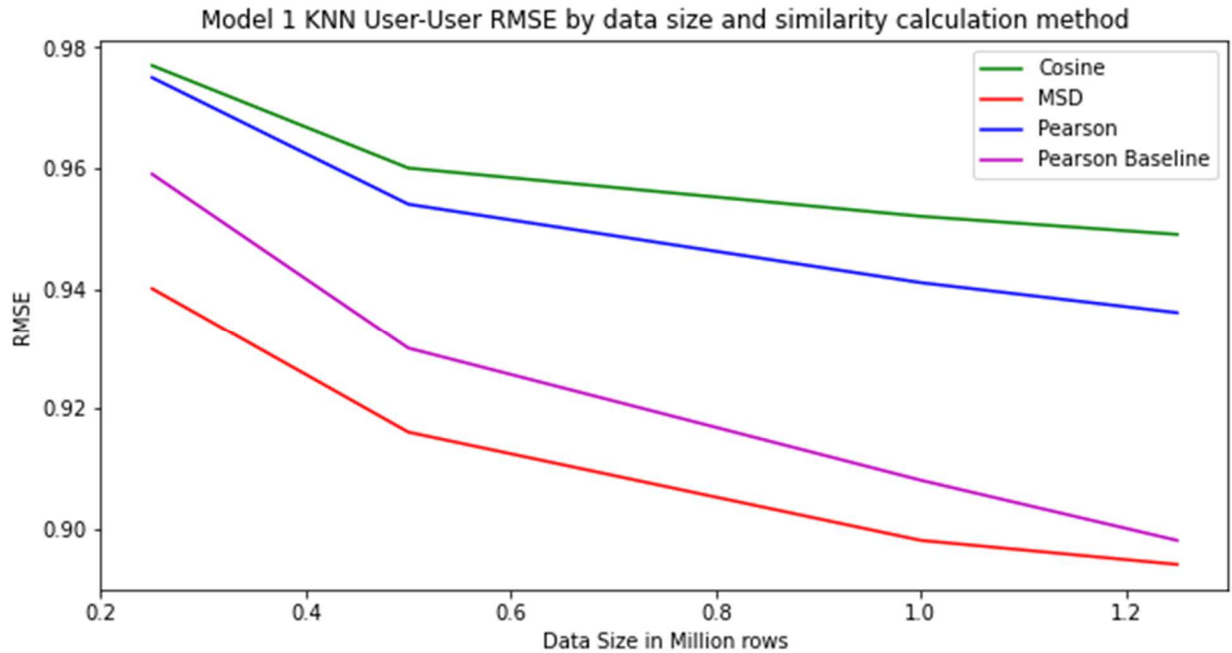

*Figure 9.* KNN User-User RMSE predictive accuracy values

Values of MAE by size of the input User Ratings dataset and similarity calculation method are displayed in Figure 10. As was the case for RMSE, there is also a monotonically decreasing value of MAE for all four similarity calculation methods as the size of the input dataset increases, with no interaction between increasing data size and similarity method. Similarly, as we observed for RMSE, it follows that increasing sample sizes will be associated with decreasing MAE, if we assume the calculation of MAE in the algorithm corresponds to error calculations in other regression-type models.



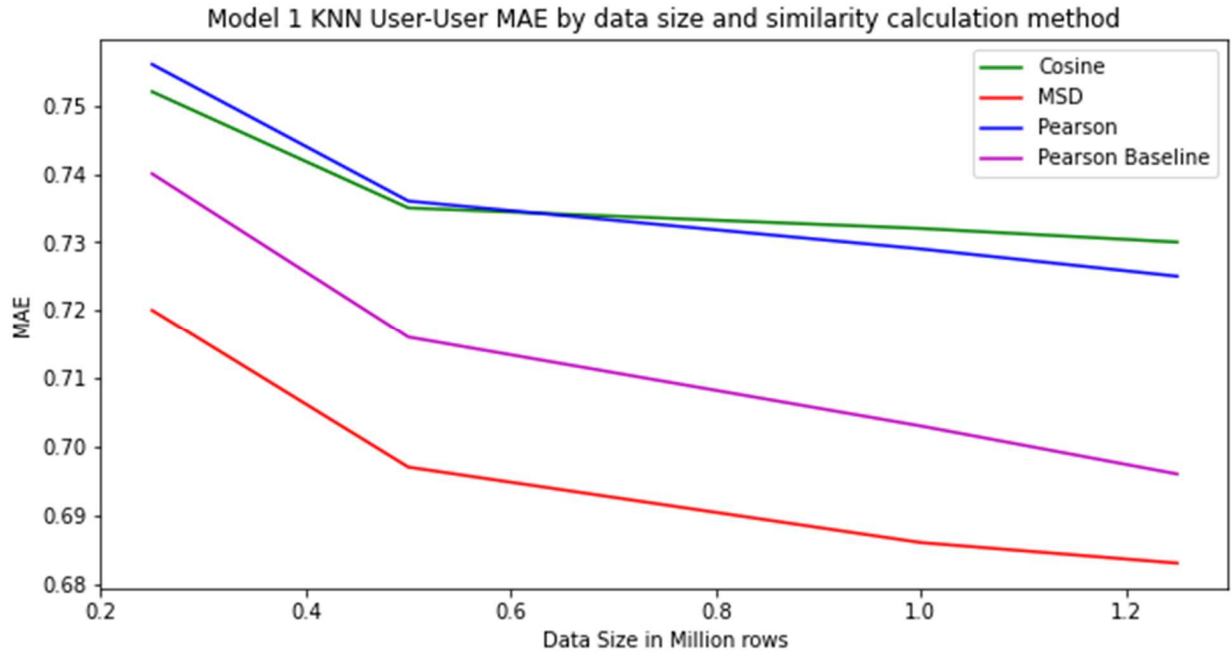

*Figure 10.* KNN User-User MAE predictive accuracy values

Apart from the input ratings dataset size, number of retained recommendations, and the

similarity calculation method no other algorithm parameters were exposed by the framework.

Execution clock time ranged from a low of 79 seconds for the 100K ratings dataset to a high of

1174 seconds for the 1.25M ratings dataset, illustrating an approximately linear relation between

increases in data size and increases in clock time. Published benchmarks (Hug, 2020) for

Surprise using cosine similarity calculation and a 5-fold cross-validation procedure show RMSE

values of .98 with a 100K ratings dataset and .92 with a 1M dataset, comparing with RMSE of

.99 (100K) and .95 (1M) for the present study.



**Experiment Two Matrix Factorization**

Results of the second experiment, using non-negative matrix factorization, are displayed in Table 7. Twelve replications were completed using Google Colab Pro. Depending upon the input data size the algorithm execution time varied from 80 to 1317 seconds and returned predictive accuracy metrics of RMSE and MAE. In this experiment a Scikit *Surprise* implementation of non-negative Matrix Factorization (Koren, Y., Bell, R. and Volinsky, C., 2009) was applied for user-based collaborative filtering. The optimization procedure in this algorithm used regularized stochastic gradient descent with a specific choice of step size to ensure non-negativity of factors, provided that the initial values are also positive.

| Experiment Run Number | Input Data Size (# Ratings) | Factors | Epochs | Similarity calculation | RMSE | MAE | Clock time (secs) |
|---|---|---|---|---|---|---|---|
| 5 | 1.5M | 15 | 100 | cosine | .8504 | .6559 | 853 |
| 6 | 1.5M | 20 | 200 | cosine | .8290 | .6387 | 1317 |
| 1 | 1M | 15 | 100 | cosine | .9168 | .7241 | 162 |
| 4 | 1M | 20 | 200 | cosine | .8402 | .6463 | 522 |
| 12 | 750K | 20 | 200 | cosine | .8476 | .6519 | 561 |
| 7 | 500K | 20 | 200 | cosine | .8624 | .6612 | 350 |
| 8 | 500K | 20 | 200 | MSD | .8624 | .6612 | 353 |
| 9 | 500K | 20 | 200 | Pearson | .8624 | .6612 | 391 |
| 10 | 500K | 20 | 200 | Pearson-baseline | .8624 | .6612 | 361 |
| 11 | 250K | 20 | 200 | cosine | .9002 | .6924 | 199 |
| 2 | 100K | 15 | 100 | cosine | .9523 | .7351 | 40 |
| 3 | 100K | 20 | 200 | cosine | .9540 | .7392 | 80 |

*Table 7.* Experiment Two Matrix factorization

Values of RMSE by size of the input User Ratings dataset are displayed in Figure 11. As can be seen, for the Matrix Factoring algorithm there is also a monotonically decreasing value of RMSE as the size of the input dataset increases. Using the 500K input dataset, four replications were completed to test for an effect of the similarity calculation method. Because results showed no effect attributable to the similarity method, the similarity method was set to cosine and not varied for replications of other data sizes. As is true for general linear and regression models, it



follows that increasing sample sizes will be associated with decreasing error. This is to be

expected given the calculation of RMSE in the matrix factorization algorithm corresponds to

error calculations in other regression-type models

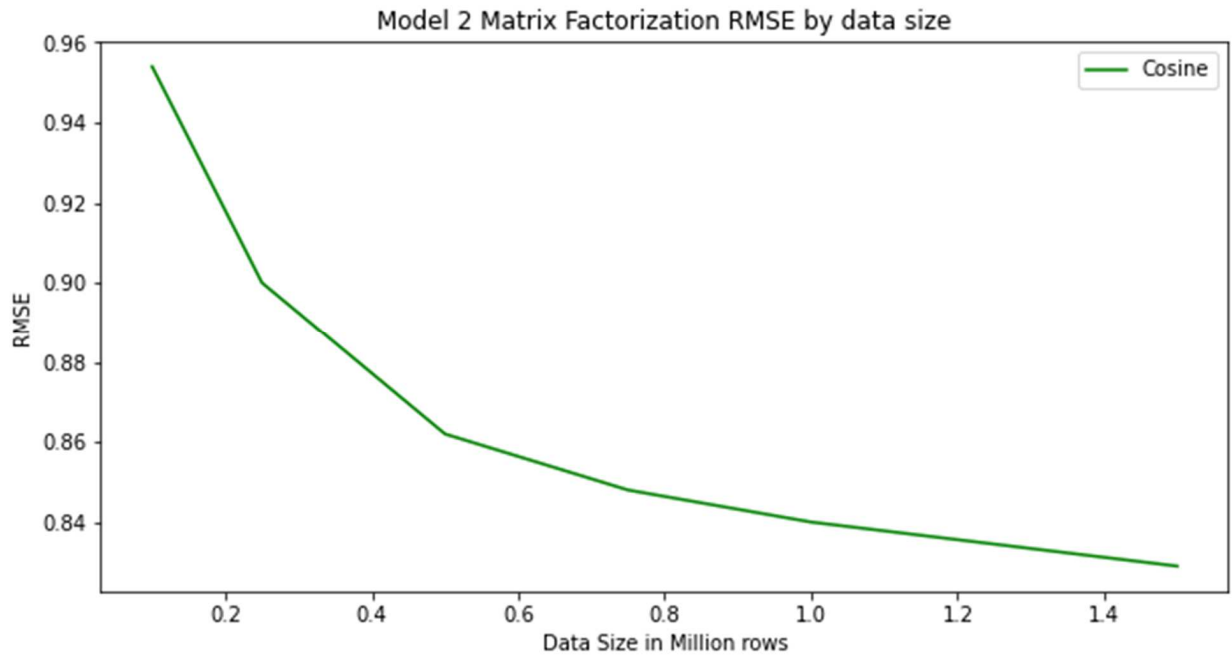

*Figure 11.* Matrix Factorization RMSE predictive accuracy values

Values of MAE by size of the input User Ratings dataset are displayed in Figure 12. As

can be seen, for MAE values obtained by the Matrix Factoring algorithm there is also a

monotonically decreasing value of MAE as the size of the input dataset increases. Using the

500K input dataset four replications were also completed to test for an effect of the similarity

calculation method upon MAE accuracy. As was true for RMSE, results showed no effect

attributable to the similarity method hence the similarity method was set to cosine and not varied

for replications of other data sizes. As was the case for RMSE, we observe decreasing values of

MAE as the sample size increases.



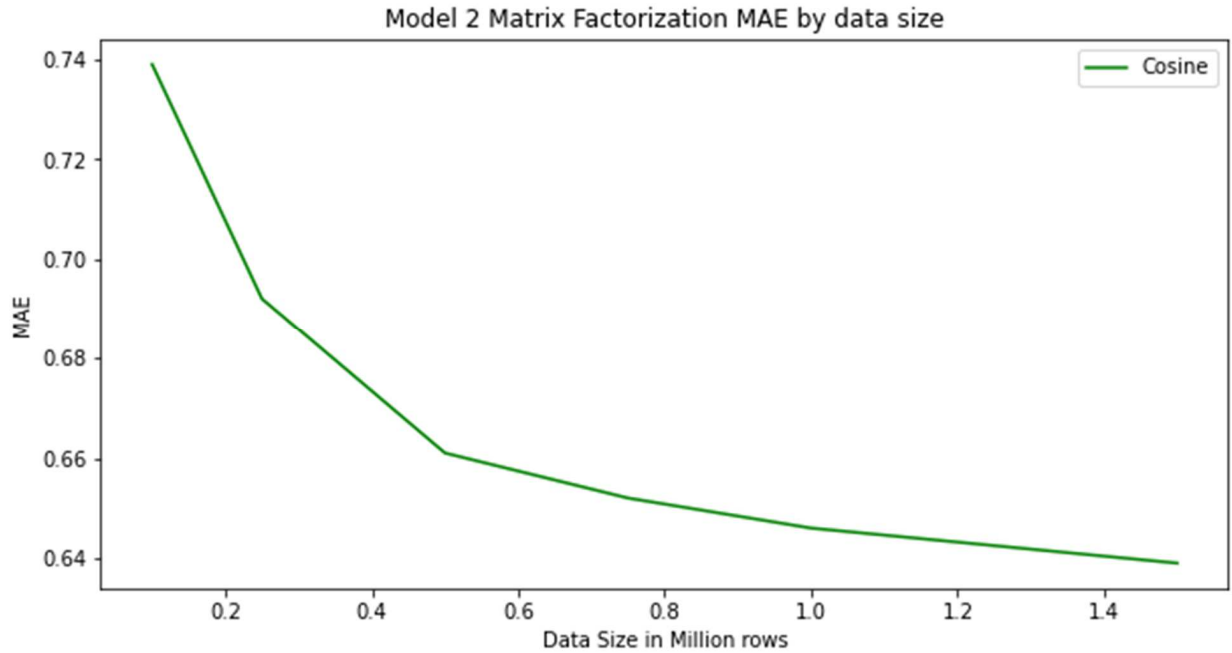

*Figure 12.* Matrix Factorization MAE predictive accuracy values

In addition to the input ratings dataset size, the number of factors retained in the matrix

reduction, and the number of calculation epochs used for stochastic gradient descent, no other

algorithm parameters were varied during the replications. Execution clock time ranged from a

low of 80 seconds for the 100K ratings dataset to a high of 1317 seconds for the 1.5M ratings

dataset, illustrating an approximately linear relation between increases in data size and increases

in clock time. Published benchmarks (Hug, 2020) for the *Surprise* implementation of non-

negative matric factorization with a 5-fold cross-validation procedure show RMSE values of .96

with a 100K ratings dataset and .92 with a 1M dataset, comparing with RMSE of .95 (100K) and

.83 (1.5M) for the present study.



**Experiment Three Singular Value Decomposition**

This experiment used the Scikit *Surprise* implementation of singular value decomposition (SVD) made notable by the Netflix recommender competition (Funk, S., 2006). The matrix minimization is performed by stepwise stochastic gradient descent with steps performed over all the ratings of the training dataset and repeated a number of times according to the value set for epochs. Baselines are initialized to .0 and both the user and item factors are randomly initialized according to a normal distribution. The SVD algorithm also allows the learning rate and regularization value to be set at run time. Default values of .005 for learning rate and .02 for regularization were held constant across all replications. An SVD++ version of the SVD algorithm was also used to make recommendation predictions with the input rating datasets handled as implicit ratings (Hu, Y., Koren, Y. and Volinsky, C., 2008). A series of ratings datasets sized from 100K ratings to 1.25M ratings were processed with both the versions of the SVD algorithm. Initially, the SVD and SVD++ algorithms were tested without grid search used to determine the optimal set of hyperparameters. Results of the SVD and SVD++ replications without grid search are shown in Table 8.

| Experiment Run Number | Input Data Size (# Ratings) | Model | RMSE | MAE | Clock Time (secs) |
|---|---|---|---|---|---|
| 3 | 1.25M | SVD | .8317 | .6363 | 12982.07 |
| 3 | 1.25M | SVD++ | .8177 | .6229 | 12982.07 |
| 1 | 1M | SVD | .8380 | .6415 | 7752.64 |
| 1 | IM | SVD++ | .8237 | .6278 | 7752.64 |
| 6 | 750K | SVD | .8469 | .6483 | 7725.20 |
| 6 | 750K | SVD++ | .8307 | .6325 | 7725.20 |
| 4 | 500K | SVD | .8606 | .6575 | 8548.49 |
| 4 | 500K | SVD++ | .8441 | .6430 | 8548.49 |
| 5 | 250K | SVD | .8855 | .6811 | 2391.98 |
| 5 | 250K | SVD++ | .8703 | .6666 | 2391.98 |
| 2 | 100K | SVD | .9039 | .6984 | 1010.90 |
| 2 | 100K | SVD++ | .8943 | .6887 | 1010.90 |

*Table 8.* Experiment Three Singular Value Decomposition without Grid Search



The grid search class in Surprise (Hug, 2020) computes accuracy metrics for an algorithm on various combinations of parameters, over a cross-validation procedure. Grid search is used to find the best set of parameters for a prediction algorithm and is analogous to GridSearchCV from scikit-learn.

Values of RMSE by size of the input User Ratings dataset are displayed in Figure 13. The RMSE values are shown for the SVD algorithm as well as the modified version of SVD++ As can be seen, for both the SVD and SVD++ algorithms there is a monotonically decreasing value of RMSE as the size of the input dataset increases. The graph shows a consistent additive difference between the RMSE values for SVD and SVD++ of approximately .0168 (1.68%) with lower values obtained for the SVD++ algorithm. As we observed with the previous models, it follows that increasing sample sizes are associated with decreasing error. The clock time for execution of the models ranged from a low of 1010 seconds for the 100K dataset to a high of 12982 seconds for the 1.25M dataset, also illustrating a linear relation between processing time for the models and increases in dataset sizes.

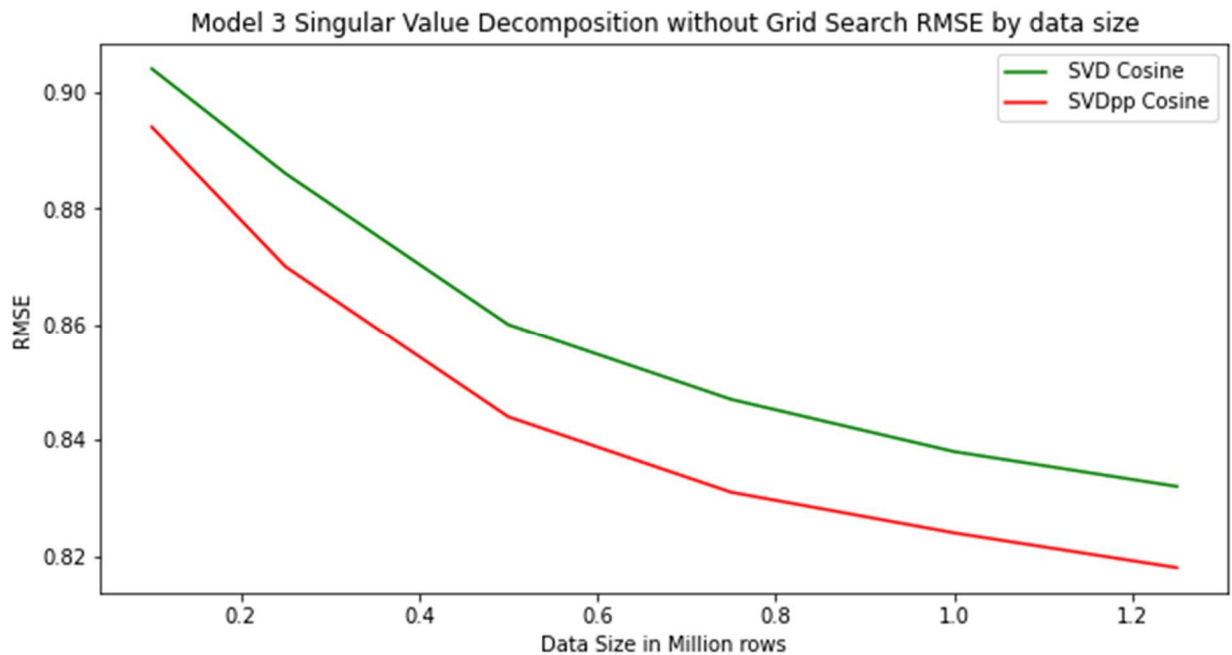



*Figure 13.* SVD without Grid Search RMSE predictive accuracy values

Values of MAE by size of the input User Ratings dataset are displayed in Figure 14. The

MAE values are shown for the SVD algorithm as well as the modified version of SVD++ As can

be seen, for both the SVD and SVD++ algorithms there is a monotonically decreasing value of

MAE as the size of the input dataset increases. The graph shows a consistent additive difference

between the MAE values for SVD and SVD++ of approximately .0213 (2.13%) with lower

values obtained for the SVD++ algorithm. We also observe that increasing sample sizes are

associated with decreasing error, and is expected since the calculation of MAE in the algorithm

corresponds to error calculations in other regression-type models. The SVD and SVD++

algorithms were calculated simultaneously within the same execution session; in these sessions

grid search hyperparameter optimization was not applied during execution of the algorithms.

Published benchmarks (Hug, 2020) for the *Surprise* implementation of SVD without grid

search parameter optimization show RMSE values of .93 with a 100K ratings dataset and .87

with a 1M dataset, compared with RMSE of .90 (100K) and .84 (1M) for the present study.

Benchmarks for the SVD++ implementation without gird search show RMSE values of .92 with

a 100K ratings dataset and .87 with a 1M dataset, compared with RMSE of .89 (100K) and .82

(1M) for the present study.



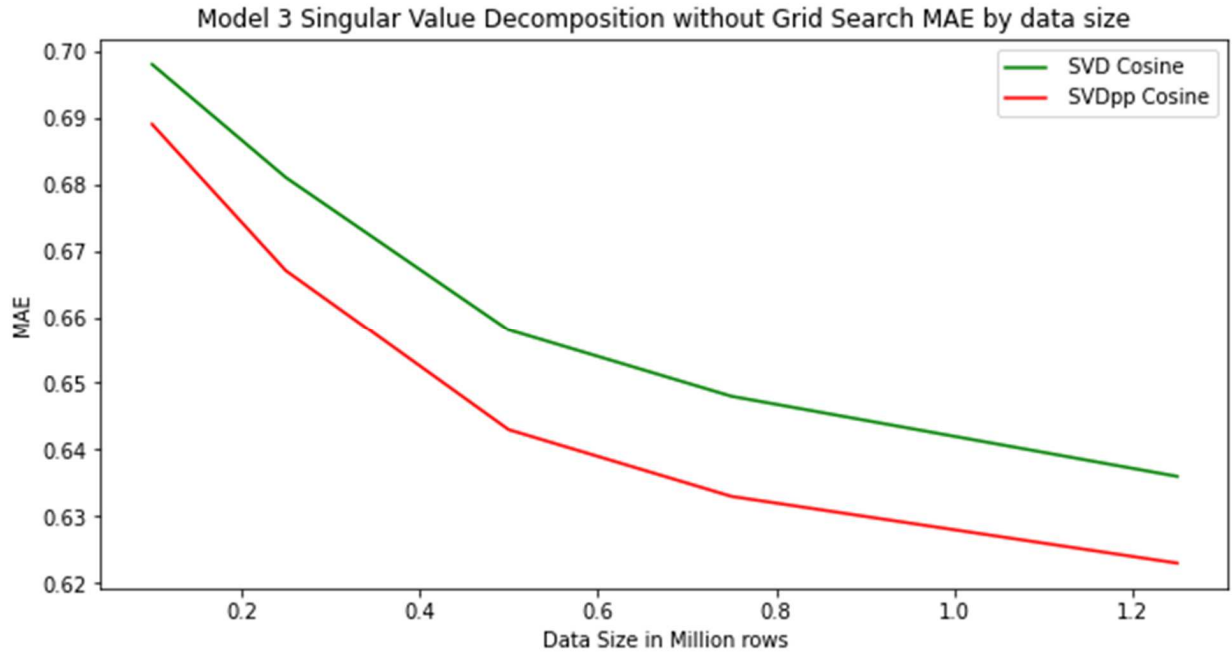

*Figure 14.* SVD without Grid Search MAE predictive accuracy values

| Experiment Run Number | Input Data Size (# Ratings) | Model | Epochs | Factors | Learning Rate | RMSE | MAE | Clock Time (secs) |
|---|---|---|---|---|---|---|---|---|
| 9 | 1.25M | SVD Tuned | 20 | 50 | .005 | .8298 | .6348 | 1886.04 |
| 9 | 1.25M | SVD Untuned | | | | .8318 | .6367 | 1886.04 |
| 7 | 1M | SVD Tuned | 20 | 50 | .005 | .8356 | .6396 | 1517.86 |
| 7 | IM | SVD Untuned | | | | .8370 | .6407 | 1517.86 |
| 12 | 750K | SVD Tuned | 20 | 50 | .005 | .8440 | .6462 | 1285.18 |
| 12 | 750K | SVD Untuned | | | | .8457 | .6474 | 1285.18 |
| 10 | 500K | SVD Tuned | 20 | 50 | .005 | .8574 | .6547 | 741.08 |
| 10 | 500K | SVD Untuned | | | | .8603 | .6576 | 742.08 |
| 11 | 250K | SVD Tuned | 20 | 50 | .005 | .8814 | .6773 | 376.17 |
| 11 | 250K | SVD Untuned | | | | .8852 | .6802 | 376.17 |
| 8 | 100K | SVD Tuned | 20 | 50 | .005 | .9002 | .6958 | 161.05 |
| 8 | 100K | SVD Untuned | | | | .9033 | .6992 | 161.05 |

*Table 9.* Experiment Three-A Singular Value Decomposition with Grid Search

Next, the SVD algorithm was tested with grid search hyperparameter optimization. Table

9 summarizes results of six joint replications of SVD tuned with Grid Search and SVD not tuned.

Figure 15 displays values of RMSE by size of the input user ratings dataset for the two SVD

algorithm variations tested in this experiment. The RMSE values are shown for the SVD



algorithm with best parameters tuned by grid search as well as SVD without best parameters

tuned. As can be seen, for both SVD variations, there are decreasing values of RMSE as the size

of the input dataset increases. The graph shows a consistent difference between the RMSE values

for the two models of approximately .003 (0.3%) with lower values obtained for SVD with grid

search. Execution time for the joint model executions ranged from a low of 161 seconds for the

100K dataset and 1886 seconds for the 1.25M dataset, these execution times also illustrate an

approximately linear relation with increasing input dataset sizes.

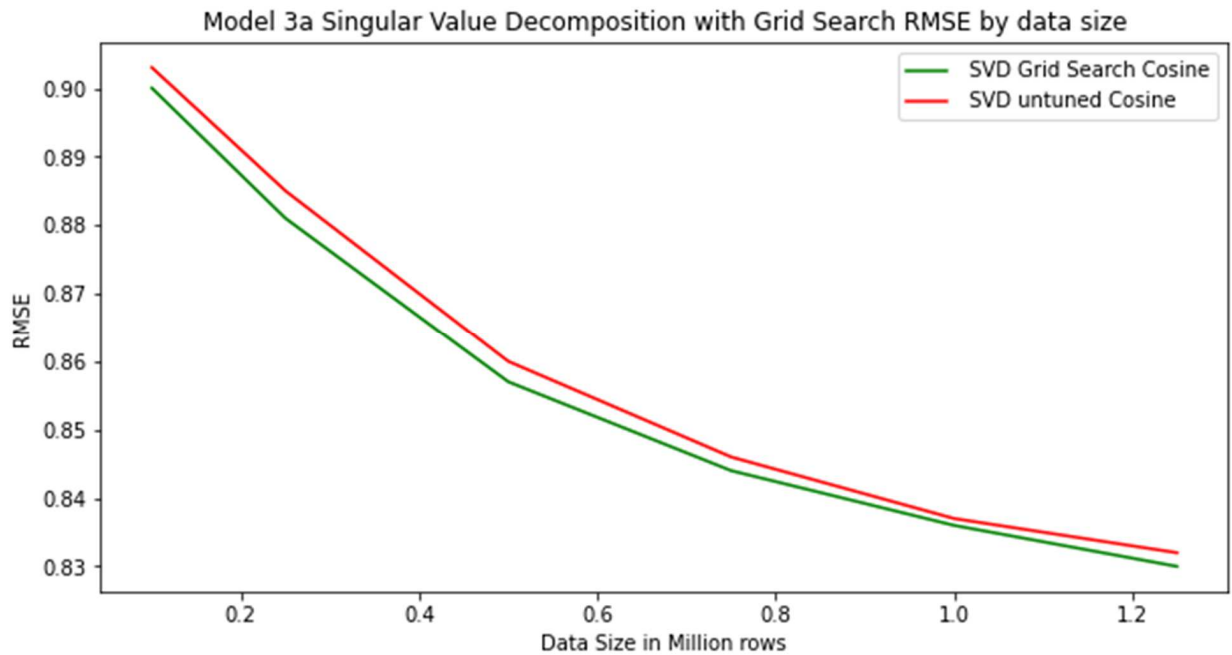

*Figure 15* SVD with Grid Search RMSE predictive accuracy values

Figure 16 displays values of MAE by size of the input user ratings dataset for the two

SVD variations tested in this experiment. The MAE values are shown for the SVD algorithm

with best parameters tuned by grid search as well as untuned SVD without best parameters

selected. As can be seen, for both models, there are decreasing values of MAE as the size of the

input dataset increases. The graph shows a consistent difference between the MAE values for



SVD with grid search and SVD without grid search of approximately .004 (0.4%) with lower

values obtained for SVD with grid search.

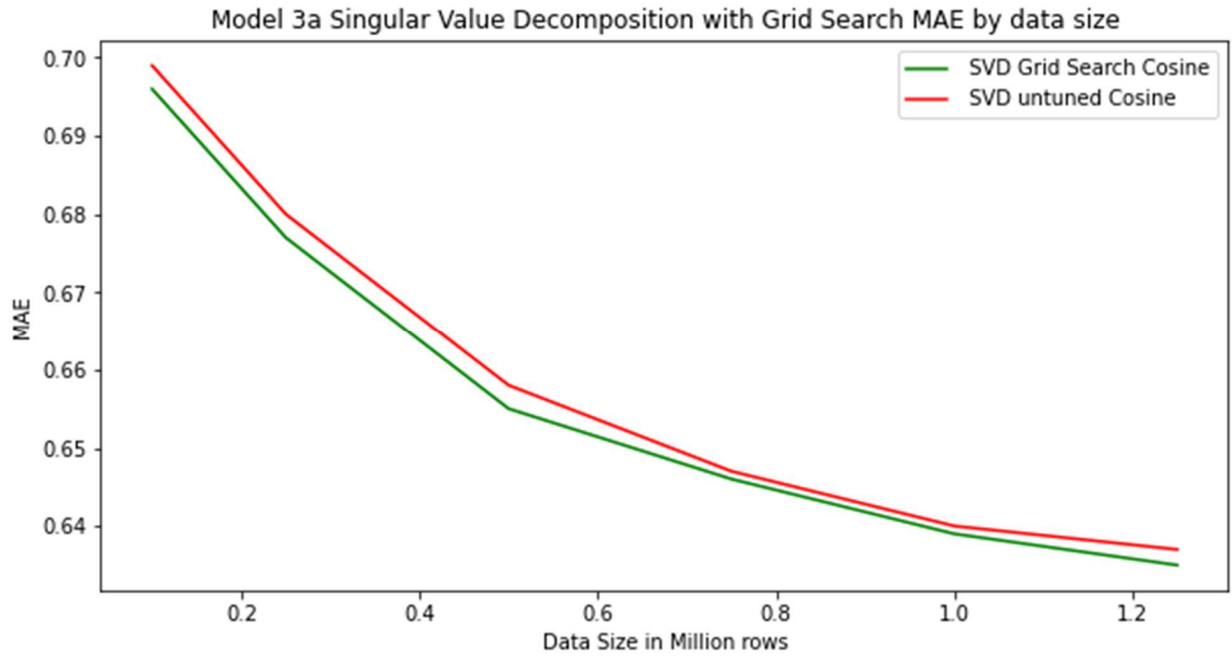

*Figure 16.* SVD with Grid Search MAE predictive accuracy values



**Experiment Four Restricted Boltzmann Machine**

The Restricted Boltzmann Machine (RBM) was the first of two deep learning algorithms tested. Two versions of the RBM were tested, one without gird search optimization and one using grid search optimization. The algorithm used here is implemented using the Surprise framework and follows the specification described in Salakhutdinov, R., Mnih, A. and Hinton, G. (2007). The Restricted Boltzmann Machine (RBM) takes several hyperparameters to control the manner in which the network learns from the input ratings and movies data. The visible dimension value is a product of the number of distinct rating values times the number of movies being considered. The processing epochs value controls the number of forward and backward iterations taken by the network as it minimizes the error between the observed rating values and the estimated predicted values while it converges on an optimal set of weights and biases. The value for hidden dimensions is the number of hidden neurons in the RBM, the distinct rating value is the number of points on the ordinal scale used to rate movies and includes half-point values. The learning rate controls how fast the model attempts to converge during each iteration, if set too low more steps may be taken than necessary and if set too high the optimal converge point may be skipped. The batch size controls the number of user ratings that are processed together during network training.

The train function in the code is called to create an RBM with weights and biases to permit the network to reconstruct any user's rating for any movie item. The function takes as input a two dimensional array with a row of rating values for every user and each row holding binary data for the number of movies times the number of rating values.

The RBM without gird search was configured with visible dimensions equal to the product of the number of distinct rating values (10) and the number of movies in the Movies input data, and 50 hidden dimensions, execution epochs were set to 20, 10 user rating values



were used in the calculations, learning rate was set to .001 or .1, and the batch size (number of

user ratings grouped during an execution) was set to 200 or 100. Results for each replication for

this RBM are shown in Table 10.

| Experiment Run Number | Input Data Size (# Ratings) | Epochs | Hidden Dimensions | Rating Values | Learning Rate | Batch Size | RMSE | MAE | Clock Time (secs) |
|---|---|---|---|---|---|---|---|---|---|
| 1 | 1M | 20 | 50 | 10 | .001 | 100 | 1.3923 | 1.1887 | 5238.04 |
| 3 | 1M | 20 | 50 | 10 | .1 | 100 | 1.3899 | 1.1864 | 4954.13 |
| 7 | 400K | 20 | 50 | 10 | .001 | 100 | 1.3105 | 1.1165 | 6646.86 |
| 6 | 300K | 20 | 50 | 10 | .001 | 100 | 1.3090 | 1.1140 | 5014.09 |
| 4 | 250K | 20 | 50 | 10 | .1 | 200 | 1.3091 | 1.1127 | 4043.43 |
| 5 | 200K | 20 | 50 | 10 | .1 | 200 | 1.3112 | 1.1144 | 2973.58 |
| 2 | 100K | 20 | 50 | 10 | .001 | 100 | 1.3257 | 1.1337 | 1431.77 |

*Table 10.* Experiment Four Restricted Boltzmann Machine without Grid Search

Values of RMSE by size of the input User Ratings dataset for the Restricted Boltzmann

Machine without grid search optimization are displayed in Figure 17. As can be seen, over the

course of the replications the lowest value of RMSE was obtained for the 300K ratings dataset

with higher values of RMSE found for the smallest datasets as well as the larger datasets. The

RMSE value obtained for the 1M ratings dataset was markedly higher than for the smaller

datasets. This curvilinear result for increasing dataset sizes stands in contrast to the RMSE

results obtained with the three previous algorithms. Notably, the values of RMSE were higher,

indicative of greater error, than values obtained for the three previous algorithms not configured

as deep learning. These results suggest model overfitting contributed to the deceased accuracy

and an inconsistent relation between data size and predictive error.



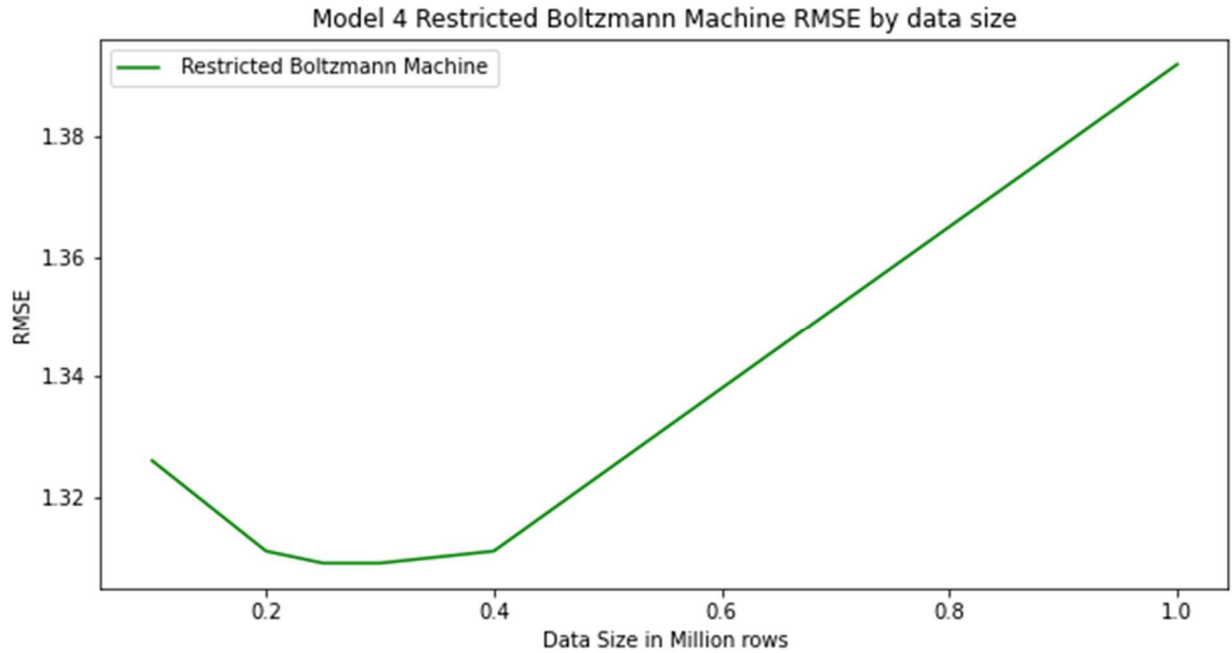

*Figure 17.* Restricted Boltzmann Machine without Grid Search RMSE predictive accuracy
values

Values of MAE by size of the input User Ratings dataset for the Restricted Boltzmann

Machine without grid search optimization are displayed in Figure 18. As can be seen, over the

course of the replications the lowest value of MAE was obtained for the 200K ratings dataset

with the highest MAE found for the smallest dataset and for the larger datasets. The MAE value

obtained for the 100K ratings dataset was markedly higher than for the intermediate and larger

datasets. This curvilinear result also stands in contrast to the MAE results obtained with the three

previous algorithms and is likely a consequence of model overfitting.

Processing time for the RBM without grid search optimization ranged from a low of

1,432 seconds for the 100K ratings dataset to a high of 6,647 seconds clock time for the 400K

dataset. Notably, the IM dataset required less clock time with 5,238 seconds.



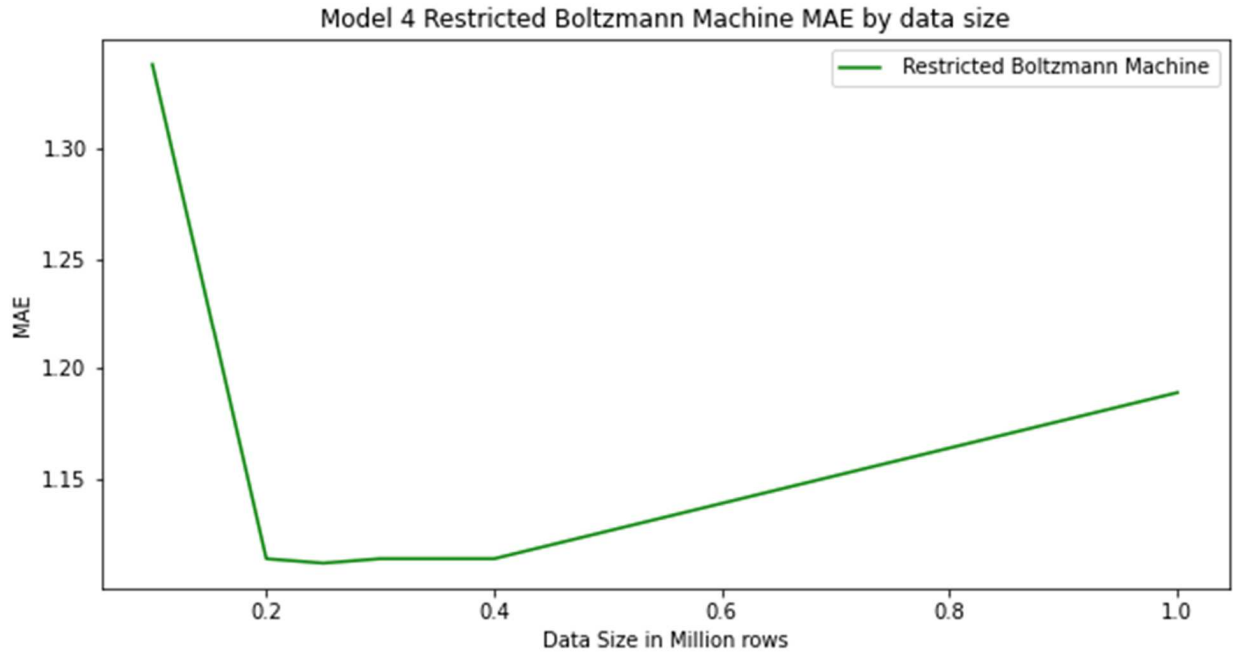

*Figure 18.* Restricted Boltzmann Machine without Grid Search MAE predictive accuracy values

The RBM with gird search was configured with visible dimensions equal to the product of the number of distinct rating values (10) and the number of movies in the Movies input data, 50 hidden dimensions, execution epochs were set to 20, 10 user rating values were used in the calculations, learning rate was set to .001 or .1, and the batch size (number of user ratings grouped during an execution) was set to 200 or 100. Results for each replication for this RBM are shown in Table 11. Notably, the values of RMSE and MAE were higher, also indicative of model overfitting, than values obtained for the three previous algorithms not configured as deep learning.



| Experiment Run Number | Input Data Size (# Ratings) | Visible Dimensions | Epochs | Hidden Dimensions | Rating Values | Learning Rate | Batch Size | RMSE | MAE | Clock Time (secs) |
|---|---|---|---|---|---|---|---|---|---|---|
| 1 | 1M | | 20 | 50 | 10 | .001 | 100 | 1.3911 | 1.1877 | 15229.38 |
| 3 | 1M | | 20 | 50 | 10 | .1 | 100 | 1.3911 | 1.1877 | 16292.11 |
| 6 | 750K | | 20 | 20 | 10 | .1 | 200 | 1.3042 | 1.1094 | 40078.32 |
| 4 | 500K | | 20 | 50 | 10 | .1 | 200 | 1.3075 | 1.1142 | 26850.53 |
| 5 | 250K | | 20 | 20 | 10 | .1 | 200 | 1.3098 | 1.1135 | 10700.41 |
| 7 | 200K | | 20 | 20 | 10 | .1 | 200 | 1.3105 | 1.1137 | 7739.32 |
| 2 | 100K | | 20 | 50 | 10 | .1 | 100 | 1.3250 | 1.1332 | 3656.39 |

*Table 11.* Experiment Four-A Restricted Boltzmann Machine with Grid Search

Values of RMSE by size of the input User Ratings dataset for the Restricted Boltzmann

Machine with grid search optimization are displayed in Figure 19. As can be seen, over the

course of the replications the lowest value of RMSE was obtained for the 750K ratings dataset

with higher values of RMSE found for the smaller datasets as well as the largest dataset. The

RMSE value obtained for the 1M ratings dataset was markedly higher than for the smaller

datasets. This curvilinear result of RMSE and MAE for increasing dataset sizes is similar to the

results obtained with the RBM without grid search optimization. Model overfitting is a likely

explanation for these results.



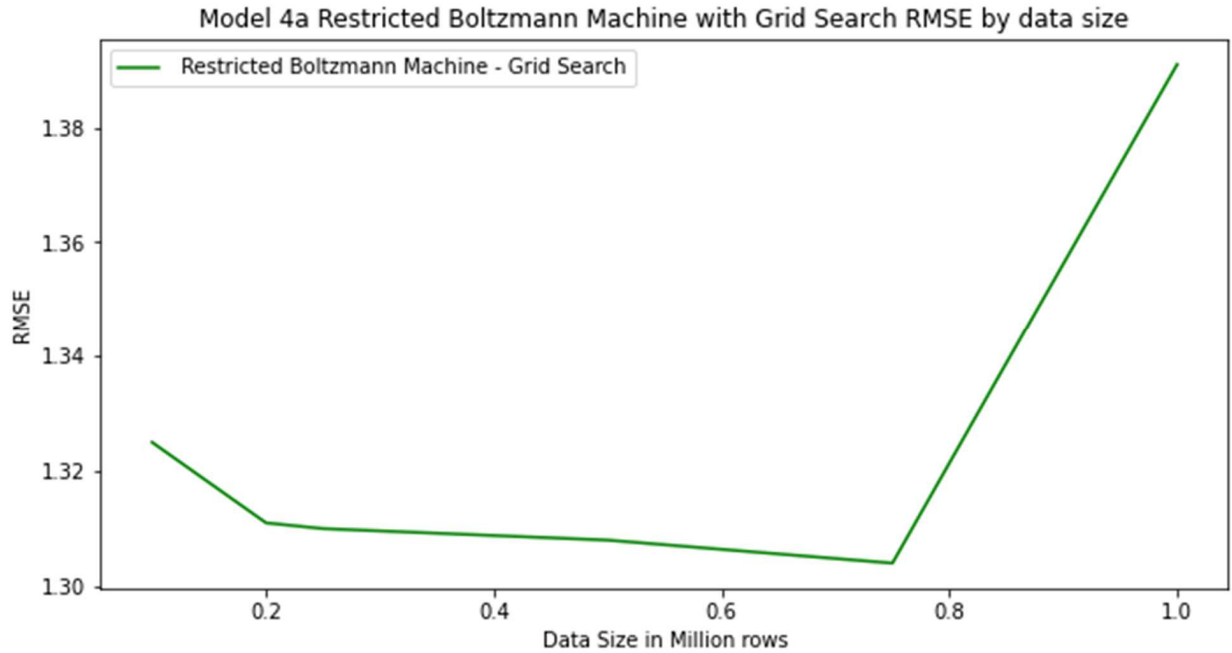

*Figure 19.* Restricted Boltzmann Machine with Grid Search RMSE predictive accuracy values

Values of MAE by size of the input User Ratings dataset for the Restricted Boltzmann

Machine with grid search optimization are displayed in Figure 20. As can be seen, over the

course of the replications the lowest value of MAE was obtained for the 400K ratings dataset

with higher values of RMSE found for the smaller datasets as well as the largest dataset. The

MAE value obtained for the 1M ratings dataset was markedly higher than for the smaller

datasets. This curvilinear result for increasing dataset sizes is similar to the MAE results obtained

with the RBM without grid search optimization.



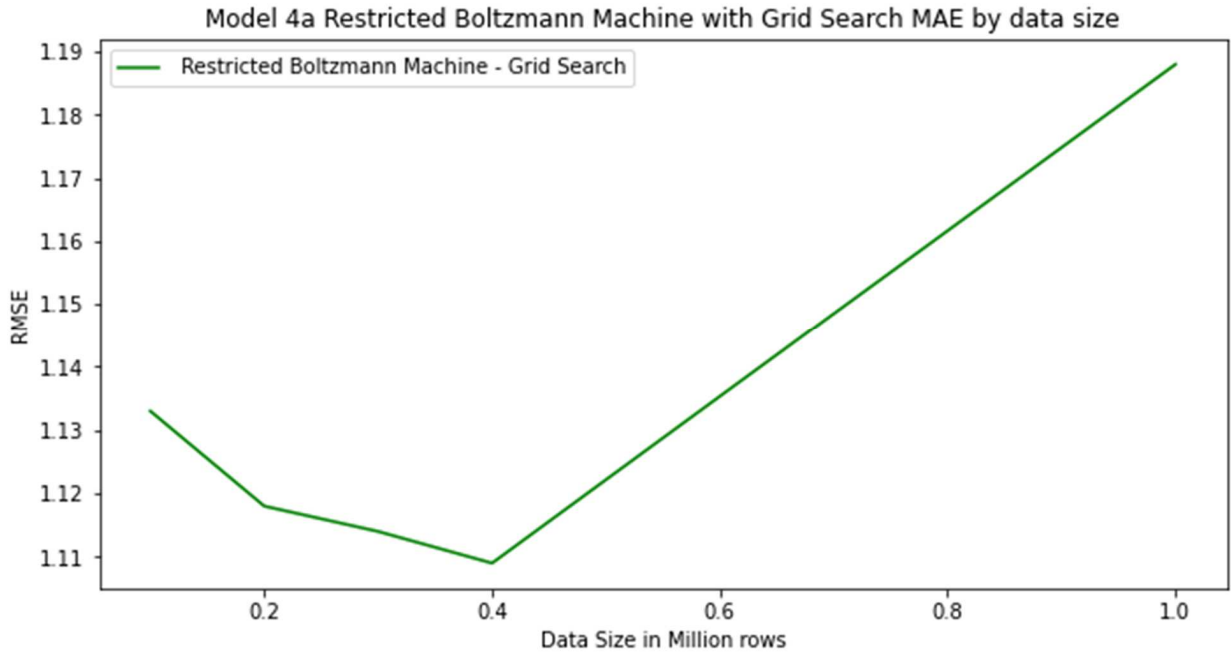

*Figure 20.* Restricted Boltzmann Machine with Grid Search MAE predictive accuracy values

Processing time for the RBM with grid search optimization ranged from a low of 3,656

seconds for the 100K ratings dataset to a high of 40,078 seconds clock time for the 750K dataset.

Again, the 1M dataset required less clock time with 16,292 seconds.



## Experiment Five Autoencoder

An Autoencoder was the second of two deep learning algorithms tested. A single model
of the autoencoder was tested and configured with visible dimensions equal to the product of the
number of distinct rating values times the number of movies being considered, 50 or 100 hidden
dimensions, and execution epochs of 20, 100, or 200. User rating values were 10, the learning
rate of gradient descent was set to .01 or .1, and the batch size (number of user ratings grouped
during an execution) was set to 200. Results for each replication for this Autoencoder are shown
in Table 12.

| Experiment Run Number | Input Data Size (Ratings) | Visible Dimensions | Epochs | Hidden Dimensions | Rating Values | Learning Rate | Batch Size | RMSE | MAE | Clock Time (secs) |
|---|---|---|---|---|---|---|---|---|---|---|
| 1 | 1M | | 100 | 100 | 10 | .01 | 200 | 2.0894 | 1.7115 | 10842.93 |
| 3 | 1M | | 50 | 50 | 10 | .1 | 200 | 2.1453 | 1.7757 | 10273.96 |
| 9 | 400K | | 20 | 20 | 10 | .1 | 200 | 2.0716 | 1.7284 | 14120.30 |
| 8 | 300k | | 20 | 20 | 10 | .1 | 200 | 2.0646 | 1.7230 | 9889.23 |
| 4 | 250K | | 20 | 20 | 10 | .1 | 200 | 2.0658 | 1.7223 | 7720.81 |
| 7 | 200K | | 20 | 20 | 10 | .1 | 200 | 2.0621 | 1.7176 | 6092.52 |
| 2 | 100K | | 200 | 100 | 10 | .1 | 200 | 2.0367 | 1.6719 | 2795.09 |
| 5 | 100K | | 20 | 20 | 10 | .1 | 200 | 2.0464 | 1.6994 | 2839.48 |
| 6 | 100K | | 20 | 50 | 10 | .1 | 200 | 2.0792 | 1.7307 | 2550.22 |

*Table 12.* Experiment Five Autoencoder

Values of RMSE by size of the input User Ratings dataset for the Autoencoder are
displayed in Figure 21. As can be seen, over the course of the replications the lowest value of
RMSE was obtained for the smallest ratings dataset with higher values of RMSE found for the
larger datasets. As was true for the RBM, the RMSE value obtained for the 1M ratings dataset
was markedly higher than for the smaller datasets. This result for increasing dataset sizes also
stands in contrast to the RMSE and MAE results obtained with the first three algorithms. Model
overfitting is also a possible explanation for the performance of this model.



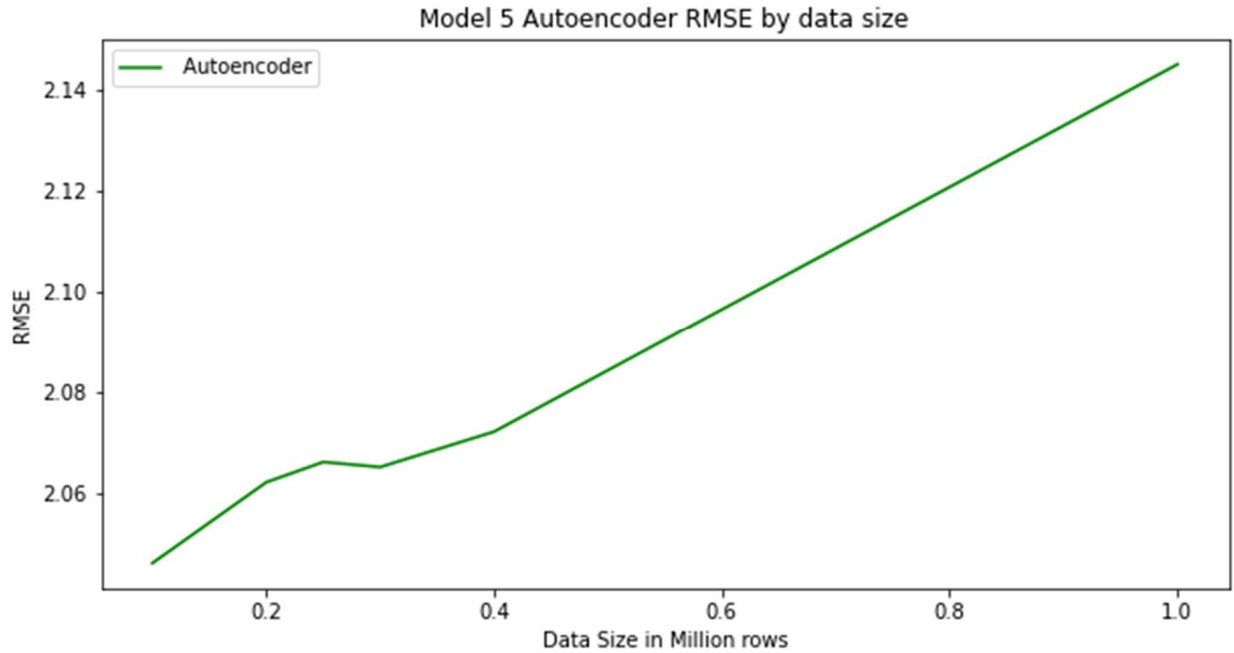

*Figure 21.* Autoencoder RMSE predictive accuracy values

Values of MAE by size of the input User Ratings dataset for the Autoencoder are
displayed in Figure 22. As can be seen, over the course of the replications the lowest value of
MAE was obtained for the smallest ratings dataset with higher values of MAE found for the
larger datasets. As was true for the RBM, the MAE value obtained for the 1M ratings dataset was
markedly higher than for the smaller datasets. This result for increasing dataset sizes stands in
contrast to the MAE results obtained with the first three algorithms.



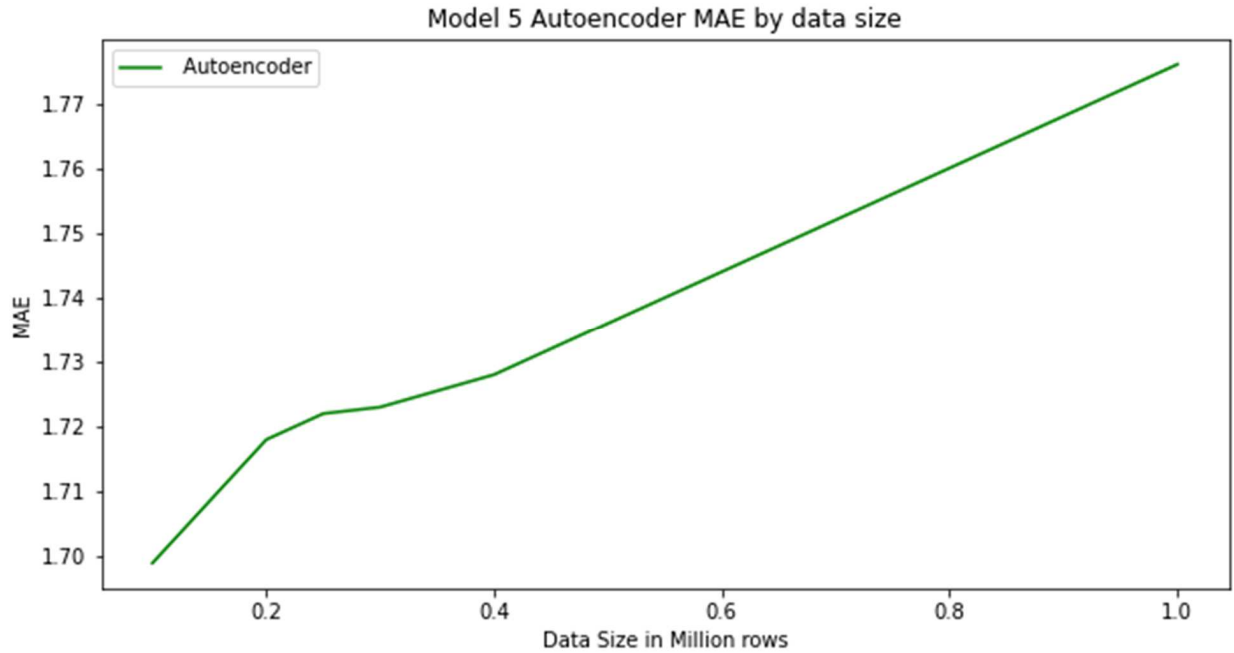

*Figure 22.* Autoencoder MAE predictive accuracy values

Processing time for the Autoencoder across the different data sizes ranged from a low of

2,550 seconds for the 100K ratings dataset to a high of 14,120 seconds clock time for the 400K

dataset. As was true of the RBM, the 1M dataset required less clock time with 10,274 seconds.

This relation between ratings dataset size and execution time is very similar to that of the RBM

algorithm.

**Summary**

Results for the five experiments showed the three conventional algorithms were

significantly more accurate than the two deep learning methods. Model overfitting was a

significant contributor to this effect, as evidenced by the fact that datasets with more ratings

produced poorer results. This finding justifies the need for further research in the use of deep

learning models for recommender systems.



# Chapter 5

# Conclusions, Implications, Recommendations, and Summary

## Conclusions and Implications

Results obtained for the first three algorithms, KNN User-user Collaborative Filtering, Non-negative Matrix Factorization, and Singular Value Decomposition were consistent with published benchmarks and results in the literature (Hug, 2020; Salakhutdinov, et al., 2007; Koren, et al., 2009). The RMSE and MAE accuracy measures obtained for these algorithms largely replicated those of previously published results. Table 13 compares RMSE benchmark values with results obtained for the three conventional algorithms.

| Model and Data Size | RMSE Benchmark (Hug, 2020) | RMSE Model | Benchmark – Model Difference |
|---|---|---|---|
| KNN User-User 100K | .98 | .99 | + .01 |
| KNN User-User 1M | .92 | .95 | + .03 |
| Matrix Factorization 100K | .96 | .95 | - .01 |
| Matrix Factorization 1M | .92 | .84 | - .08 |
| SVD 100K | .93 | .90 | - .03 |
| SVD 1M | .87 | .84 | - .03 |
| SVD++ 100K | .92 | .89 | - .03 |
| SVD++ 1M | .86 | .82 | - .04 |

*Table 13*. Comparison of RMSE Benchmark Values and Results for Three Models

The present results also corresponded well to published findings regarding the relation between the size of the input ratings datasets used in the analyses and improvements of RMSE and MAE when the dataset sizes are larger. This effect was confirmed for all three algorithms and their variations; simply stated, sample size matters. Use of grid search parameter optimization also produced incremental improvements in RMSE and MAE. Until very recently matrix factorization and singular value decomposition (SVD) have been among the most frequently used recommender algorithms for production applications in the commercial sector



(Ricci, et al., 2015). The research objective of the present investigation was to compare the predictive accuracy of recommender algorithms and loss functions of RMSE and MAE were selected as outcome metrics. The findings for these three algorithms replicate results in the literature.

Outcomes obtained for two deep learning (DL) algorithms, the Restricted Boltzmann Machine and the Autoencoder showed weaker results and greater loss functions than the three previous models in terms of RMSE and MAE values; the RMSE and MAE values were consistently higher and indicative of greater predictive error for the RBM and Autoencoder than for the non-DL algorithms. The relation between ratings dataset sizes and RMSE and MAE loss values for these two deep learning models was also not similar to the previous models with lowest accuracy measures being obtained for intermediate sized ratings datasets. The accuracy of DL models is highly determined by model configuration, such as the number of units in the hidden layer (Sedhain, et al., 2015), the number of hidden layers, and the number of model training epochs (Zhang, et al., 2019).

In recent work Jannach, D., Moreira, G., and Oldridge, E. (2020) point out that DL models are being successfully applied by notable companies in the commercial sector, such as Google, Pinterest, and Facebook. However, DL models are not consistently winning Recommender Systems competitions when matched head-to-head with more traditional non-DL methods. Several explanations are offered for the discrepancy. In many cases the DL researchers competing in the matches may not have access to the level of computational resources such as multiple GPU or TPU availability and high levels of memory required for DL models while applied practitioners in industry are not so constrained. DL researchers in the competitions are often from the academic sector and pursuing more general theory and architecture-driven general solutions rather than the more domain-centric or proprietary solutions implemented in



commercial applications. The commercial domain focus may rely on extensive feature engineering, specifically tuned to the data at hand and at very high dimensionality, in turn requiring computational resources at a level prohibitive to many academic researchers.

**Model Overfitting**

Model overfitting (Goodfellow, et. al., 2016, Chapter 5) is a likely contributor to the increased values of RMSE observed for the deep learning models in this study. Regression and neural network models are prone to overfitting when the model is too sensitive to noise from outliers in the training data. This is likely to happen in a regression model when a large number of collinear parameters are introduced in the model and similar results can adversely affect a poorly tuned neural network model or a wide network model with many hidden nodes or layers. As a result, the model may find a solution for outliers in the training data but fail to produce adequate (i.e., sufficiently low RMSE) predications when applied to test data. When overfit, the model will produce poor results for successive samplings from the same sample space and as a consequence will poorly generalize and have less utility as a predictor.

A model with a very large number of parameters, or network nodes, has a high degree of complexity. Such a model may be able to make very accurate predictions for cases in the training data but this complexity can make generalization of the model to new data less reliable due to high variance in the model. This is discussed in the machine learning and statistical literature as the bias-variance tradeoff (Belkin, M., Hsu, D., Ma, S. and Mandal, S., 2019; Hastie, et al., 2017). The high variance in a model can render it quite sensitive to small fluctuations and hence less able to generalize to a wider range of cases, even if these cases are representative.



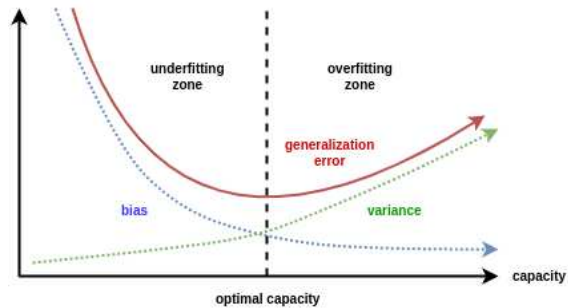

*Figure 23.* The Bias-Variance Tradeoff

As is shown in Figure 23 (Fortmann-Roe, S., 2012), the bias-variance tradeoff can be

visualized as a graph of two functions, a bias function that decreases as the model capacity

increases, and a variance function that increases with model capacity (Goodfellow, et al., 2016,

pp. 127). Capacity can be viewed as a model's ability to fit a variety of functions; a model with a

low capacity may not fit cases in the training dataset while a model with a high capacity, and

high complexity, might overfit due to settling on values of the training data that are not

representative of the sample from which they are drawn. An optimal point can be found between

decreasing model bias and increasing model variance where generalization error will be a

minimum. This minimum will coincide with an optimal capacity but be under-fit if the model

capacity is below the optimum and over-fit when it is above the optimum. Deviations from the

optimum can happen during model training if non-optimal solutions are obtained, such as settling

on local minima during gradient descent.

However, recent work suggests the classic bias-variance tradeoff may not always hold.

Neal, B., Mittal, S., Baratin, A., Tantia, V., Scicluna, M., Lacosta-Julien, S. and Mitliagkas, I.

(2019) report results showing both bias and variance can simultaneously decrease as the number

of parameters grow to large values in models with high network widths. Belkin, et al. (2019)

observe that increasing model capacity beyond the optimal point on the classical bias-variance

curve can result in improved model performance. When the number of features in the model are



much smaller than the sample size the training error will be close to the test error but as the number of features approaches an interpolation threshold (the optimal point on the bias-variance curve) features weakly present in the data are forced to fit the training data, resulting in classical overfitting. Belkin et al., discuss examples showing that as the model continues to become more complex past the interpolation threshold the training error can improve.

In practice, steps can be taken when developing deep learning models to avoid both under or overfitting. Regularization refers to a body of techniques used to reduce generalization error in learning models. Many of these techniques are discussed in Goodfellow, et al. (2016). Representative methods include $L^1$ parameter norm penalties, $L^2$ parameter penalties, and early stopping. Parameter norm penalties limit the capacity (complexity) of models, including linear regression, neural networks, or logistic regression, by adding a norm penalty to the objective function. We have seen how reduced complexity can reduce overfitting and if the penalty is appropriate the model will still have sufficient capacity and low predictive error on test cases.

- An $L^2$ parameter norm regularization introduces weight decay into the model. $L^2$ regularization is also known as ridge regression or Tikhonov regularization (Tikhonov, et al., 1995). Adding weight decay modifies the learning rule to shrink the weight vector by a constant factor on each model iteration prior to the SGD update.

- A related parameter norm regularization is $L^1$ regularization, another weight decay strategy. In comparison to $L^2$ regularization, $L^1$ regularization produces a sparse solution and sets a number of parameters to their optimal values of zero. This property of $L^1$ regularization also enables its use as a feature selection procedure. The LASSO (least absolute shrinkage and selection operator, Tibshirani, 1995)



combines an $L^1$ adjustment with a linear model and causes some of the model

weights to become zero, indicating the corresponding features may be eliminated.

- Early stopping is a strategy that exploits a behavior of large models in which

  training error decreases over time but validation set error begins to rise again after

  reaching a minimum. As a result a model with better validation set error can be

  reached by returning to the parameter setting at the time point with the lowest

  validation set error. At each iteration when the validation set error improves a

  copy of the model parameters is stored. The algorithm stops when no parameters

  have improved over the stored values.

## Recommendations

Recommender algorithms have seen considerable research interest within the machine

learning community. A good body of work exists focusing upon non-DL approaches to

recommenders and many applications of the technology have been scaled to large commercial

implementations. In recent years great attention has been devoted to DL methods by both the

academic community as well as practitioners developing commercial deployments. The DL

methods have been successful in the commercial setting where resources to support the

computational demands are available. As discussed previously, DL methods in formal

competitions alongside non-DL methods have not consistently performed as well, despite

success with deployments in the commercial setting. In the context of the work reported here, we

saw notably lower accuracy metrics achieved for the two DL models compared to the three non-

DL models.

Moving forward in this vein of research a number of strategies and modifications may be

attempted:



- Pursue possible access to greater computational resources. The present investigations reported herein were executed on the Google Colab Plus platform and while this afforded TPU processing as well as 35GB memory allocation the two DL models failed on datasets over 1M in size and also on some trails with smaller data sizes. The typical failure happened when execution of the code exceeded the 35GB Colab memory size. Other execution environments, such as short-term rental of an AWS (Amazon Web Services) allocation may offer a solution. To avoid execution failures on Google Colab Plus for the DL models, a series of smaller ratings datasets were created from the original 20M MovieLens data source but access to greater computing resources could be expected to eliminate the need for smaller datasets. This is likely a material benefit because the DL models will perform with greater accuracy with more complex configurations and use of larger data.

- As is well documented, the DL models are very sensitive to tuning and often prone to overfitting. Model overfitting may be reduced by manipulation of regularization parameters in the models. The non-DL methods are less prone with a number of stable implementations and frameworks in place. In future experiments a number of DL configuration options will benefit from systematic parametric manipulations, such as number of visible and hidden layers, node count per layer, regularization functions, learning rate, factors, and batch size. The framework used in the present investigation implemented well-known non-DL algorithm classes but the two DL algorithms, the RBM and the Autoencoder, were built from an investigatory implementation.



- Attempt to identify other existing but well performing implementation code for
  the DL algorithms, or related algorithms, to be used in the successive
  experiments. Nonetheless, the DL code used in this investigation did implement
  classes for TensorFlow, Keras, and Python large numeric processing.

**Summary**

This dissertation sought to deploy well-designed experiments to evaluate alternate models
for recommendation systems on specific tasks. Given training data consisting of a set of objects
with specified features, a set of users with specified characteristics, and ratings assigned by users
to objects, the goal is to evaluate models that can predict the rating assigned by a user, with
specified characteristics, to an object with specified features.

The investigation used three conventional recommender algorithms: Knn Nearest
Neighbor (KNN) User-User collaborative filtering, Feature-based biased matrix factorization
trained with alternating least squares (NMF), and Feature-based filtering using singular value
decomposition (SVD). In addition, two deep-learning inspired algorithms were be used: a
Restricted Boltzmann Machine (RBM) for collaborative filtering and a multilayer autoencoder
network. The predictive accuracies of the models were evaluated based on the mean absolute
error (MAE) and root mean square error (RMSE), appropriate for regression related models. The
time to train the model and the time required for a trained model to predict a rating were will also
recorded. The models were evaluated using various sized input sources of ratings data such as
the *MovieLens* data sets available from GroupLens.com, Kaggle.com, and other sources. The
publicly available MovieLens (ML) datasets are collected and distributed by the GroupLens
Project (Kluver, D., Ekstrand, M. and Konstan, A., 2018) at the University of Minnesota
(http://grouplens.org/). The MovieLens data sources contain information of movie users, movie



items, and user preference content in the form of movie ratings and descriptive content data;

these data sources also include time stamps.

Drawing upon the many algorithms that have been developed for recommender

applications, three conventional recommender algorithms were used in this investigation: KNN

User-user Collaborative Filtering, Non-negative Matrix Factorization (NMF) and Singular Value

decomposition (SVD). These three algorithms are well established in the recommender system

literature and were selected for being representative of commonly investigated recommender

algorithms. In addition, two more-recently developed deep learning recommender algorithms,

the Restricted Boltzmann Machine (RBM), and an Autoencoder (AutoRec), were also

investigated. These two algorithms are also frequently found in the recent literature for deep

learning approaches in recommender systems. The research goal in each case was to measure the

predictive accuracy in terms of Root Mean Square Error (RMSE) and Mean Absolute Error

(MAE) of each of the five algorithms with a variety of user-rating dataset sizes and manipulation

of other selected parameters of the five algorithms.

To investigate the predictive accuracy of each algorithm a series of experiments were

executed using the Google Colab Plus environment with multiple replications completed for each

of the five algorithms. Google Colab Plus allocated 35GB memory and offered access to TPU

(Tensor Processing Unit) capability. During the course of the investigation 74 experimental

replications were executed using Python 3.6 code and ran to completion. The Colab environment

managed Tensorflow and Keras, plus other Python modules and dependencies needed to

implement the conventional and deep learning algorithms. In total, the experimental sessions ran

for 69 hours of clock time across several weeks.

The first model, *KNN User-User Collaborative filtering*, included 17 replications with

execution times between 79 and 1174 seconds. A series of input user ratings datasets were



processed with 100K, 250K, 500K, 1M, and 1.25M rows of user movie ratings drawn from the

MovieLens 20M ratings dataset. With all but the 100K dataset, four similarity calculation

methods were used (cosine, MSD: Mean Squared Difference, Pearson, and Pearson-baseline).

Values of RMSE for the 100K Movielens dataset was .99 compared to a 100K benchmark (Hug,

2020) of .98 and MAE of .77 for 100K Movielens with compared to a benchmark MAE of .77.

For the 1M Movielens dataset the best RMSE value was .89 with MSD similarity compared to a

benchmark RMSE of .92 and a best MAE value of .68 with MSD similarity compared to a

benchmark MAE of .72. There was a consistent decrease for both RMSE and MAE as the size of

the ratings datasets increased.

The second model, *Non-negative Matrix Factorization (NMF)*, included 12 replications

with execution times between 40 and 1317 seconds. A series of input user ratings datasets were

processed with 100K, 250K, 500K, 750K, 1M, and 1.5M rows of user movie ratings drawn from

the MovieLens 20M ratings dataset. For the NMF algorithm similarity calculation was not

relevant. Values of RMSE for the 100K Movielens dataset was .95 compared to a 100K

benchmark (Hug, 2020) of .96 and MAE of .74 for 100K Movielens with compared to a

benchmark MAE of .76. For the 1M Movielens dataset the best RMSE value was .84 compared

to a benchmark RMSE of .92 and a best MAE value of .64 compared to a benchmark MAE of

.72. There was a consistent decrease for both RMSE and MAE as the size of the ratings datasets

increased.

The third model, *Singular Value Decomposition (SVD, SVD++)* was executed in two

variations, one with grid search (GS) hyperparameter tuning and one without. SVD without GS

included 6 joint replications of SVD and SVD++ with execution times between 1010 and 12982

seconds. A series of input user ratings datasets were processed with 100K, 250K, 500K, 750K,

1M, and 1.25M rows of user movie ratings drawn from the MovieLens 20M ratings dataset.



Values of RMSE for SVD with the 100K Movielens dataset was .90 compared to a 100K benchmark (Hug, 2020) of .93 and MAE of .69 for 100K Movielens with compared to a benchmark MAE of .73. For the 1M Movielens dataset the best SVD RMSE value was .84 compared to a benchmark RMSE of .93 and a best MAE value of .64 compared to a benchmark MAE of .68. For SVD there was a consistent decrease of both RMSE and MAE as the size of the ratings datasets increased. Values of RMSE for SVD++ with the 100K Movielens dataset was .89 compared to a 100K benchmark (Hug, 2020) of .92 and MAE of .68 for 100K Movielens with compared to a benchmark MAE of .72. For the 1M Movielens dataset the best SVD++ RMSE value was .82 compared to a benchmark RMSE of .86 and a best MAE value of .63 compared to a benchmark MAE of .67. For SVD++ there was also a consistent decrease of both RMSE and MAE as the size of the ratings datasets increased.

SVD with GS included 6 replications of SVD with execution times between 161 and 1886 seconds. A series of input user ratings datasets were processed with 100K, 250K, 500K, 750K, 1M, and 1.25M rows of user movie ratings drawn from the MovieLens 20M ratings dataset. Values of RMSE for SVD with GS on the 100K Movielens dataset was .90 compared to a 100K benchmark (Hug, 2020) of .93 and MAE of .69 for 100K Movielens with compared to a benchmark MAE of .73. For the 1M Movielens dataset the best SVD with GS RMSE value was .83 compared to a benchmark RMSE of .93 and a best MAE value of .64 compared to a benchmark MAE of .68. For SVD with GS there was a consistent decrease of both RMSE and MAE as the size of the ratings datasets increased.

The fourth model, *Restricted Boltzmann Machine (RBM)* was also executed in two variations, one with grid search (GS) hyperparameter tuning and one without. RBM without GS included 6 replications of RBM with execution times between 1431 and 6647 seconds. A series of input user ratings datasets were processed with 100K, 200K, 250K, 300K, 400K, and 1M rows



of user movie ratings drawn from the MovieLens 20M ratings dataset. The RBM was configured with visible dimensions equal to the number of ratings levels times the number of movies being rated, 50 hidden dimensions, 20 processing epochs, learning rate of .1 or .001, and a batch size of 100. Values of RMSE for this RBM without GS with the 100K Movielens dataset was 1.33 and MAE of 1.13 for 100K Movielens. For the 1M Movielens dataset the RBM RMSE value was 1.38 and a MAE value of 1.18. A comparative RBM benchmark from the original Boltzmann Paper (Salakhutdinov, et al., 2007) was an RMSE of .91 for their RBM with 20 training epochs, also using a 100K ratings dataset. For RBM there was a curvilinear relation between values of both RMSE and MAE as the size of the ratings datasets increased with the best RMSE and MAE values obtained for intermediate-sized ratings datasets.

Values of RMSE of this RBM with GS for the 100K Movielens dataset was 1.32 and MAE of 1.13. For the 1M Movielens dataset the RBM RMSE value was 1.39 and a MAE value of 1.18. For RBM with GS there was also a curvilinear relation between values of both RMSE and MAE as the size of the ratings datasets increased with the best RMSE and MAE values obtained for intermediate-sized ratings datasets.

The fifth model, *Autoencoder (AutoRec)* was also executed with a single model variation. The AutoRec included 9 replications with execution times between 2550 and 14,120 seconds. A series of input user ratings datasets were processed with 100K, 200K, 250K, 300K, 400K, and 1M rows of user movie ratings drawn from the MovieLens 20M ratings dataset. The Autorec was configured with visible dimensions equal to the number of ratings levels times the number of movies being rated; 20, 50, or 100 hidden dimensions; 20, 50, 100 or 200 training epochs; learning rate of .1 or .01; and a batch size of 200. Values of RMSE for this AutoRec with the 100K Movielens dataset, 200 training epochs and 100 hidden dimensions was 2.03 and MAE of 1.67. For the 1M Movielens dataset the Autorec with 100 training epochs and 100 hidden



dimensions, the RMSE value was 2.08 and a MAE value of 1.71. A comparative AutoRec

benchmark (Sedhain, 2015) on the 1M dataset was an RMSE of .87 for their model with 500

hidden dimensions. For the Autorec there was an inconsistent relation between values of both

RMSE and MAE as the size of the ratings datasets increased with the best RMSE and MAE

values obtained for the model with 200 training epochs and 100 hidden dimensions.

These investigations replicated much of the large body of previous work with

conventional recommender models using KNN User-user Collaborative Filtering, Matrix

Factorization, and Singular value decomposition. These methods and related variations are

among the most widely used recommender algorithms for commercial production applications.

Much recent work has investigated deep learning models for recommender algorithms.

Replication of recommender results with deep learning models has been more difficult and in

many cases findings in the published literature fail to show deep learning models exceeding the

predictive results of non-DL models. Deep learning algorithms are more likely to be individually

"crafted" with tuning of hyperparameters and often will not generalize well to different datasets.

Explanation for the greater predictive error observed with the deep learning models in this

investigation was offered in terms of model overfitting and the bias-variance tradeoff. Altering of

model regularization parameters and further tuning of the deep learning model configurations

may improve performance of the models.



# References


Adagun, O. & Kosko, B. (2020). Bidirectional backpropagation. *IEEE Transactions on Systems, Man, and Cybernetics: Systems*. *50*(5).

Adomavicius, G. & Tuzhilin, A. (2005). Toward the next generation of recommender systems: A survey of the state of the art and possible extensions. *IEEE Transactions on Knowledge and Data Engineering*, *17*(6), pp. 734-749.

Adomavicius, G. & Zhang, J. (2010). On the stability of recommendation algorithms. In *RecSys '10*, pp. 47-54. Association for Computing Machinery.

Adomavicius, G. & Zhang, J. (2014). Improving stability of recommender systems: A meta-algorithmic approach. *IEEE Transactions on Knowledge and Data Engineering.*

Aslanian, E., Radmanesh, M. & Jalili, M. (2016). Hybrid recommender systems based on content feature relationship. IEEE Transactions

Bengio, Y. (2009). Learning deep architectures for AI. *Foundational Trends in Machine Learning*. *2*(1), pp. 1-127.

Batmaz, Z., Yurekll, A., Bilge, A. & Kaleli, C. (2019). A review on deep learning for recommender systems: Challenges and remedies. *Artificial Intelligence Review*, *52*, pp. 1-37.

Behnel, S., Bradshaw, R., Citro, C., Dalcin, L., Seljebotn, D. & Smith, K. (2011). Cython: The best of both worlds. *Computing in Science and Engineering*, *13*(2), pp. 31-39.

Belkin, M., Hsu, D., Ma, S. & Mandal, S. (2019). Reconciling modern machine-learning practice and the classical bias-variance trade off. *Proceedings of the National Academy of Sciences*, *116*(32), pp. 15849-15854.

Breese, J., Heckerman, D., & Kadie, C. (1998). Empirical Analysis of Predictive Algorithms for Collaborative Filtering. In *Proceedings of the 14$^{th}$ Conference on Uncertainty in Artificial Intelligence*, pp. 43-52.

Breiman, L. (1996). Bagging predictors. *Machine Learning*, *24*(2), pp. 123-140.

Burke, R. (2002). Hybrid recommender systems: Survey and experiments. *User Modeling and User-Adopted Interaction*, *12*(4), pp. 331-370

Cai, X., Han, J. and Yang, I. (2017). Generative adversarial network based heterogeneous bibliographic network representation for personalized citation recommendation.

Cao, S., Yang, N. & Liu, Z. (2017). Online news recommender based on stacked auto-encoder. In *Proceedings of the ICIS*. 721-726.





Chen, H., Chiang, R.H. & Storey, V. (2012). Business intelligence and analytics: From big data to big impact. *MIS Quarterly*, *36*(4), pp. 1165-1188.

Chen, J., Zhang, H., He, X., Nie, L., Liu, W. and Chua, T. (2017). Attentive collaborative filtering: Multimedia recommendations with item and component-level attention. In *Proceedings of the SIGIR*. ACM.

Cheng, H., Koe, L., Harmsen, J., Shaked, T., Chandra, T., Aradhye, H., Anderson, G., Corrado, G., Chai, W., Ispir, M. & others. (2016). Wide and deep learning for recommender systems. In *Proceedings of the RecSys*, pp. 7-10.

Coba, L., Symeonidis, P. & Zanker, M. (2018). Replicating and improving Top-N recommendations in open source packages. In *WIMS '18: 8ᵗʰ Annual International Conference on Web Intelligence, Mining, and Semantics*. ACM

Ciresan, D., Meir, U., Gambardella, L. and Sehmidhuber, J. (2010). Deep, big, simple neural nets for handwritten digit recognition. *Neural Computation*, *22*(12), pp. 3207-3220.

Dacrema, M., Cremonesi, P & Jannach, D. (2019). Are we really making much progress? A worrying analysis of recent neural recommendation approaches. In *Thirteenth ACM Conference on Recommender Systems (Recsys '19)*. ACM, New York, NY.

Da'u, A. & Salim, N. (2020). Recommendation system based on deep learning methods: a systematic review and new directions. *Artificial Intelligence Review*, *53*, pp. 2709-2748.

Deng, L. (2014a). A tutorial survey of architectures, algorithms, and applications for deep learning. *APSIPA Transactions on Signal and Information Processing Vol. 3*.

Deng. L. & Yu, D. (2014b). Deep learning methods and applications. *Foundations and Trends in Signal Processing*, *7*(3-4), pp. 197-387.

Deng, S., Huang, L., Xu, G., Wu, X. and Wu, Z. (2017). On deep learning for trust-aware recommendations in social networks. *IEEE Transactions on Neural Network Learning Systems, 28*(5), pp. 1164-1177.

Donahue, J., Hendricks, A., Guadarrama, S., Rohrbach, M., Venugopalan, S., Saenko, K. & Darrell, T. (2015). Long-term recurrent convolutional networks for visual recognition and description. In: *Proceedings of the 28ᵗʰ IEEE conference on computer vision and pattern recognition*. Boston, MA, USA, pp. 2625-2634.

Du, Y., Yao, C., Huo, S. and Liu, J. (2017). A new item-based deep network structure using a restricted Boltzmann machine for collaborative filtering. *Technology Electronic Engineering, 18*(5), pp. 658-666.

Ebesu, T. & Fang, Y. (2017). Neural semantic personalized ranking for item cold-start recommendation. *Information Retrieval Journal*, *22*(18), pp. 233-239.





Ekstrand, M. D. (2019). *The LKPY Package for Recommender Systems Experiments*. Retrieved from https://arxiv.org/abs/1809.

Ekstrand, M. D. & Riedl, J.T. (2012). When recommenders fail: Predicting recommender failure for algorithm selection and combination. In *RecSys '12*, pp. 233-236. Association for Computing Machinery.

Fahlman, S., Hinton, G. and Sejnowski, T. (1983). Massively parallel architectures for AI: NETL, thistle, and Boltzmann machines. In *Proceedings of the National Conference on Artificial Intelligence AAAI-83.*

Fan, W., Ma, Y., Yin, D., Wang, J., Tang, J. & Li, Q. (2019). Deep social collaborative filtering. In *proceedings of Recsys '19 September, 2019*, Copenhagen, Denmark.

Fortmann-Roe, S. (2012). The bias-variance tradeoff. Retrieved from https://scholar.google.com/citations?user=Dt0YMpQAAAAJ&hl=en&oi=sra, October, 2012.

Fouss, F. & Sacrens, M. (2008). Evaluating performance of recommender systems: An experimental comparison. In *Proceedings of the 2008 IEEE/WIC/ACM international Conference on Web Intelligence and Intelligent Agent Technology.*

Funk, S. (2006). Try this at home. Retrieved from http://sifter.org/~simon/journal/20061211.html.

Gedikli, F. (2013). *Recommender Systems and the Social Web*. Wiesbaden, Germany, Springer Verlag.

Gedikli, F., Jannach, D., & Ge, M. Z. (2014). How should I explain? A comparison of different explanation types for recommender systems. *International Journal of Human-Computer Studies, 72*(4), 367-382. doi: 10.1016/j.ijhcs.2013.12.007

Gelfend, A. & Smith, A. (1990). Sampling-based approaches to calculating marginal densities. *Journal of the American Statistical Association*, vol. *85*, pp. 398-409.

Georgiev, K. & Nakov, P. (2013). A non-iid framework for collaborative filtering with restricted Boltzmann Machines. In: *Proceedings of the 30$^{th}$ international conference on machine learning*, pp. 1148-1156.

Gong, Y. and Zhang, Q. (2016). Hashtag recommendation using attention-based convolutional neural network. In *Proceedings of the IJCAI*, pp. 2782-2788.

Goodfellow, I., Pouget-Abadie, J., Mirza, M., Xu, B., Warde-Farley, D., Ozair, S., Courville, A. & Bengio, Y. (2014). Generative adversarial networks. In: *Proceedings of the International Conference on Neural Information Processing Systems (NIPS 2014)*. pp. 2672–2680.

Goodfellow, I., Bengio, Y. & Courville, A. (2016). *Deep Learning*. MIT Press, Cambridge, MA.




Green, D. & Swets, J. (1966). *Signal Detection Theory and Psychophysics*. New York: Wiley.

Gunawardana, A. & Shani, G. (2009). A survey of accuracy evaluation metrics of recommendation tasks. *Journal of Machine Learning Research*, *10*, pp. 2935-2962.

Hand, D. & Till, R. (2001). A simple generalisation of the area under the ROC curve for multiple class classification problems. *Machine Learning*, *45*(2), pp. 171-186.

Hastie, T., Tibshirani, R. & Friedman, J. (2009). *The Elements of Statistical Learning: Data Mining, Inference, and Prediction.* Springer, New York, NY.

He, X., Liao, L., Zhang, H., Nie, L., Hu, X. & Chua, T. (2017). Neural collaborative filtering. In *Proceedings WWW '17*, pp. 173-182.

He, X. & Tat-Seng, C. (2017). Neural factorization machines for sparse predictive analytics. In *Proceedings of the SIGIR*, ACM, pp. 255-364.

He, X., He, Z., Du, X. and Chua, T. (2018). Adversarial personalized ranking for recommendations. In *Proceedings of the SIGIR 2018*, pp. 355-364.

Hebb, D. (1949). *The Organization of Behavior*. New York: Wiley.

Herlocker, J., Konstan, J. & Reidl, J. (2002). An empirical analysis of design choices in neighborhood-based collaborative filter algorithms. *Information Retrieval*, *5*(4), pp. 287-310.

Herlocker, J., Konstan, J., Terveen, L. & Riedl, J. (2004). Evaluating collaborative filtering recommender systems. *ACM Transactions on Information Systems*, *22*(1), pp. 5-53.

Hidasi, B., Karatzoglou, A., Baltrunas, L. and Tikk, D. (2015). Session-based recommendations with recurrent neural networks. In *Proceedings of the International Conference on Learning Representations.*

Hinton, G. & Sejnowski, T. (1986). Learning and relearning in Boltzmann Machines. In *Parallel Distributed Processing: Explorations in the Microstructures of Cognition*. Vol. 1. MIT Press, Cambridge, MA.

Hinton, G., Osindero, S. & Teh, Y. (2006). A fast learning algorithm for deep belief nets. *Neural Computation. 18* (7).

Hinton, G. (2009). Deep belief networks. *Scholarpedia. 4 (5):* 5947.

Hornik, K., Stinchcombe, M. & White, H. (1989). Multilayer feedforward networks are universal approximators. *Neural Networks*, *2*, pp. 359-366.

Hu, L., Cao, J., Xu, G., Cao, L., Gu Z. & Cao, W. (2014). Deep modeling of group preferences for group-based recommendation. In: *Proceedings of the 28th AAAI conference on artificial intelligence*. Quebec City, Quebec, Canada, pp.1861-1867.

Hu, B., Shi, C., Zhao, W.X., & Yu, P.S. (2018). Leveraging meta-path based context for top-n recommendation with a neural co-attention model. In *Proceedings KDD '18*. pp. 1531-1540.



Hu, Y., Koren, Y. & Volinsky, C. (2008). Collaborative filtering for implicit feedback datasets. *8th IEEE International Conference on Data Mining*. Pisa, Italy.

Hug, N. (2020). Surprise: A Python library for recommender systems. *Journal of Open Source Software*, *5*(52).

Ioannidis, J. (2005). Why most published research findings are false. *PLoS Med, 2*(8).

Isinkaye, F., Folajimi, Y. & Ojokoh, B. (2015). Recommendation systems: Principles, methods and evaluation. *Egyptian Information Journal*, *16*(3), pp. 261-273.

Jannach, D., Lerche, L., Gedikli, F., & Bonnin, G. (2013). What recommenders recommend – An analysis of accuracy, popularity, and sales diversity effects: User modeling, adaptation, and personalization. *Lecture Notes in Computer Science.* 7899. Springer, Berlin, Heidelberg. Pp. 25-37.

Jannach, D. & Ludewig, M. (2017). When recurrent neural networks meet the neighborhood for session-based recommendation. In *Proceedings of the RecSys (RecSys'17)*. ACM, Como, pp. 306-310.

Jannach, D., Moreira, G. & Oldridge, E. (2020). Why are deep learning models not consistently winning recommender systems competitions yet? *RecSys Challenge '20, September 26, 2020, Virtual Event, Brazil*. ACM.

Jannach, D., Zanker, M., Felfernig, A. & Friedrich, G. (2011). *Recommender systems: An introduction*. Cambridge, Cambridge University Press.

Jouppi, N., Young, C., Patil, N & Patterson, D.  (2018). Motivation for and evaluation of the first tensor processing unit. *IEEE Micro, 38* (3).

Kane, F. (2018). *Building Recommender Systems*. Sundog Software LLC.

Khalid, S., Khalil, T. and Nasreen, S. (2014). A Survey of Feature Selection and Feature Extraction Techniques in Machine Learning. *Science and Information Conference 2014*, August 27-29, London, UK

Kim, D., Park, C., Oh, J., Lee, S. & Yu, H. (2016). Convolutional matrix factorization for document context-aware recommendation. *Proceedings of the 10th ACM Conference on Recommender Systems – RecSys'16*, pp. 233-240.

Kluver, D., Ekstrand, M. & Konstan, A. (2018). Rating-Based Collaborative Filtering: Algorithms and Evaluation. In *Social Information Access*. Springer, Cham, 344–390. https://doi.org/10.1007/978-3-319-90092-6_10

Kotkov, D., Wang, S. & Veijalainen, J. (2016). A survey of serendipity in recommender systems. *Knowledge-based Systems, 111*, pp. 180-192.

Koren, Y. (2010). Factor in the neighbors: Scalable and accurate collaborative filtering. *ACM Transactions on Knowledge Discovery*, *4*(1), doi: 10.1145/1644873.1644874

Koren, Y., Bell, R. & Volinsky, C. (2009). Matrix factorization techniques for recommender systems. *IEEE Computer*, *42*(8), pp. 30-37.



Krishnamurthy, B., Puri, N. & Goel, R. (2016). Learning vector-space representations of items for recommendations using word embedding models. *Proceedings of Computer Science*, *80*, pp. 2205-2210.

Kumar, V., Kbattar, D., Gupta, S. and Varma, V. (2017). Deep neural architecture for news recommendation. In: *Working notes of CLEF 2017 conference and labs of the Evaluation Forum*. Dublin, Ireland.

Kunaver, M. & Pozrl, T. (2017). Diversity in recommender systems – a survey. *Knowledge-based Systems, 123*: 154-162.

Le Roux, N. and Bengio, Y. (2008). Representational power of restricted Boltzmann machines and deep belief networks. *Neural Computation, 20*(6), pp.1631-1649.

Lee, J., Abu-El-Haija, S., Varadarajan, B. and Natsev, A. (2018). Collaborative deep metric learning for video understanding. In *Proceedings of the KDD'18*. ACM, pp. 481-490.

Li, Xiaopeng & She, James. (2017). Collaborative variational autoencoder for recommender systems. In *Proceedings KDD '17*, pp. 305-314.

Li, Y., Liu, T., Jiang, J. and Zhang, L. (2018). Hashtag recommendation with topical attention-based LSTM. In *Proceedings of the COLING*, pp. 943-952.

Lin, J. (2018). The neural hype and comparisons against weak baselines. *ACM SIGIR Forum*, *52*, pp. 40-51.

Liu, X., Ouyang, Y., Rong, W. and Xiong, Z. (2015). Item category aware conditional restricged Boltzmann based recommendation. ICONIP

Liu, Q. & Wang, J. (2011). A one-layer dual recurrent neural network with a Heaviside step activation function for linear programming with its linear assignment application. *Artificial Neural Networks and Machine Learning – ICANN 2011 Pt II*. Lecture Notes in Computer Science, Vol 6792.

Lu, J., Wu, D., Mao, M., Wang, W. & Zhang, G. (2015). Recommender system application developments: A survey. *Decision Support Systems*.

Ludewig, M., Mauto, N., Latifi, S. & Jannach, D. (2019). Performance comparison of neural and non-neural approaches to session-based recommendation. In *Thirteenth ACM Conference on Recommender Systems (RecSys '19)*, ACM, New York.

Macsassy, A. & Provost, F. (2007). Classification in networked data: A toolkit and univariate case study. *Journal of Machine Learning Research*, *8*, pp. 935-983.

Mao, X., Li., Q., Xie, H., Lau, R., Wang, Z. & Smolley, S. (2017). Least Squares Generative Adversarial Networks. *The IEEE International Conference on Computer Vision (ICCV)*, pp. 2794-2802

McCulloch, W. & Pitts, W. (1943). A logical calculus of the ideas immanent in nervous activity. *Bulletin of Mathematical Biophysics*, *5*, 155-133.




McLaughlin & Herlocher (2004). A collaborative filtering algorithm and evaluation metric that accurately model user experience. In *Proceedings of the 27th annual international ACM SIGIR Conference on Research and Development in Information Retrieval. SIGIR '04* ACM.

Melville, P., Mooney, R. and Nagarajan, R. (2002). Content-boosted collaborative filtering for improved recommendations. In *Proceedings of the 18th National Conference on Artificial Intelligence (AAAI '02)*, pp. 187-192. Edmonton, Canada.

Minisky, M. & Pappert, S. (1969). Perceptrons: An introduction to computational geometry. *Science*, *165*(3895).

Neal, B., Sarthak, M., Baratin, A., Tantia, V., Scicluna, M., Lacoste-Julien, S. and Mitliagkas, I. (2019). A modern take on the bias-variance tradeoff in neural networks. *arXiv*:1810.08591v4

Neville, J. & Jensen, D. (2000). Iterative classification in relational data. *Workshop on Learning Statistical Models from Relational Data*, pp. 42-49.

Newell, A., Shaw, J. C., & Simon, H. A. (1958). Elements of a theory of human problem solving. *Psychological Review*, *65*, 151-166.

Noffsinger, W.B. (1989). A functional, contingency-based approach to machine learning. In *Proceedings of the Second Annual Florida Artificial Intelligence Research Symposium (FLAIRS)*, April, 1989, Orlando, Florida.

Paradarami, T., Bastian, N. & Wightman, H. (2017). A hybrid recommender system using artificial neural networks. *Expert Systems Applications*, *83*, pp. 300-313.

Park, D. H., Kim, H. K., Choi, I. Y., & Kim, J. K. (2012). A literature review and classification of recommender systems research. *Expert Systems with Applications, 39*(11), 10059-10072. doi: 10.1016/j.eswa.2012.02.038

Paterek, A. (2007). Improving regularized singular value decomposition for collaborative filtering. In *KDD Cup and Workshop 2007*.

Pavlov, D. & Pennock, D. (2002). A maximum entropy approach to collaborative filtering in dynamic, sparse, high-dimensional domains. In *Advances in Neural Information Processing Systems*, pp. 1441-1448. MIT Press, Cambridge, USA.

Pennock, D., Horvitz, E., Lawrence, S. & Giles, C. (2000). Collaborative filtering by personality diagnosis: a Hybrid memory and model-based approach. In *Proceedings of the 16th Conference on Uncertainty in Artificial Intelligence (UAI '00)*, pp. 473-480.

Pfeiffer, T. & Hoffmann, R. (2009). Large-scale assessment of the effect of popularity on the reliability of research. *PLoS One, 4*(6).




Polatidis N., Kapetanakis S., Pimenidis E. & Kosmidis K. (2018). Reproducibility of
        experiments in recommender systems evaluation. In: Iliadis L., Maglogiannis I.,
        Plagianakos V. (Eds) *Artificial Intelligence Applications and Innovations. AIAI 2018.
        IFIP Advances in Information and Communication Technology*, vol *519*. Springer.

Resnik, P., Iacovou, N., Suchak, M., Bergstrom, P. & Riedl, J. (1994). GroupLens: An open
        architecture for collaborative filtering of netnews. In ACM CSCW '94, pp. 175-186.
        Association for Computing Machinery.

Ricci, F., Rokach, L., Shapira, B. & Kantor, P.B. (2015). *Recommender Systems Handbook*.
        Springer.

Rosenblatt, F. (1958). The perceptron: A probabilistic model for information storage and
        organization in the brain. *Psychological Review, 65*(6), 386-408.

Rumelhart, D., Hinton, G. & Williams, R. (1986). Learning representations by back-propagating
        errors. *Nature*, Vol. *323*, pp. 533-536.

Rumelhart, D. & McClelland, J. (1986). *Parallel Distributed Processing: Explorations in the
        Microstructure of Cognition. Vol 1: Foundations*. MIT Press: Cambridge, MA.

Salakhutdinov, R, Mnih, A. & Hinton, G. (2007). Restricted Boltzmann machines for
        collaborative filtering. In *Proceedings of the 24th International Conference on Machine
        Learning*, Covallis, OR.

Sarwar, B., Karypis, G., Konstan, J. & Riedl, J. (2000). Application of dimensionality reduction
        in recommender system: A case study. *Technical Report TR00-043*. University of
        Minnesota Computer Science and Engineering.

Sarwar, B., Karypis, G., Konstan, J. & Riedl, J. (2001). Item-based collaborative filtering
        recommendation algorithms. In *ACM WWW '01*, pp. 285-295. Association for Computing
        Machinery.

Schmidhuber, J. (2015). Deep learning in neural networks: an overview. *Neural Networks*, *61*,
        pp. 85-117.

Sedhain, S., Menon, A., Sanner, S. & Xie, L. (2015). AutoRec: Autoencoders meet collaborative
        filtering. *WWW 2015 Companion*. ACM, New York, NY.

Seo, S., Huang, J., Yang, H., and Liu, Y. (2017). Representation learning of users and items for
        review rating prediction using attention-based convolutional neural network. In
        *Proceedings of the MLRec*.

Shalom, O.S., Jannach, D., & Guy, L. (2019). First workshop on the impact of Recommender
        Systems at ACM RecSys2019. In *Thirteenth ACM Conference on Recommender Systems*,
        Copenhagen, Denmark. ACM, New York, NY, USA.
        https://doi.org/10.1145/3298689.3347060



Shani, G. & Gunawardana, A. (2015). Evaluating recommendation systems. In Ricci, F., Rokach, L., Shapira, B. & Kantor, P. *Recommender Systems Handbook*. Springer.

Skinner, B.F. (1938). *The Behavior of Organisms: An Experimental Analysis*. Oxford, England: Appleton-Century.

Su, X. & Khoshgoftaar, T. M. (2009). A survey of collaborative filtering techniques. *Advances in Artificial Intelligence*, *Vol.* 2009, article ID 421425. doi: 10.1155/2009/421425.

Tibshirani, R. (1995). Regression shrinkage and selection via the LASSO. *Journal of the Royal Statistical Society B*, *58*, pp. 267-288.

Tikhonov, A., Goncharsky, A., Stepanov, V. & Yagola, A. (1995). *Numerical Methods for the Solution of Ill-Posed Problems.* Netherlands: Springer Netherlands.

Uria, B., Murray, I., & Larochelle, H. (2014). A deep and tractable density estimator. *Journal of Machine Learning Research*, *32*(1), pp. 467–475.

Uria, B., Cote, M., Gregor, K., Murray, I. and Larochelle, H. (2016). Neural autoregressive distribution estimation. *Journal of Machine Learning Research*, *17*, pp. 1-37.

Van den Oord, A., Dieleman, S., & Schrauwen, B. (2013). Deep content based music recommendation. In *Proceedings of the NIPS*. Curran Associates, Inc., pp. 2643-2651.

Vincent, P., Larochelle, H., Lajote, I., Bengio, Y. & Manzagol, P. (2010). Stacked denoising autoencoders: Learning useful representations in a deep belief network with a local denoising criterion. *Journal of Machine Learning Research*, *11*, pp. 3371-3408.

Wang, J. & Kowagoe, K. (2018). A recommender system for ancient books, pamphlets and paintings in the Ritsumeikan Art Research Center database. I*CCAE 2018: Proceedings of the 2018 10th International Conference on Computer and Automation Engineering*, pp. 53–57

Wang, H., Wang, N. & Yeung, D. (2015). Collaborative deep learning for recommender systems. In *Proceedings KDD '11*, pp. 1235-1244.

Wang, X. & Wang, Y. (2014). Improving content-based and hybrid music recommendation using deep learning. In: *Proceedings of MM'14*, ACM, pp. 627-636.

Wang, J., Yu, L., Zhang, W., Gong, Y., Xu, Y., Wang, B., Zhang, P. & Zhang, D. (2017). IRGAN: A Minimax Game for Unifying Generative and Discriminative Information Retrieval Models. *SIGIR '17: Proceedings of the 40th International ACM SIGIR Conference on Research and Development in Information Retrieval*, pp. 515–524



Wang, S., Wang, Y., Tang, J., Shu, K., Ranganath, S. and Liu, H. (2017). What your images reveal: Exploring visual contents for point-of-interest recommendation. In *Proceedings of the WWW 2017*, pp. 391-400.

Wang, Q., Lian, D. & Wang, H. (2018). Neural memory streaming recommender networks with adversarial training. KDD 2018, pp. 2467-2475.

Wu, Y., DuBois, C., Zheng, A., & Ester, M. (2016). Collaborative denoising autoencoders for top-n recommender systems. In *Proceedings of the 9$^{th}$ ACM international conference on Web search and data mining*, San Francisco, CA, USA, pp 153-162,

Yang, W., Lu, K., Yang, P. & Lin, J. (2019). Critically examining the "neural hype": Weak baselines and the additivity of effectiveness gains from neural ranking models. In *42$^{nd}$ Int'l ACM SIGIR Conference on Research and Development in Information Retrieval.* ACM, New York, NY.

Zhang, S., Yao, L. & Tay, Y. (2019). Deep learning recommender system: A survey and new perspectives. *ACM Computing Surveys. 52*(1), Article 5.

Zheng, L., Noroozi, V. & Yu, P. (2017). Joint deep modeling of users and items using reviews for recommendation. In *Proceedings of the WSDM, 2017*, pp. 425-434.

Zhao, X., Xia, L., Ding, Z., Yin, D. & Tang, J. (2018). Deep reinforcement learning for page-wise recommendations. *ArXiv preprint arXiv: 1802.02343*.

Zhao, X., Zhang, L., Ding, Z., Xia, L., Tang, J. and Yin, D. (2018). Recommendations with negative feedback via pair-wise deep reinforcement learning. *ArXiv preprint arXiv: 1802.06501*

Zhou, Y., Wilkinson, D., Schreiber, R. & Pan, R. (2008). Large-scale parallel collaborative filtering for the Netflix Prize. In *Algorithmic Aspects in Information and Management*. DOI: 10 1007/978-3-540-68880-8_32.